\documentclass[a4paper,11pt]{article}
\usepackage[margin=1in]{geometry}
\usepackage{epsfig}
\usepackage{cite}
\usepackage[flushleft]{threeparttable}
\usepackage{graphicx}
\DeclareGraphicsExtensions{%
    .png,.PNG,%
    .pdf,.PDF,%
    .jpg,.mps,.jpeg,.jbig2,.jb2,.JPG,.JPEG,.JBIG2,.JB2}
\usepackage{epstopdf}
\usepackage{color}
\UseRawInputEncoding
\usepackage[nottoc]{tocbibind}
\usepackage[english]{babel}
\usepackage{tabularx}
\usepackage{xparse}
\usepackage{braket}
\usepackage{amsmath}
\DeclareMathOperator{\sgn}{sgn}
\usepackage{mathrsfs}
\usepackage{hyperref}
\usepackage{grfext}
\usepackage{comment}
\usepackage{amssymb}
\usepackage{caption}
\usepackage{mathtools}
\usepackage{subcaption}
\usepackage{algorithm2e}
\usepackage{float}
\usepackage{subcaption}
\usepackage{multirow}
\restylefloat{table}
\usepackage{booktabs}
\usepackage[dvipsnames]{xcolor}
\usepackage{tikz}
\usepackage{pdflscape}
\usepackage{lscape}
\usepackage{tablefootnote}
\date{\today}
\def\unit{\leavevmode\hbox{\small1\kern-3.6pt\normalsize1}}

\def\gtwid{\mathrel{\raise.3ex\hbox{$>$\kern-.75em\lower1ex\hbox{$\sim$}}}}
\def\ltwid{\mathrel{\raise.3ex\hbox{$<$\kern-.75em\lower1ex\hbox{$\sim$}}}}
\def\gev{{\rm \, Ge\kern-0.125em V}}
\def\tev{{\rm \, Te\kern-0.125em V}}

\def    \be            {\begin{equation}}
\def    \ee            {\end{equation}}
\def    \bea           {\begin{eqnarray}}
\def    \eea           {\end{eqnarray}}

\def\a{\alpha}
\def\b{\beta}

\def\d{\delta}
\def\n{\nu}

\def\nn{\nonumber}
\def\d{\delta}
\def\D{\Delta}
\def\s{\sigma}
\def\r{\rho}
\def\t{\theta}

 \normalsize

 \normalsize

\newcommand{\bmat}{\left(\begin{array}}
\newcommand{\emat}{\end{array}\right)}

\begin{document}
\renewcommand{\thefootnote}{\fnsymbol{footnote}}
\vspace{.3cm}
\title{\Large\bf Texture of Two Vanishing Subtraces in Neutrino Mass Matrix and Current Experimental Tests} \author
{ \hspace{-2.cm} \it \bf A. Ismael$^{1,2}$\thanks{ahmedEhusien@sci.asu.edu.eg},  E. I. Lashin$^{1,2}$\thanks{slashin@zewailcity.edu.eg, elashin@ictp.it} and  N. Chamoun$^{3}$\thanks{nidal.chamoun@hiast.edu.sy},
 \\\hspace{-2.cm}
\footnotesize$^1$  Department of Physics, Faculty of Science, Ain Shams University, Cairo 11566,  Egypt.  \\\hspace{-2.cm}
\footnotesize$^2$ Centre for Fundamental Physics, Zewail City of Science and
Technology, 
 \footnotesize  6 October City, Giza 12578, Egypt.  \\\hspace{-2.cm}
\footnotesize$^3$ Physics Department, HIAST, P.O.Box 31983, Damascus, Syria.
}
\date{\today}
\maketitle
\begin{abstract}
We present a full phenomenological and analytical study for the neutrino mass matrix characterized by two vanishing $2\times2$ subtraces. We update one past result in light of the recent experimental data. Out of the fifteen possible textures, we find seven cases can accommodate the experimental data instead of eight ones in the past study.
We also introduce few symmetry realizations for viable and nonviable textures based on non-abelian ($A_4$ or $S_4$) flavor symmetry within type II seesaw scenario.
\end{abstract}
\maketitle
\vskip 0.3cm \hrule \vskip 0.5cm'
\section{Introduction}
The fact that neutrinos are massive was the first firm sign of physics beyond standard model. Many flavor models for neutrino mass matrix were conceived, motivated by phenomenological data on neutrino oscillations, and the examination of specific textures became a traditional approach to the flavor structure in the lepton sector.

Zero textures were studied extensively \cite{Frampton_2002, Xing_2002, Fritzsch_2011, Merle_2006}, but other forms of textures were equally studied, such as zero minors \cite{Lavoura_2005, Lashin_2008}, and partial $\mu-\tau$ symmetry textures \cite{Lashin_2014}. The main motivation for such a phenomenological approach is simplicity and
predictive power, especially when the texture, under study, has a small number of free parameters but, nonetheless, leads to simple relations and interesting predictions for observables. Most of the textures with only one constraint were found able to accommodate data, such as the one vanishing element (or minor or subtrace) textures \cite{Lashin_2012, Lashin_2009, PhysRevD.103.035020}. Textures with more constraints are more restrictive and many fail to be viable \cite{Xing_2002, Fritzsch_2011, Lavoura_2005, Lashin_2008, Dev_2013, Liu_2013, Dev_2020, Meloni_2014, Alcaide_2018}. One way to constrain the number of free parameters in the neutrino mass matrix is to work in the subspace of vanishing nonphysical phases, which has the additional benefit of simplifying the resulting formulae \cite{ismael2021texture}.

In \cite{Alhendi_2008}, a particular texture of vanishing two subtraces was studied. Therein, analytical expressions for the measurable neutrino observables were derived, and numerical analysis was done to show that $8$ patterns,  out of the $15$ independent ones, can  accommodate data. However, new bounds on experimental data have appeared since then, in particular the non-vanishing value of the smallest mixing angle \cite{daya}, and the objective of this work is just to re-examine the texture of vanishing two-subtraces in light of the new data. Moreover, we shall carry out a complete numerical analysis, where all the free parameters are scanned within their experimentally accepted ranges, in contrast to \cite{Alhendi_2008} where slices of the parameter space were chosen by picking up some admissible values for the mixing angles, some of which correspond to the now obsolete value $5^o$ for the smallest mixing angle, and scanning over the remaining few free parameters\footnote{The procedure in \cite{Alhendi_2008} consisted of fixing the solar mass squared difference, and $\theta_{13}$, to their best-fit experimental measurements, then picking up the value of $\delta$ which, for given ($\theta_{12}, \theta_{23}$) chosen within their allowed experimental ranges, would make $R_\n$ acceptable, noting the very sensitive dependence of $R_\n$ on $\delta$.}. Seven patterns are found to accommodate data, with one of them allowing for two types of hierarchies, another one accepting normal hierarchy (NH) alone, and five textures accommodating only inverted hierarchy (IH) type.

Furthermore, we present in this work a strategy for justifying the textures viability/unviability by studying the correlations, by which we mean certain formulae involving the correlated observables under study, in the `unphysical' subspace of vanishing solar mass squared difference $\d m^2$. Actually, even if experimental data preclude a vanishing ratio of solar to atmospheric squared mass differences $R_\nu$, however its value is of order $10^{-2}$, small enough to approximate the full true correlations by those resulting from imposing a vanishing $R_\nu$ where the analytical expressions become easier to handle and would allow to test directly the viability. Actually, we shall distinguish between various correlations of different precision levels. The utmost precision level corresponds to the full numeric correlation, possibly written as an expansion showing an approximate truncation term, with no approximations to $\d m^2$ at all. Then comes the intermediate precision level correlation which originates from equalling to zero the exact expression of $\d m^2$, and it can be presented in a truncated form as well. The least precision level correlation, which we shall consider in our work, results when putting equal to zero an approximate truncated part of the $\d m^2$ exact expression. We could see that the resulting correlations are in many cases very similar to the true non-approximate correlations, which are the ones depicted in the presented figures.

We motivate our study as follows: First, the two zero textures can be seen as two $1\times1$ vanishing subtrace. Therefore, the two vanishing $2\times2$ subtrace textures can be considered as a nontrivial generalization of the two zero textures. Second, like the two zero textures, the two vanishing trace conditions put four real constraints on the neutrino mass matrix, which leave only five free parameters. Third, our model is very predictive concerning the CP-violating phases. There exist strong restrictions on them at all $\sigma$-levels with either hierarchy types for all cases. Fourth, one can look at the two vanishing subtraces texture, comprising 15 cases,  as a special case of the broader class characterized by two anti-equalities between elements, which would contain 105 cases. Actually, the authors of \cite{Dev_2013} have studied the class of 65 textures characterized by two equalities, with no condition on the nonphysical phases, so some of the studied textures, there, are equivalent physically to the case where some equalities are replaced by anti-equalities. However, in our study, we shall study neutrino patterns whose equivalent matrices with vanishing nonphysical phases have the form of two vanishing subtraces, providing thus more constraints on the studied texture, and the study can not be considered related to that of \cite{Dev_2013}.

The simple results one obtains are suggestive of some nontrivial symmetries or other underlying dynamics. While abelian symmetries are simple and were used abundantly within type I and type II seesaw scenarios (e.g. see \cite{PhysRevD.103.035020,ismael2021texture} and references therein), non-Abelian discrete symmetries are considered a far richer and more interesting choice
for the flavor sector. Model builders have tried to derive experimental values of
quark/lepton masses and mixing angles by assuming non-Abelian discrete flavor
symmetries of quarks and leptons. In particular, lepton mixing has been intensively
discussed in the context of non-Abelian discrete flavor symmetries, as seen, e.g., in
the reviews \cite{Altarelli_2010, Ishimori_2010}.
We present two examples of non-abelian symmetries within type II seesaw scenario, the first one based on the alternating $A_4$ group (even permutations of four objects) leading to a texture where the two vanishing traces in question lie on the diagonal, then present a second example based on the symmetric group $S_4$ (permutations of four objects) where the two traces in question do not lie on the diagonal.

The plan of the paper is as follows. We present the notations in section 2, followed in section 3 by the texture definition. In section 4, we present the simple viability check strategy based on imposing a vanishing solar mass squared difference and studying the consequent correlations. In section 5, we present the numerical analysis of all the seven (out of fifteen) viable patterns, where for each one we present the analytical results and the correlation plots. In section 6, we present symmetry realizations for two cases, and end up with conclusions and summary in section 7. In appendix A, we present the analysis of the eight failing patterns, whereas in appendix B we state the analytical formulae for the Majorana phases for the viable patterns. Appendix C is devoted to the group theory of $A_4$, whereas in the last appendix D, we present the group theory basics for ($S_n,n=1,\ldots,4$) useful to understand the corresponding realization.

\section{Notations}
In the `flavor' basis, where the charged lepton mass matrix is diagonal, and thus the observed neutrino mixing matrix comes solely from the neutrino sector, we have
\begin{equation}
V^{\dagger}M_{\nu}V^{*}= \left( \begin {array}{ccc} m_{1}&0&0\\ \noalign{\medskip}0&m_{2}&0
\\ \noalign{\medskip}0&0&m_{3}\end {array} \right),
\end{equation}
with ($m_{i}, i=1,2,3$) real and positive neutrino masses. We adopt the parameterization where the third column of $V$ is real, and work in the subspace of vanishing nonphysical phases. The lepton mixing matrix $V$ contains three mixing angles and three CP-violating phases. It can be written as a product of the Dirac mixing matrix U (consisting of three mixing angles and a Dirac phase) and a diagonal matrix P (consisting of two Majorana phases). Thus, we have
\bea
P^{\mbox{\tiny Maj.}} = \mbox{diag}\left(e^{i\rho},e^{i\sigma},1\right)\,, U \; = R_{23}\left(\t_{23}\right)\; R_{13}\left(\t_{13}\right)\; \mbox{diag}\left(1,e^{-i\d},1\right)\; R_{12}\left(\t_{12}\right)\,  \; ,\label{defOfU}\\
V = U\;P^{\mbox{\tiny Maj.}}  {\footnotesize = \left ( \begin{array}{ccc} c_{12}\, c_{13} e^{i\rho} & s_{12}\, c_{13} e^{i\sigma}& s_{13} \\ (- c_{12}\, s_{23}
\,s_{13} - s_{12}\, c_{23}\, e^{-i\delta}) e^{i\rho} & (- s_{12}\, s_{23}\, s_{13} + c_{12}\, c_{23}\, e^{-i\delta})e^{i\sigma}
& s_{23}\, c_{13}\, \\ (- c_{12}\, c_{23}\, s_{13} + s_{12}\, s_{23}\, e^{-i\delta})e^{i\rho} & (- s_{12}\, c_{23}\, s_{13}
- c_{12}\, s_{23}\, e^{-i\delta})e^{i\sigma} & c_{23}\, c_{13} \end{array} \right )},
\label{defv}
\eea
where $R_{ij}(\theta_{ij})$ is the rotation matrix through the mixing angle $\theta_{ij}$ in the ($i,j$)-plane, ($\delta,\rho,\sigma$) are three CP-violating phases, and we denote ($c_{12}\equiv \cos\theta_{12}...)$.

The neutrino mass spectrum is divided into two categories: Normal hierarchy ($\textbf{NH}$) where $m_{1}<m_{2}<m_{3}$, and Inverted hierarchy ($\textbf{IH}$) where $m_{3}<m_{1}<m_{2}$. The solar and atmospheric neutrino mass-squared differences, and their ratio $R_{\nu}$, are defined as follows.
\begin{equation}
\delta m^{2}\equiv m_{2}^{2}-m_{1}^{2},~~\Delta m^{2}\equiv\Big| m_{3}^{2}-\frac{1}{2}(m_{1}^{2}+m_{2}^{2})\Big|,     \;\;
R_{\nu}\equiv\frac{\delta m^{2}}{\Delta m^{2}}.\label{Deltadiff}
\end{equation}
with data indicating ($R_{\nu}\ll1$). Two parameters which put bounds on the neutrino mass scales, by the nuclear experiments on beta-decay kinematics and
neutrinoless double-beta decay, are the
effective electron-neutrino mass:
\begin{equation}
\langle
m\rangle_e \; = \; \sqrt{\sum_{i=1}^{3} \displaystyle \left (
|V_{e i}|^2 m^2_i \right )} \;\; ,
\end{equation}
and the effective Majorana mass term
$\langle m \rangle_{ee} $:
\begin{equation} \label{mee}
\langle m \rangle_{ee} \; = \; \left | m_1
V^2_{e1} + m_2 V^2_{e2} + m_3 V^2_{e3} \right | \; = \; \left | M_{\n 11} \right |.
\end{equation}
Cosmological observations put bounds on the  `sum'
parameter $\Sigma$:
\be
\Sigma = \sum_{i=1}^{3} m_i.
\ee
The last measurable quantity we shall consider is the Jarlskog rephasing invariant \cite{PhysRevLett.55.1039}, which measures CP-violation in the neutrino oscillations defined by:
\begin{equation}\label{jg}
J = s_{12}\,c_{12}\,s_{23}\, c_{23}\, s_{13}\,c_{13}^2 \sin{\delta}
\end{equation}

The allowed experimental ranges of the neutrino oscillation parameters at different $\sigma$ error levels as well as the best fit values are listed in Table(\ref{TableLisi:as}) \cite{de_Salas_2021}.
\begin{table}[h]
\centering
\scalebox{0.8}{
\begin{tabular}{cccccc}
\toprule
Parameter & Hierarchy & Best fit & $1 \sigma$ & $2 \sigma$ & $3 \sigma$ \\
\toprule
$\delta m^{2}$ $(10^{-5} \text{eV}^{2})$ & NH, IH & 7.50 & [7.30,7.72] & [7.12,7.93] & [6.94,8.14] \\
\midrule
 \multirow{2}{*}{$\Delta m^{2}$ $(10^{-3} \text{eV}^{2})$} & NH & 2.51 & [2.48,2.53] & [2.45,2.56] & [2.43,2.59] \\
 \cmidrule{2-6}
           & IH & 2.48 & [2.45,2.51] & [2.42,2.54] & [2.40,2.57]\\
\midrule
$\theta_{12}$ ($^{\circ}$) & NH, IH & 34.30 & [33.30,35.30] & [32.30,36.40] & [31.40,37.40] \\
\midrule
\multirow{2}{*}{$\theta_{13}$ ($^{\circ}$)}  & NH & 8.53 & [8.41,8.66] & [8.27,8.79] & [8.13,8.92] \\
\cmidrule{2-6}
    & IH & 8.58 & [8.44,8.70] & [8.30,8.83]& [8.17,8.96]\\
\midrule
\multirow{2}{*}{$\theta_{23}$ ($^{\circ}$)}  & NH & 49.26 & [48.47,50.05]& [47.37,50.71] & [41.20,51.33] \\
\cmidrule{2-6}
      & IH & 49.46 & [48.49,50.06]  & [47.35,50.67] & [41.16,51.25]   \\
\midrule
\multirow{2}{*}{$\delta$ ($^{\circ}$)}  & NH & 194.00 & [172.00,218.00] & [152.00,255.00] & [128.00,359.00] \\
\cmidrule{2-6}
 & IH & 284.00 & [256.00,310.00] & [226.00,332.00] & [200.00,353.00]   \\
\bottomrule
\end{tabular}}
\caption{\footnotesize The experimental bounds for the oscillation parameters at 1-2-3$\sigma$-levels, taken from the global fit to neutrino oscillation data \cite{de_Salas_2021} (the numerical values of $\Delta m^2$ are different from those in the reference which uses the definition $\Delta m^2 = \Big| \ m_3^2 - m_1^2 \Big|$ instead of Eq. \ref{Deltadiff}). Normal and Inverted Hierarchies are
respectively denoted by NH and IH}.
\label{TableLisi:as}
\end{table}

For the non-oscillation parameters, we adopt the upper limits, which are obtained by KATRIN and Gerda experiment for $m_{e}$ and $m_{ee}$ \cite{Aker_2019,Agostini_2019}\footnote{In \cite{Katrin2022}, a more recent lower upper bound of $0.9 \textrm{ eV}$ for $m_e$ was reported. However, the results of Tab. \ref{Predictions} show that adopting this newer limit will not change the conclusions and viability of the patterns.} . However, we adopt for $\Sigma$ the results of Planck 2018 \cite{Planck} from temperature information with low energy by using the simulator SimLOW.
\begin{equation}\label{non-osc-cons}
\begin{aligned}
\Sigma~~~~~&<0.54~\textrm{ eV},\\
 m_{ee}~~&<0.228~\textrm{ eV},\\
 m_{e}~~~&<1.1~\textrm{ eV}.
\end{aligned}
\end{equation}
Actually, for the parameter $\Sigma$, we did not opt to take neither the most constraining cosmological bound ($\Sigma<0.09~\textrm{ eV}$) \cite{Valentino_2021} using data from Supernovae Ia luminosity distances with Baryon Acoustic Oscillation observations and determinations of the growth rate parameter, nor the strict constraint of Planck 2018 combining baryon acoustic oscillation data in $\Lambda CDM$ cosmology, which amounts to ($\Sigma<0.12~\textrm{ eV}$) \cite{Planck}, nor the bound ($\Sigma<0.2~\textrm{ eV}$) stated in the PDG Live \footnote{https://pdglive.lbl.gov/DataBlock.action?node=S066MNS} originating from fits assuming various cosmological considerations, for the following reasons. First, since this constraint, as we shall see, proves to be quite severe ruling out many viable patterns, then by relaxing this `cosmological' constraint, we give more weight to colliders' data in testing our particle physics model. Second, we note that, even cosmology wise, this bound assumes some cosmological assumptions which are not anonymous. For example, in \cite{Chacko_2020} it was argued that adopting the assumption of decaying neutrino into invisible dark radiation, on time scales of the age of the universe, will alleviate the $\Sigma$ bound and push it higher up to ($\Sigma<0.9~\textrm{ eV}$). Having said this though, we shall comment on the effect of adopting the severe bounds, coming from cosmology, where at least some patterns remain viable.

For simplification and clarity purposes regarding the analytical formulae, we henceforth denote, in line with the notations of the past study \cite{Alhendi_2008}, the mixing angles as follows.
\begin{equation}
\theta_{12}\equiv\theta_{x},~~\theta_{23}\equiv\theta_{y},~~\theta_{13}\equiv\theta_{z}.\label{redefine}
\end{equation}
However, we shall keep the standard nomenclature in the tables and figures for rapid consultation purposes.

\section{Textures with two traceless submatrices}
The matrix $M_{\nu}$ is $3\times3$ complex symmetric matrix. Thus, it has 6 independent $2\times 2$ submatrices. When they are taken into pairs, we obtain 15 possibilities. Each location at the ($i,j$)-entry of the $3\times3$ symmetric neutrino mass matrix $M_{\nu}$ determines, by deleting the $i^{th}$-line and $j^{th}$-column, a $2\times2$ submatrix, denoted by ${\boldmath C}_{ij}$. We are considering the texture characterized by two traceless such submatrices, which are shown in Table \ref{Textures}.
\begin{table*}[h]
\begin{center}
\begin{threeparttable}
\begin{tabular}{ | m{4.5em} | c  | m{5.5cm} |c| }
\hline
\hspace{0.3cm}Texture& Symbol\tnote{\S}
and Viability in \cite{Alhendi_2008}  &  \hspace{+0.7cm}Independent constraints & Current Viability
 \\
 \hline
$(\textbf{C}_{33},\textbf{C}_{13})$& $D_1$ ,  \checkmark &$M_{ee}+M_{\mu\mu}=0,~M_{e\mu}+M_{\mu\tau}=0$ & {\bf IH}\\
\hline
$(\textbf{C}_{22},\textbf{C}_{33})$& $D_2$  ,  \checkmark& $M_{ee}+M_{\mu\mu}=0,~M_{ee}+M_{\tau\tau}=0$ & {\bf NH}, {\bf IH}\\
\hline
$(\textbf{C}_{11},\textbf{C}_{12})$& $D_3$ ,  \checkmark & $M_{\mu\mu}+M_{\tau\tau}=0,~M_{e\mu}+M_{\tau\tau}=0$ & {\bf IH} \\
\hline
$(\textbf{C}_{13},\textbf{C}_{23})$& $N_1$ ,  \checkmark & $M_{e\mu}+M_{\mu\tau}=0,~M_{ee}+M_{\mu\tau}=0$ & {\bf NH} \\
\hline
$(\textbf{C}_{13},\textbf{C}_{12})$& $N_2$, $\times$ & $M_{e\mu}+M_{\mu\tau}=0,~M_{e\mu}+M_{\tau\tau}=0$ & \\
\hline
$(\textbf{C}_{33},\textbf{C}_{23})$& $I_1$ , $\times$ &$M_{ee}+M_{\mu\mu}=0,~M_{ee}+M_{\mu\tau}=0$ & \\
\hline
$(\textbf{C}_{33},\textbf{C}_{12})$& $I_2$ ,  \checkmark & $M_{ee}+M_{\mu\mu}=0,~M_{e\mu}+M_{\tau\tau}=0$ & {\bf IH} \\
\hline
$(\textbf{C}_{13},\textbf{C}_{11})$& $I_3$ ,  \checkmark& $M_{e\mu}+M_{\mu\tau}=0,~M_{\mu\mu}+M_{\tau\tau}=0$ & {\bf IH}\\
\hline
$(\textbf{C}_{11},\textbf{C}_{23})$& $I_4$  ,  \checkmark & $M_{\mu\mu}+M_{\tau\tau}=0,~M_{ee}+M_{\mu\tau}=0$ &\\
\hline
$(\textbf{C}_{22},\textbf{C}_{23})$&  $I_5$  , $\times$ &$M_{ee}+M_{\tau\tau}=0,~M_{ee}+M_{\mu\tau}=0$ &\\
\hline
$(\textbf{C}_{33},\textbf{C}_{11})$& $I_6$ , $\times$ & $M_{ee}+M_{\mu\mu}=0,~M_{\mu\mu}+M_{\tau\tau}=0$ & \\
\hline
$(\textbf{C}_{22},\textbf{C}_{12})$& $I_7$ ,  \checkmark & $M_{ee}+M_{\tau\tau}=0,~M_{e\mu}+M_{\tau\tau}=0$ & {\bf IH} \\
\hline
$(\textbf{C}_{22},\textbf{C}_{11})$& $I_8$ , $\times$ & $M_{ee}+M_{\tau\tau}=0,~M_{\mu\mu}+M_{\tau\tau}=0$ & \\
\hline
$(\textbf{C}_{22},\textbf{C}_{13})$& no symbol, $\times$ & $M_{ee}+M_{\tau\tau}=0,~M_{e\mu}+M_{\mu\tau}=0$ & \\
\hline
$(\textbf{C}_{23},\textbf{C}_{12})$& no symbol, $\times$ & $M_{ee}+M_{\mu\tau}=0,~M_{e\mu}+M_{\tau\tau}=0$ &\\
\hline
\end{tabular}
\caption{The fifteen possible textures of two traceless submatrices. In the last column, we stated the current viability with the accommodated hierarchy type, to be contrasted with that of \cite{Alhendi_2008} in that $I_4$ ceases now to be allowed.}
\begin{tablenotes}
\item{\S}{\small
The symbols (D, N, I) corresponded, respectively, in \cite{Alhendi_2008} to (`degenerate' ($m_1 \sim m_2 \sim m_3$), normal, inverted) ordering type, for some candidate benchmark points taken in each pattern with `testable' $\t_x$ tuned to accommodate allowed values of ($R_\n, \theta_z, \theta_y$), whereas no such $\t_x$ was possible in the last two patterns.}
\end{tablenotes}
\label{Textures}
\end{threeparttable}
\end{center}
\end{table*}

The two vanishing-trace conditions are written as
\begin{align}
M_{\nu~ab}+M_{\nu~cd}=0,\nonumber\\
M_{\nu~ij}+M_{\nu~kl}=0.\label{Traceconds}
\end{align}
where $(ab)\neq(cd)$ and $(ij)\neq(kl)$. We write Eq. (\ref{Traceconds} in the terms of the V matrix elements as
\begin{align}
\sum_{m=1}^{3}&(U_{am}U_{bm}+U_{cm}U_{dm})\lambda_{m}=0,\nonumber\\
\sum_{m=1}^{3}&(U_{im}U_{jm}+U_{km}U_{lm})\lambda_{m}=0,\label{conds}
\end{align}
where
\begin{equation}
\lambda_1=m_1e^{2i\rho},~~\lambda_{2}=m_2e^{2i\sigma},~~\lambda_3=m_3.
\end{equation}
By writing Eq. \ref{conds} in a matrix form, we obtain
\begin{equation}
\left( \begin {array}{cc} A_1&A_2\\ \noalign{\medskip}B_1&B_2\end {array}
 \right)\left( \begin {array}{c} \frac{\lambda_1}{\lambda_3}\\ \noalign{\medskip}\frac{\lambda_2}{\lambda_3}\end {array} \right)=-\left( \begin {array}{c} A_3\\ \noalign{\medskip}B_3\end {array} \right),
\end{equation}
where
\begin{align}
A_m=&U_{am}U_{bm}+U_{cm}U_{dm},\nonumber\\
B_m=&U_{im}U_{jm}+U_{km}U_{lm},~~~~~m=1,2,3.\label{Coff}
\end{align}
By solving Eq. \ref{conds}, we obtain
\begin{align}
\frac{\lambda_1}{\lambda_3}=&\frac{A_3B_2-A_2B_3}{B_1A_2-A_1B_2},\nonumber\\
\frac{\lambda_2}{\lambda_3}=&\frac{A_1B_3-A_3B_1}{B_1A_2-A_1B_2}.
\end{align}
Therefore, we get the mass ratios and the Majorana phases in terms of the mixing angles and the Dirac phase
\begin{align}
m_{13} \equiv \frac{m_1}{m_3}=&\bigg|\frac{A_3B_2-A_2B_3}{B_1A_2-A_1B_2}\bigg|,\nonumber\\
m_{23} \equiv \frac{m_2}{m_3}=&\bigg|\frac{A_1B_3-A_3B_1}{B_1A_2-A_1B_2}\bigg|,\label{ratio}
\end{align}
and
\begin{align}
\rho=&\frac{1}{2}\mbox{arg}\bigg(\frac{A_3B_2-A_2B_3}{B_1A_2-A_1B_2}\bigg),\nonumber\\
\sigma=&\frac{1}{2}\mbox{arg}\bigg(\frac{A_1B_3-A_3B_1}{B_1A_2-A_1B_2}\bigg).
\end{align}
The neutrino masses are written as
\begin{equation}
m_3=\sqrt{\frac{\delta m^2}{m_{23}^2-m_{13}^2}},~~m_1=m_3\times m_{13},~~m_2=m_3\times m_{23}.\label{spectrum}
\end{equation}
As we see, we have five input parameters corresponding $(\theta_{x},\theta_{y},\theta_{z},\delta,\delta m^2)$, which together with four real constraints in Eq. \ref{conds} allows us to determine the nine degrees of freedom in $M_{\nu}$.

\section{$R_\n$-roots as a simple strategy for viability cheking}
In \cite{Alhendi_2008}, one noted the sensitivity of $R_\n$ to $\d$, in that imposing the `small' allowed values of $R_\n$ singled out two corresponding $\d$'s symmetric with respect to $\pi$. To see this, we note that the expression of the $U$'s involving $e^{i\d}$ (c.f. Eq. \ref{defOfU}) means that doing the transformation
\bea \label{symd}
\d \rightarrow 2\pi - \d ,
\eea
would correspond to complex conjugating the A's and B's, and so the ratios ($m_{13}, m_{23}$), and consequently $R_\n$, remain invariant as we have from eq. \ref{ratio}:
\bea
\label{Rnu}
R_\n &=& \frac{m_{23}^2 - m_{13}^2}{\left| 1-\frac{1}{2} (m_{13}^2 + m_{23}^2 ) \right|} \approx 10^{-2}.
\eea

Also, since $R_\n <<1$ is a very restrictive constraint on the allowed points, one can approximate the allowed parameter space with that corresponding to vanishing $R_\n$, i.e. to a zero for the $(m_2^2-m_1^2)$-expression. This means that any allowed point ($\t_x,\t_y,\t_z,\d,\d m$) would lie in the vicinity of the point ($\t_x,\t_y,\t_z,\d,\d m=0$). In our textures, $R_\n$, being a function of the A's and B's and thus of $U$'s, is a function of the angles ($\t_x, \t_y,\t_z$ and $\d$), so a root of $R_\n$, or equivalently of ($m_{23}^2-m_{13}^2$), would impose  functional relations between these angles corresponding to correlations quite approaching the real ones. We shall see that the zeros of ($m_{23}^2-m_{13}^2$) play a decisive role in determining the correlation between the mixing and Dirac phase angles, and that  would reflect on all other correlations depending on these angles, like those of $\rho$ and $\sigma$.

In fact, imposing, in the textures under discussion, a zero for ($m_{23}^2-m_{13}^2$) leads to determining $\d$ as a function of the angles $(\t_x, \t_y, \t_z)$. Taking into consideration that the range of variability for the allowed $\t_z$ is quite tight, we can fix it to its best fit value $\t_z \approx 8.5^o$, and obtain $\d$ as a function of $(\t_x, \t_y)$. Drawing two curves obtained by fixing $(\t_x)$ to its extreme values, one gets the approximative correlation region between $\d$ and $\t_y$ delimited by the former two curves. Exchanging the roles of $\t_x$ and $\t_y$ leads to the correlation $(\d, \t_x)$.

\begin{figure}[hbtp]
\hspace*{-3.5cm}
\includegraphics[width=22cm, height=16cm]{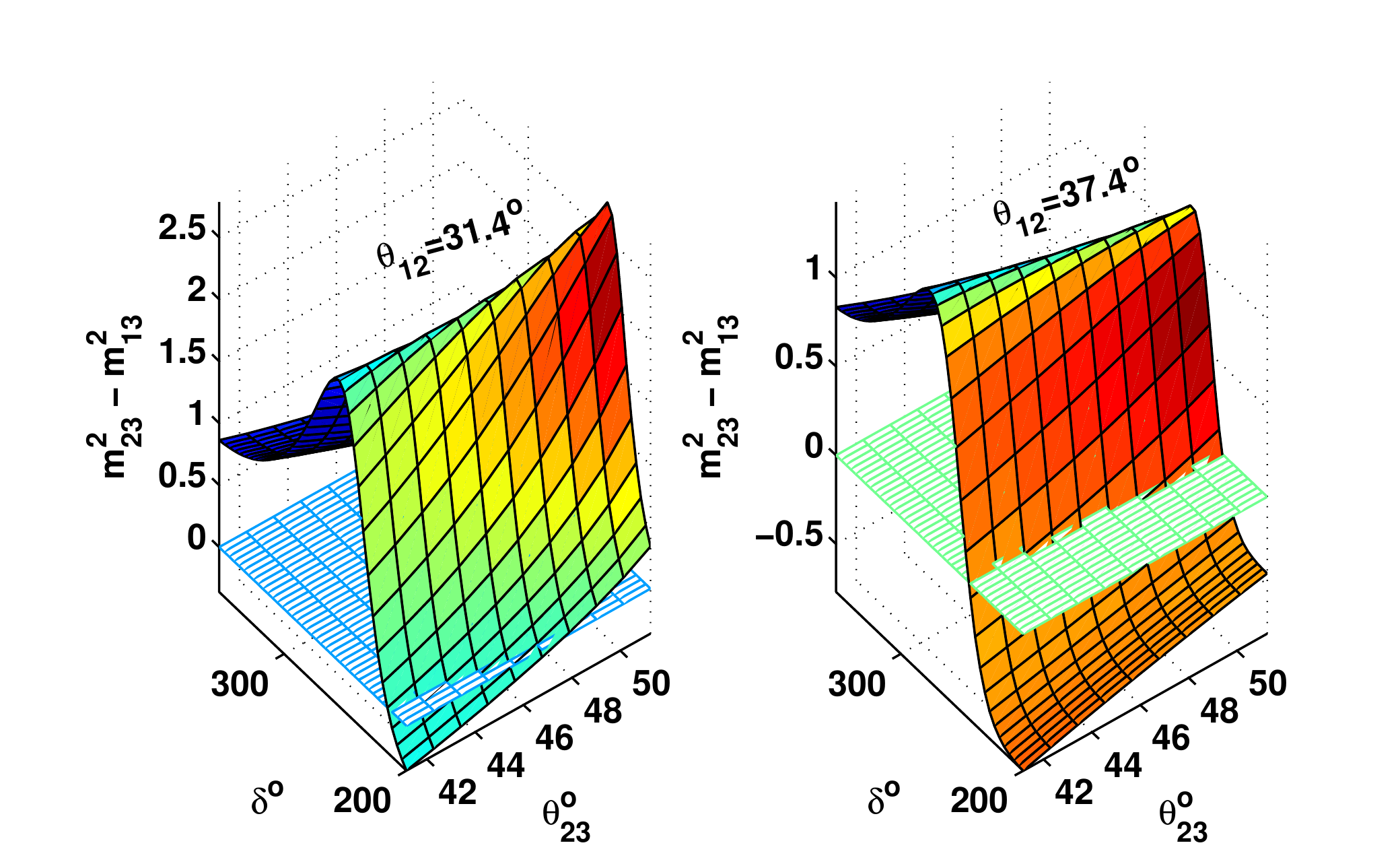}
\caption{Intersection of the zero surface ($m_1=m_2$) with that of ($m_{23}^2-m_{13}^2$) (as a $2$-dim surface in $\d, \t_y$ after fixing $\t_x)$ gives an approximate correlation ($\t_y,\d$), in the texture $(C_{22}, C_{12})$, whose delimiting curves correspond to extreme values for $\t_x$}.
\label{C22_C12-delta_theta_y}
\end{figure}

 We illustrate this in Fig. (\ref{C22_C12-delta_theta_y}) where in the left (right) part, for one pattern to be studied later, we take the minimum (maximum) allowed value of $\t_x=\t_{12}=31.4^o (37.4^o)$. Then, the surface of ($m_{23}^2-m_{13}^2$) as an expression in ($\d, \t_y=\t_{23}$) intersects with the surface ($m_{23}^2-m_{13}^2=0$) in a curve representing the correlation ($\d, \t_y$) for the considered $(\t_z, \t_x)$. Thus, the extreme corresponding `intersection' curves of ($\d, \t_y=\t_{23}$) delimit the corresponding correlation region.

Another remark applies here in that if the zeros of ($m_{23}^2-m_{13}^2$) imply ($m_{13} = 1$), then the corresponding pattern is failing and can not accommodate data. This comes because one can not here get the good order of magnitude for $|m_3^2 - \frac{1}{2}(m_1^2+m_2^2)| \equiv \D m^2 \approx 10^{-3}$, since, up to order  $10^{-5} \approx (m_2^2 - m_1^2)$, we have $m_3=m_1$ leading to $\D m^2= \frac{1}{2} (m_2^2-m_1^2) = O(10^{-5})$ which can not be amended to $10^{-3}$. We shall see that two patterns are failing due to this remark. In general, one can plug any expression resulting from imposing zeros of ($m_{23}^2 - m_{13}^2$) into the expression of $m_{13}$ to deduce the hierarchy type.

In practice, we should distinguish between various kinds of correlations at successive levels of precision. First, there are the ``full'' correlations, where no approximation was used, and all experimental constraints were taken into consideration in the numerical scanning. One can take successive terms up to a certain order in the expansion of these ``full'' correlations in powers of $s_z$ to get ``truncated full'' correlations. Second, come the approaching correlations resulting from equating the exact expression of the squared mass difference ($m_{23}^2 - m_{13}^2$) to zero, which we would call ``exact'' correlations. These correlations, formulae involving the observables, can in their turn be expanded to some order in powers of $s_z$ to get ``truncated exact'' correlations.  Third, the squared mass difference expression may form a complicated analytical expression of $(\t_x, \t_y,\t_z, \d)$, so one might resort to expanding this expression in increasing powers of $s_z$, and get ``approximate'' correlations resulting from putting equal to zero the expansion, up to a fixed order in $s_z$, of the ($m_{23}^2 - m_{13}^2$)-expression. We illustrate these different correlations in Table (\ref{correlations}), where the first column corresponds to the ``physical world'' with ($\d m^2\neq 0$), whereas the second column corresponds to the ``nonphysical world'' where ($\d m^2 =0$) leading to ``exact'' or ``approximate'' correlations. Concretely, for the measurable quantity $m_{13}$, suppose we have the following mathematical expressions (where, to fix the ideas, we assume the expansion is done up to ${\cal O}(s_z)$):
\bea
\label{explanation}
m_{13}=f=f_0+\mathcal{O}(s_z) &,& m_{23}^2-m_{13}^2 = g = g_0  +\mathcal{O}(s_z) ,\nn \\
 \left(g=0\Rightarrow m_{13}=h=h_0+\mathcal{O}(s_z)\right)
 &,&
 \left(g_0=0\Rightarrow m_{13}=k=k_0+\mathcal{O}(s_z)\right),
\eea
then we have ($k_0=h_0$) and the following meanings:
\bea
 \mbox{``Full'' means: } m_{13}=f &,&
\mbox{``Truncated Full'' means: } m_{13}=f_0 ,\nn \\
  \mbox{``Exact'' means: } m_{13}=h
&,&  \mbox{``Truncated Exact'' means: } m_{13}=h_0=k_0, \nn \\
  \mbox{``Approximate'' means: } m_{13}=k.
\eea

We checked that the ``exact'' correlations are very near the ``full'' ones in all patterns, whereas the ``approximate'' correlations may represent a non-negligible deviation unless one expands up to sufficiently high order in $s_z$.
\begin{table}[h]
\begin{center}
\begin{tabular}{  c | c  | c  }
\hline
\hspace{0.3cm}Observable expression & $\d m^2 \neq 0$  & $\d m^2 = 0$ \\
 using & (Physical World) & (Non-physical World)\\
 \hline
complete formula  &  ``Full'' & ``Exact''\\
\hline
expansion up to  & ``Truncated Full'' & ``Approximate''/\\
a certain order in $s_z$ &  & ``Truncated Exact'' \\
\hline
\end{tabular}
\caption{Various precision-level correlations.}
\label{correlations}
\end{center}
\end{table}

\section{Numerical results}
In this section, we introduce the numerical and analytical results for all the seven viable two-vanishing-subtraces cases, and present the corresponding correlations graphs (we shall justify the non-viability of the remaining eight cases in appendix (App. \ref{appFailing}). For each texture, we give the analytical expression for the coefficients  A's and B's of Eq. \ref{Coff}, and the leading expansion of the parameter $R_\nu$. Due to cumbersomeness, we do not present the expressions of the other observables ($m_{13}, m_{23}, \rho, \sigma, m_{ee}, m_e$), some of which appear in \cite{Lashin_2008}, but rather make use of the roots of the $\d m^2$-expression to find approximate formulae allowing to interpret their correlations. However, for completeness, we put in an appendix (App. \ref{appCP}) the leading orders of the Majorana phases ($\rho, \sigma$), whose detailed study helps in justifying their relatively tight extents, eventhough, as said above, we shall use the $R_\nu$-roots strategy in order to interpret, when possible, the general feature of the CP violation phases' narrow ranges.

As mentioned before, the free parameter space is fifth-dimensional corresponding, say, to the three mixing angles ($\theta_{x},\theta_{y},\theta_{z}$), the Dirac phase $\delta$, and the solar neutrino mass difference $\delta m^2$.
We throw N points of order $(10^7-10^{10})$ in the 5-dimensional parameter space ($\theta_{x},\theta_{y},\theta_{z},\delta,\delta m^2$), and check first the type of the mass hierarchy through the Eqs. (\ref{ratio},\ref{spectrum}). Second, we test the experimental bounds of $\Delta m^2$ besides those of Eq. \ref{non-osc-cons} in order to determine the experimentally allowed regions. We notice from Table (\ref{TableLisi:as}) that the experimental bounds of the neutrino oscillation parameters are different, except for $\theta_{x}$ and $\delta m^2$, in the two hierarchy cases, and so, we have to repeat the sampling for each hierarchy case. The various predictions for the ranges of the neutrino physical parameters ($\theta_{x},~\theta_{y},~\theta_{z},~\delta,~\rho,~\sigma,~m_1,~m_2,~m_3
,~m_{ee},~m_{e},~J$) at all $\sigma$ error levels with either hierarchy type are introduced in the Table \ref{Predictions}.

We find that out of the fifteen possible textures, only seven ones can accommodate the experimental data. Only the texture $(\textbf{C}_{22},\textbf{C}_{33})$ is viable for both normal and inverted hierarchies, whereas the texture $(\textbf{C}_{13},\textbf{C}_{23})$ can accommodate the data only for normal hierarchy, and the textures $(\textbf{C}_{33},\textbf{C}_{13})$, $(\textbf{C}_{22},\textbf{C}_{33})$, $(\textbf{C}_{11},\textbf{C}_{12})$, $(\textbf{C}_{33},\textbf{C}_{12})$, $(\textbf{C}_{11},\textbf{C}_{13})$, $(\textbf{C}_{22},\textbf{C}_{12})$ are viable for inverted hierarchy only. All cases can accommodate data at all three $\s$-levels except the textures $(\textbf{C}_{22},\textbf{C}_{33})$ in normal ordering and $(\textbf{C}_{11},\textbf{C}_{12})$, which is of inverted type, accommodating data only at the $3\s$ level, and the textures, of inverted type, $(\textbf{C}_{33},\textbf{C}_{12})$ and $(\textbf{C}_{22},\textbf{C}_{12})$ which fail at the $1\s$-level.

 We also find that neither $m_1$ for normal hierarchy nor $m_3$ for inverted hierarchy does approach a vanishing value, so there are no signatures for the singular textures. From Table \ref{Predictions}, we see that the allowed ranges for $\theta_{y}$ are strongly restricted for the texture $(\textbf{C}_{22},\textbf{C}_{33})$ in normal ordering and in the texture $(\textbf{C}_{11},\textbf{C}_{12})$, which is of inverted ordering, at the 3-$\sigma$ level. There exist acute restrictions on the allowed ranges of the CP-violating phases $(\delta,\rho,\sigma)$ at all $\sigma$-levels with either hierarchy type for all textures. We note from Eq. \ref{jg} that the J parameter depends strongly on $\delta$ ($J\propto \sin\delta$) because of the tight allowed experimental ranges of the mixing angles, which makes $J$-variations, due to these angles' changes,  tiny compared with those resulting from $\delta$-changes. The allowed values of the J parameter at the 1-$\sigma$ level for the texture $(\textbf{C}_{13},\textbf{C}_{23})$, which is of normal ordering, are negative. Therefore, the corresponding Dirac phase $\delta$ lies in the third or fourth quarters. Table \ref{Predictions} also reveals that $m_{ee}<0.04 \textrm{ eV}$ for the textures $(\textbf{C}_{13},\textbf{C}_{23})$, $(\textbf{C}_{33},\textbf{C}_{12})$, $(\textbf{C}_{13},\textbf{C}_{11})$ and $(\textbf{C}_{22},\textbf{C}_{12})$. However, it has a bit higher upper bound $m_{ee}<0.17$ for the remaining cases.

 If we adopt the tightest bound of the sum parameter ($\Sigma<0.09~ \textbf{eV}$), or the strict Planck 2018 bound ($\Sigma<0.12~ \textbf{eV}$), we find that only the texture $(\textbf{C}_{22},\textbf{C}_{33})$ in normal ordering can accommodate the data, in line with the conclusions of \cite{Valentino_2021} that this low bound highly compromises the viability of the inverted mass ordering. Five patterns remain viable ($(\textbf{C}_{22},\textbf{C}_{33})$ with $(\textbf{C}_{13},\textbf{C}_{23})$ in normal ordering,  and $(\textbf{C}_{22},\textbf{C}_{12})$ with $(\textbf{C}_{33},\textbf{C}_{12})$ and with $(\textbf{C}_{11},\textbf{C}_{13})$ in inverted ordering) if we assume the constraint $\Sigma<0.2~ \textbf{eV}$ as Tab. (\ref{sigma_predictions}) shows. However, as said earlier, we shall take the relaxed Planck 2018 bound $\Sigma<0.54~ \textbf{eV}$ as we believe that by relaxing cosmological bounds, laboratory data are made to carry more weight than non-anonymously agreed upon cosmological data.

 \begin{table}[h]
\begin{center}
\begin{tabular}{ c || p {20mm}  | p {17mm} | p {17mm} | p {17mm} | p {17mm}| p {17mm}| p {17mm} }
\hline
Texture& $(\textbf{C}_{22},\textbf{C}_{33})$   &
 $(\textbf{C}_{22},\textbf{C}_{12})$  &
  $(\textbf{C}_{11},\textbf{C}_{12})$ &
    $(\textbf{C}_{33},\textbf{C}_{12})$ &
     $(\textbf{C}_{33},\textbf{C}_{13})$ &
       $(\textbf{C}_{11},\textbf{C}_{13})$ &
        $(\textbf{C}_{13},\textbf{C}_{23})$  \\
 \hline
       Ordering &{\bf N \& I}&{\bf I}& {\bf I}& {\bf I}& {\bf I}& {\bf I}& {\bf N} \\
 \hline
 \hline
 Predictions & { \bf N}: $\Sigma \in$ \newline $[0.089,0.211]$ &
  $0.192 $  \newline $\leq \Sigma \leq$  &
 $0.532 $ \newline $\leq \Sigma \leq$   &
 $0.125 $ \newline $\leq \Sigma \leq$    &
 $0.291 $ \newline $\leq \Sigma \leq$    &
 $0.151 $  \newline $\leq \Sigma \leq$  &
 $0.163 $  \newline $\leq \Sigma \leq$   \\
 & { \bf I}:  $\Sigma \in$ \newline  $[0.467,0.539]$ & $ 0.205$ &
  $0.540$ &
   $0.136$ &
    $0.330$ &
     $ 0.163$ &
      $ 0.213$  \\
\hline
$\Sigma < 0.09$ & $\checkmark$ \newline $\times$ & $\times$& $\times$& $\times$& $\times$&$\times$ & $\times$\\
\hline
$\Sigma < 0.2$ & $\checkmark$ \newline $\times$ & $\checkmark$ & $\times$& $\checkmark$& $\times$ & $\checkmark$& $\checkmark$\\
\hline
\end{tabular}
\caption{The parameter $\Sigma$ predictions, evaluated in eV, in the viable patterns adopting a relaxed Planck 2018 constraint ($\Sigma < 0.54~ \textbf{eV}$). Second (Third) line shows the corresponding viability adopting a strictest (more constrained) bound from \cite{Valentino_2021} (PDG live). }
\label{sigma_predictions}
\end{center}
\end{table}

We introduce 15 correlation plots for each viable texture, in any allowed hierarchy type, generated from the accepted points of the neutrino physical parameters at the 3-$\sigma$ level. The red (blue) plots represent the normal (inverted) ordering. The first and second rows represent the correlations between the mixing angles and the CP-violating phases. The third row introduces the correlations amidst the CP-violating phases, whereas the fourth one represents the correlations between the Dirac phase $\delta$ and each of $J$, $m_{ee}$ and $m_2$ parameters respectively. The last row shows the degree of mass hierarchy plus the ($m_{ee}, m_2$) correlation.

In order to interpret the numerical results, we write down, for each pattern of the fifteen possible ones, the complete analytical expression of ($\d m^2=m_{23}^2-m_{12}^2$), possibly written as an expansion in $s_z$ when the exact expression turns out to be too complicated, and analyze its zeros analytically and numerically. By assuming these zeros, we can justify the viability/nonviability of the pattern and its ordering hierarchy type, say by examining respectively the resulting ($R_\n, m_{13}$). Moreover, whenever the texture is viable, by assuming the zeros of the complete (leading order of the) $\d m^2$-expression we get ``exact" (``approximate'')  correlation spectrum properties which would provide some explanations for the distinguishing features in the corresponding ``full'' correlation plots presented at the 3-$\sigma$ level, such as those involving ($\rho, \sigma$).

Finally, we reconstruct $M_{\nu}$ with either allowed hierarchy type for each viable texture from the one representative point at the 3-$\sigma$ level in the 5-dimensional parameter space. The point is chosen to be as close as possible to the best fit values for mixing and Dirac phase angles.

\begin{landscape}
\begin{table}[h]
 \begin{center}
\scalebox{0.75}{
{\tiny
 \begin{tabular}{c|c|c|c|c|c|c|c|c|c|c|c|c}
  \hline
 \hline
\multicolumn{13}{c}{\mbox{Pattern} $(C_{22},C_{33})\equiv (M_{\nu~11} + M_{\nu~33}=0,~M_{\nu~11} + M_{\nu~22}=0)$} \\
\hline
\hline
  \mbox{quantity} & $\theta_{12}^{\circ}\equiv \t_x^o$ & $\theta_{23}^{\circ} \equiv \t_y^o$& $\theta_{13}^{\circ}\equiv \t_z^o$ & $\delta^{\circ}$ & $\rho^{\circ}$ & $\sigma^{\circ}$ & $m_{1}$ $(10^{-1} \text{eV})$ & $m_{2}$ $(10^{-1} \text{eV})$ & $m_{3}$ $(10^{-1} \text{eV})$ & $m_{ee}$ $(10^{-1} \text{eV})$
 & $m_{e}$ $(10^{-1} \text{eV})$ & $J$ $(10^{-1})$\\
 \hline
 \multicolumn{13}{c}{\mbox{Normal  Hierarchy}} \\
 \cline{1-13}
 $1~\sigma$ &$\times$& $\times$ &$\times$ &$\times$ &$\times$ &
  $\times$& $\times$ &$\times$ & $\times$ &$\times$ &
  $\times$ & $\times$ \\
 \hline
 $2~\sigma$ &$\times$& $\times$ &$\times$ &$\times$ &$\times$ &
  $\times$& $\times$ &$\times$ & $\times$ &$\times$ &
  $\times$ & $\times$ \\
 \hline
 $3~ \sigma$ & $31.40 - 37.40$ & $44.87 - 45.13$ & $8.13 - 8.92$& $128.07 - 252.50 \cup 288.40 - 358.93$ & $74.61 - 105.36$ & $74.64 - 105.34$ & $0.17 - 0.65$ & $0.19 -0.65$ & $0.53 - 0.82$ & $0.16 - 0.62$ & $0.19 - 0.65$ & $-0.35 - 0.28$ \\
 \hline
 \multicolumn{13}{c}{\mbox{Inverted  Hierarchy}} \\
 \cline{1-13}
 $1~\sigma$ & $33.30 - 35.30$  &$48.49 - 50.06$ &$8.44 - 8.70$ &$268.68 - 269.25$ &$90.69 - 91.08$ & $87.72 - 88.57$ &$1.64 - 1.71$ &$1.64 - 1.71$ &$1.56 - 1.64$ &$1.56 - 1.64$ & $1.64 - 1.71$ &$-0.34 - -0.32$  \\
 \hline
 $2~\sigma$ & $32.30 - 36.39$ & $47.35 - 50.66$ & $8.30 - 8.83$ & $268.35 - 269.53$ & $90.46 - 91.26$ & $87.29 - 89.09$ & $1.60 - 1.76$ & $1.60 - 1.76$ & $1.52 - 1.68$ & $1.52 - 1.68$ & $1.60 - 1.75$ & $-0.35 - -0.31$  \\
 \hline
 $3~\sigma$ & $31.40 - 37.39$ & $41.19 - 44.97 \cup 45.03 - 51.24$ & $8.17 - 8.96$ & $267.36 - 269.80 \cup 270.20 - 272.53$ & $87.42 - 89.78 \cup 90.22 - 92.68$ & $87.11 - 92.66$ & $1.57 -1.82$ & $1.58 - 1.82$ &$1.49 - 1.75$ & $1.50 - 1.74$ & $1.57 -  1.82$& $-0.36 -  -0.30$  \\
 \hline
  \hline
 \multicolumn{13}{c}{\mbox{Pattern} $(C_{22},C_{12})\equiv (M_{\nu~11} + M_{\nu~33}=0,~M_{\nu~12}+M_{\nu~33}=0) $} \\
\hline
\hline
 \mbox{quantity} & $\theta_{12}^{\circ}\equiv \t_x^o$ & $\theta_{23}^{\circ}\equiv \t_y^o$& $\theta_{13}^{\circ}\equiv \t_z^o$ & $\delta^{\circ}$ & $\rho^{\circ}$ & $\sigma^{\circ}$  & $m_{1}$ $(10^{-1} \text{eV})$ & $m_{2}$ $(10^{-1} \text{eV})$ & $m_{3}$ $(10^{-1} \text{eV})$ & $m_{ee}$ $(10^{-1} \text{eV})$ & $m_{e}$ $(10^{-1} \text{eV})$ & $J$ $(10^{-1})$\\
 \hline
 \multicolumn{13}{c}{\mbox{Inverted  Hierarchy}} \\
 \cline{1-13}
 $1~\sigma$ &$\times$& $\times$ &$\times$ &$\times$ &$\times$ &
  $\times$& $\times$ &$\times$ & $\times$ &$\times$ &
  $\times$ & $\times$    \\
 \hline
  $2~\sigma$ & $34.40 - 36.40$ & $47.35 - 50.66$ & $8.30 - 8.83$ & $226.00 - 236.25$ & $102.59 - 106.02$ & $32.23 - 39.70$ & $0.71 - 0.74$ & $0.71 - 0.74$ & $0.51 - 0.54$ & $0.30 - 0.33$ & $0.70 - 0.73$ & $-0.29 - -0.23$\\
 \hline
 $3~\sigma$ & $31.40 - 37.39$ & $41.16 - 51.25$ & $8.17 - 8.96$ & $200.00 - 244.53$ & $95.33  - 107.77$ & $14.34  - 46.26$ & $0.70 - 0.74$ & $0.70 - 0.75$ & $0.50 - 0.55$ & $0.30 - 0.37$ & $0.70 - 0.74$ & $-0.32 - -0.10$  \\
 \hline
  \hline
\multicolumn{13}{c}{\mbox{Pattern} $(C_{11},C_{12})\equiv (M_{\nu~22}+M_{\nu~33}=0,~M_{\nu~21}+M_{\nu~33}=0) $} \\
\hline
\hline
  \mbox{quantity} & $\theta_{12}^{\circ}\equiv \t_x^o$ & $\theta_{23}^{\circ}\equiv \t_y^o$& $\theta_{13}^{\circ}\equiv \t_z^o$ & $\delta^{\circ}$ & $\rho^{\circ}$ & $\sigma^{\circ}$ & $m_{1}$ $(10^{-1} \text{eV})$ & $m_{2}$ $(10^{-1} \text{eV})$ & $m_{3}$ $(10^{-1} \text{eV})$ & $m_{ee}$ $(10^{-1} \text{eV})$
 & $m_{e}$ $(10^{-1} \text{eV})$ & $J$ $(10^{-1})$\\
 \hline
 \multicolumn{13}{c}{\mbox{Inverted  Hierarchy}} \\
 \cline{1-13}
 $1~\sigma$ &$\times$& $\times$ &$\times$ &$\times$ &$\times$ &
  $\times$& $\times$ &$\times$ & $\times$ &$\times$ &
  $\times$ & $\times$ \\
 \hline
 $2~\sigma$ &$\times$& $\times$ &$\times$ &$\times$ &$\times$ &
  $\times$& $\times$ &$\times$ & $\times$ &$\times$ &
  $\times$ & $\times$ \\
 \hline
 $3~\sigma$ & $31.40 - 37.39$ & $51.16 - 51.25$ & $8.17 - 8.96$ & $262.79 - 268.92$ & $5.71 - 9.54$ & $168.06 - 173.00$ & $1.79  - 1.82$ & $1.80 - 1.82$ &$1.73 - 1.75$ & $1.73 - 1.75$ & $1.79 - 1.82$ & $-0.35 - -0.30$  \\
 \hline
  \hline
\multicolumn{13}{c}{\mbox{Pattern} $(C_{33},C_{12})\equiv M_{\nu~11} + M_{\nu~22}=0,~M_{\nu~12}+M_{\nu~33}=0) $} \\
\hline
\hline
 \mbox{quantity} & $\theta_{12}^{\circ}\equiv \t_x^o$ & $\theta_{23}^{\circ}\equiv \t_y^o$& $\theta_{13}^{\circ}\equiv \t_z^o$ & $\delta^{\circ}$ & $\rho^{\circ}$ & $\sigma^{\circ}$  & $m_{1}$ $(10^{-1} \text{eV})$ & $m_{2}$ $(10^{-1} \text{eV})$ & $m_{3}$ $(10^{-1} \text{eV})$ & $m_{ee}$ $(10^{-1} \text{eV})$
 & $m_{e}$ $(10^{-1} \text{eV})$ & $J$ $(10^{-1})$\\
 \hline
 \multicolumn{13}{c}{\mbox{Inverted  Hierarchy}} \\
 \cline{1-13}
 $1~\sigma$ &$\times$& $\times$ &$\times$ &$\times$ &$\times$ &
  $\times$& $\times$ &$\times$ & $\times$ &$\times$ &
  $\times$ & $\times$  \\
 \hline
  $2~\sigma$ &$32.30 -36.39$ & $47.35 -50.67 $ & $8.30 - 8.83$ & $231.94 - 248.79$ &$100.59 - 105.16$ & $38.47 - 51.29$ & $0.52 - 0.54$ & $0.53 - 0.55$ & $0.20 - 0.22$ & $0.30 - 0.33$ & $0.52 - 0.54$ & $-0.33 - -0.25$    \\
 \hline
 $3~\sigma$ & $31.40 - 37.39$ & $41.16 - 51.25$ & $8.17 - 8.96$ & $226.89  -254.21$ & $99.40 - 106.17$ & $34.65 - 57.53$ & $0.52 - 0.55$ & $0.53 - 0.56$ & $0.19 - 0.23$ & $0.30 - 0.37$ & $0.52 - 0.55$& $-0.35 - -0.22$  \\
 \hline
 \hline
\multicolumn{13}{c}{\mbox{Pattern} $(C_{33},C_{13})\equiv (M_{\nu~11}+M_{\nu~22}=0,~M_{\nu~12}+M_{\nu~23}=0)$} \\
\hline
\hline
  \mbox{quantity} & $\theta_{12}^{\circ} \equiv \t_x^o$  & $\theta_{23}^{\circ} \equiv \t_y^o$& $\theta_{13}^{\circ}\equiv \t_z^o$ & $\delta^{\circ}$ & $\rho^{\circ}$ & $\sigma^{\circ}$ & $m_{1}$ $(10^{-1} \text{eV})$ & $m_{2}$ $(10^{-1} \text{eV})$ & $m_{3}$ $(10^{-1} \text{eV})$ & $m_{ee}$ $(10^{-1} \text{eV})$
 & $m_{e}$ $(10^{-1} \text{eV})$ & $J$ $(10^{-1})$\\
 \hline
 \multicolumn{13}{c}{\mbox{Inverted  Hierarchy}} \\
 \cline{1-13}
 $1~\sigma$ &$33.30 - 35.30$ & $48.49 - 50.06$ & $8.44 - 8.70$ & $264.35 - 265.70$ & $93.91 - 94.53$ & $80.10 - 81.43$ & $1.06 - 1.10$ & $1.07 - 1.10$ & $0.94 - 0.98$ & $0.99 -  1.03$ & $ 1.06 - 1.10$ & $ -0.34 - -0.32 $ \\
 \hline
 $2~\sigma$ &$32.30 - 36.39$ & $47.35 - 50.67$ & $8.30 - 8.83$ & $263.62 - 266.47$ & $93.59 - 94.83$ & $79.40 -  82.26$ & $1.05 - 1.12$ & $1.05 - 1.12$ & $0.93 - 1.00$ & $0.97 - 1.05$& $ 1.05 - 1.12$ & $-0.35 - -0.31$  \\
 \hline
 $3~\sigma$ & $31.40 - 37.39$ & $41.17 - 51.24$ & $8.17 - 8.96$ & $262.85 - 267.59$ & $92.56 -  95.16$ & $78.67 - 84.28$ & $1.00 - 1.13$ & $1.01 - 1.14$ & $0.88 - 1.02$ & $0.95 - 1.07$ & $1.00 -  1.13$ & $-0.36 - -0.30$  \\
 \hline
  \hline
 \multicolumn{13}{c}{\mbox{Pattern} $(C_{13},C_{11})\equiv M_{\nu~12} + M_{\nu~23}=0,~M_{\nu~22}+M_{\nu~33}=0) $} \\
\hline
\hline
 \mbox{quantity} & $\theta_{12}^{\circ}\equiv \t_x^o$ & $\theta_{23}^{\circ}\equiv \t_y^o$& $\theta_{13}^{\circ}\equiv \t_z^o$ & $\delta^{\circ}$ & $\rho^{\circ}$ & $\sigma^{\circ}$  & $m_{1}$ $(10^{-1} \text{eV})$ & $m_{2}$ $(10^{-1} \text{eV})$ & $m_{3}$ $(10^{-1} \text{eV})$ & $m_{ee}$ $(10^{-1} \text{eV})$
 & $m_{e}$ $(10^{-1} \text{eV})$ & $J$ $(10^{-1})$\\
 \hline
 \multicolumn{13}{c}{\mbox{Inverted  Hierarchy}} \\
 \cline{1-13}
 $1~\sigma$ & $33.39 - 35.30$ & $48.49 - 50.06$ & $8.44 - 8.70$ & $301.96 - 309.99$ & $165.23 - 167.75$ & $46.81 - 52.37$ & $0.59 - 0.60$ & $0.60 - 0.61$ & $0.33 - 0.34$ & $0.33 - 0.34$ &
  $0.59 - 0.60$ & $-0.29 - -0.25$ \\
 \hline
  $2~\sigma$ & $32.30 -36.39$ & $47.35 - 50.67$ & $8.30 - 8.83$ & $297.60- 315.86$ & $163.81 -  169.35$ & $43.84 - 56.57$ & $0.59 - 0.60$ & $0.59 - 0.61$ & $0.33 - 0.34$ & $0.32 - 0.34$ & $0.58 - 0.60$ & $-0.31 - -0.22$    \\
 \hline
 $3~\sigma$ & $31.40 - 37.39$ & $41.17 - 51.24$ & $8.17 - 8.96$ & $292.68 - 321.16$ & $162.53  - 170.80$ & $39.98 - 60.41$ & $0.58 - 0.62$ & $0.59 - 0.63$ & $0.32 - 0.37$ & $0.32 - 0.36$ & $0.58 - 0.62$ & $-0.33 -  -0.19$  \\
 \hline
 \hline
\multicolumn{13}{c}{\mbox{Pattern} $(C_{13},C_{23})\equiv (M_{\nu~12} + M_{\nu~23}=0,~M_{\nu~11}+M_{\nu~23}=0) $} \\
\hline
\hline
 \mbox{quantity} & $\theta_{12}^{\circ}\equiv \t_x^o$ & $\theta_{23}^{\circ}\equiv \t_y^o$& $\theta_{13}^{\circ}\equiv \t_z^o$ & $\delta^{\circ}$ & $\rho^{\circ}$ & $\sigma^{\circ}$  & $m_{1}$ $(10^{-1} \text{eV})$ & $m_{2}$ $(10^{-1} \text{eV})$ & $m_{3}$ $(10^{-1} \text{eV})$ & $m_{ee}$ $(10^{-1} \text{eV})$
 & $m_{e}$ $(10^{-1} \text{eV})$ & $J$ $(10^{-1})$\\
 \hline
 \multicolumn{13}{c}{\mbox{Normal  Hierarchy}} \\
 \cline{1-13}
 $1~\sigma$ & $33.30 - 34.78$ & $48.87 - 50.05$ & $8.41 -8.66$ & $202.90 - 217.99$ & $96.82 - 101.58$ & $16.20 - 26.78$ & $0.57 - 0.61$ &$0.58 - 0.62$ & $0.76 - 0.79$ &$0.23 - 0.24$ & $0.58 - 0.62$ & $-0.21 - -0.13$  \\
 \hline
 $2~\sigma$ & $32.30 - 36.39$ & $47.37 - 50.71$ & $8.27 - 8.79$ & $152.02 -232.55$ & $81.53 - 106.09$ & $0.02 - 36.72 \cup 160.00 - 179.95$ & $0.55 - 0.63$ & $0.56 - 0.64$ & $0.74 - 0.81$  & $0.23 - 0.24$ & $0.56 - 0.64$ & $-0.28 - 0.16$  \\
 \hline
 $3~\sigma$ & $31.40 - 37.39$ & $41.20 - 51.32$ & $8.13 - 8.92$ & $128.01 - 242.79$ & $73.50 - 108.48$ & $0.07 - 44.68 \cup 142.50 - 179.93$ & $0.47 - 0.65$ & $0.47 - 0.66$ & $0.68 - 0.83$ & $0.21 - 0.24$ & $0.47 - 0.66$ & $-0.31 - 0.28$  \\
 \hline
  \hline
 \end{tabular}
 }}
 \end{center}
 \caption{The various predictions for the ranges of the neutrino physical parameters for the seven viable textures at all $\sigma$ levels.}
\label{Predictions}
 \end{table}
\end{landscape}
\newpage

\subsection{Texture($\textbf{C}_{22},\textbf{C}_{33}$)$\equiv$ ($M_{ee}+M_{\tau\tau}=0$, $M_{ee}+M_{\mu\mu}=0$) }
The A's and B's are given by
\begin{align}
A_1=&c_x^2c_z^2+(-c_xc_ys_z+s_xs_ye^{-i\delta})^2,~A_2=s_x^2c_z^2+(-s_xc_ys_z-c_xs_ye^{-i\delta})^2,~A_3=s_z^2+c_y^2c_z^2\nonumber\\
B_1=&c_x^2c_z^2+(-c_xs_ys_z-s_xc_ye^{-i\delta})^2,~B_2=s_x^2c_z^2+(-s_xs_ys_z+c_xc_ye^{-i\delta})^2,~B_3=s_z^2+s_y^2c_z^2\label{coeffC22C33}
\end{align}
We have the following truncated expression for $R_\n$:
\bea
\label{c33c22-approx-R}
R_\n &=& \frac{2t_{2y}}{s_{2x}c_{\delta}}s_z+\mathcal{O}(s_z^2)
\eea
From Table \ref{Predictions}, we see that ($\textbf{C}_{22},\textbf{C}_{33}$) texture is not viable at the 1-2-$\sigma$ levels for normal ordering. We find that the mixing angles $(\theta_{x},\theta_{z}$) extend over their allowed experimental ranges with either hierarchy type. However, there exists a strong restriction on $\theta_{y}$ in the case of normal ordering to lie in the interval $[44.87^{\circ},45.13^{\circ}]$. For normal ordering, there exists a mild forbidden gap $[252.5^{\circ},288.4^{\circ}]$ for $\delta$, whereas the phases $\rho$ and $\sigma$ are restricted to the interval $[74^{\circ},106^{\circ}$]. For inverted ordering, we find a tight forbidden gap for $\delta$ and $\rho$ around $270^{\circ}$ and $90^{\circ}$ respectively at the 3-$\sigma$ level. The phases $\delta$, $\rho$, and $\sigma$ are tightly restricted at all $\sigma$-levels, they are bound to the intervals $[267.36^{\circ},272.53^{\circ}]$, $[87.42^{\circ},92.68^{\circ}]$, and $[87.11^{\circ},92.66^{\circ}]$ respectively at the 3-$\sigma$-level. Table (\ref{Predictions} also shows that neither $m_1$ for normal hierarchy nor $m_3$ for inverted hierarchy does reach zero at all error levels. Thus, the singular mass matrix is not predicted for this texture at all $\sigma$ levels.

For normal ordering plots, we see a tight forbidden gap for $\theta_{y}$ around 45$^{\circ}$ together with a mild forbidden region for the phase $\delta$. We find a strong linear correlation between $\rho$ and $\sigma$. One also notes the sinusoidal relations for $(\rho,\delta)$ and $(\sigma,\delta)$ correlations. We also find a moderate mass hierarchy where $0.32\leq m_{13} \leq0.79$ besides a quasi-degeneracy characterized by $1.01\leq\frac{m_2}{m_1}\leq1.13$.

For inverted ordering plots, We find  narrow disallowed regions for $\rho$ and $\delta$ around 90$^{\circ}$ and $270^{\circ}$ respectively. We also see a narrow forbidden gap for $\theta_{y}$ around 45$^{\circ}$ as in normal ordering. We notice a quasi degeneracy characterized by $m_1\approx m_2\approx m_3$.

In order to justify these observations, we compute the  mass-squared-difference full and approximate expressions, and we use the abbreviation ``Num" (``Den") for ``Numerator" (``Denominator"):
\bea
\label{c33c22-full-ms-difference}
m_{23}^2-m_{13}^2 &=& \frac{\mbox{Num}(m_{23}^2-m_{13}^2)}{\mbox{Den}(m_{23}^2-m_{13}^2)}: \\
\mbox{Num}(m_{23}^2-m_{13}^2) &=& 4 c_z^2 c_{2y} \left(\frac{1}{2} s_{2y} s_{2x} (c_z^2-2) s_z c_\d + c_{2y}c_{2x} s_z^2\right),\nn\\
\mbox{Den}(m_{23}^2-m_{13}^2) &=& -s_{2x}^2 s_{2y}^2 (1+c_z^2) s_z^2 c_\d^2 + s_{2x} s_{2y}c_{2y} c_{2x} (2+c_z^2) s_z c_\d -\frac{1}{4} s_{2x}^2 s_{2y}^2 c_z^4 s_z^2 -c_{2y}^2 c_{2x}^2,\nn
\\
\label{c33c22-approx-ms-difference}
m_{23}^2-m_{13}^2 &=&\frac{2 t_{2y} t_{2x} c_\d s_z}{c_{2x}} +\mathcal{O}(s_z^2).
\eea
Few remarks are in order here. First, we see from Eqs. (\ref{c33c22-approx-ms-difference}) that we should have $t_{2y} c_\d >0$ in order to meet the constraint $m_2>m_1$, and thus we should have:
\bea
\t_y < \frac{\pi}{4} \Rightarrow  \d > \frac{3 \pi}{2} &,& \t_y > \frac{\pi}{4} \Rightarrow \d < \frac{3 \pi}{2},
\eea
which is observed in the correlations between $\d$ and $\t_y$ in both {\bf NH} and {\bf IH}.
Second, from Eq. (\ref{c33c22-approx-R}, we see that both $\t_y = \frac{\pi}{4}$ and $\d=\frac{3 \pi}{2}$ are separately forbidden, as each will give a too high value for $R_\n$ unable to be brought back to the small experimental order of magnitude $10^{-2}$. This explains why we observe a narrow gap around ($\t_y = \frac{\pi}{4}$) and around ($\d=\frac{3\pi}{2}$) in the relevant correlations for both {\bf NH} and {\bf IH}.

Third, and as was stated before, studying the zeros of ($m_{23}^2-m_{13}^2$) would put us near the allowed points in the parameter space. From Eq. (\ref{c33c22-full-ms-difference}), we see that this corresponds to two regimes.
\begin{itemize}
\item $c_{2y} \approx 0 \Rightarrow \t_y \approx \frac{\pi}{4} \Rightarrow ${\bf NH}:

In this regime, we get
\bea
m_{13}^2 =  m_{23}^2 &=& \frac{(c_z^2-2)^2}{c_z^4+4(1+c_z^2)c_\d^2}\approx \frac{1}{1+8 c_\d^2} <1 \label {D2-ypi4-m13},\\
\rho \approx \sigma &=& \frac{1}{2}\tan^{-1}\bigg(\frac{s_{2\delta}}{1+2c_{\delta}^2}\bigg)+ \frac{\pi}{2}+\mathcal{O}(s_z) \label {D2-ypi4-rho},\\
m_{ee}&=& \frac{m_3}{\sqrt{1+8c_\d^2}}+\mathcal{O}(s_z^2)\label {D2-ypi4-mee}.
\eea
From Eq. (\ref{D2-ypi4-m13}), this regime corresponds to {\bf NH}. In this regime, $\d$ can take any of this texture allowed values, except the narrow band around $3\pi/2$. The correlations, resulting from Eq. (\ref{D2-ypi4-rho}), of $\r$ and $\s$ with respect to $\d$  are observed in the corresponding {\bf NH} plots. Likewise, Eq. (\ref{D2-ypi4-mee})
justifies the shape of the correlation ($m_{ee}, \d$) observed in this {\bf NH} regime.

\item $\left(\frac{1}{2} s_{2y} s_{2x} (c_z^2-2) s_z c_\d + c_{2y}c_{2x} s_z^2 \approx 0 \right) \dagger\Rightarrow ${\bf IH}:

Here we have
\bea
m_{13}^2 = m_{23}^2 &=&  1+ \frac{s_z^2}{s_y^2 c_y^2 c_z^4}> 1 \label{D2-ynpi4-m13},
\eea
so this regime corresponds to {\bf IH}. In this regime, we find, by putting ($c_z \approx 1$) in the approximative regime-defining constraint ($\dagger$), the following:
\bea c_\d &\approx& \frac{2 s_z}{t_{2y} t_{2x}} \ll 1,
\eea
and so from the allowed values of $\d \in [200^o, 353^o]$ (see Table \ref{TableLisi:as} in the {\bf IH} case) we see that $\d$ should be around the value $3\pi/2$ without hitting it, whereas no restrictions over $\t_y$ apart from disallowing the value $\pi/4$. This is what we observe in the relevant {\bf IH} correlation plots.

Now, with $\d \approx 3 \pi/2$ one finds
\bea
\r \approx \s &=& \frac{\pi}{2}+\mathcal{O}(s_z) \label{D2-ynpi4-rho},
\eea
which we observe in the ``full'', i.e. non-approximate, numerical results. Thus, we could interpret the very tight ranges of the CP violation phases with ($\d \sim 3\pi/2, \r \sim \pi/2 \sim\s$).  Finally, with Eq. (\ref{D2-ynpi4-m13}), we find a quasi degenerate spectrum ($m_{13} \approx m_{23} \approx 1$) and that $m_{ee}$ matches this common mass scale, which we observe in the plots.
\end{itemize}

%
Finally, we reconstruct the neutrino mass matrix for a representative point. For normal ordering, the representative point is taken as following:
\begin{equation}
\begin{aligned} \label{22-33_normal_spectrum}
(\theta_{12},\theta_{23},\theta_{13})=&(34.0696^{\circ},45.1044^{\circ},8.4838^{\circ}),\\
(\delta,\rho,\sigma)=&(195.8781^{\circ},95.4176^{\circ},95.3851^{\circ}),\\
(m_{1},m_{2},m_{3})=&(0.0180\textrm{ eV},0.0199\textrm{ eV},0.0530\textrm{ eV}),\\
(m_{ee},m_{e})=&(0.0171\textrm{ eV},0.0200\textrm{ eV}),
\end{aligned}
\end{equation}
the corresponding neutrino mass matrix (in eV) is
\begin{equation} \label{22-33_normal_recons}
M_{\nu}=\left( \begin {array}{ccc} -0.0167 - 0.0034i &  0.0080 + 0.0003i &  0.0067 + 0.0004i\\ \noalign{\medskip}0.0080 + 0.0003i &   0.0167 + 0.0034i &  0.0348 - 0.0035i
\\ \noalign{\medskip}0.0067 + 0.0004i &  0.0348 - 0.0035i &  0.0167 + 0.0034i\end {array} \right).
\end{equation}
For inverted ordering, the representative point is taken as following:
\begin{equation}
\begin{aligned} \label{22-33_inverted_spectrum}
(\theta_{12},\theta_{23},\theta_{13})=&(34.3118^{\circ},49.2414^{\circ},8.4204^{\circ}),\\
(\delta,\rho,\sigma)=&(269.0126^{\circ},90.8634^{\circ},88.2051^{\circ}),\\
(m_{1},m_{2},m_{3})=&(0.1733\textrm{ eV},0.1735\textrm{ eV},0.1659\textrm{ eV}),\\
(m_{ee},m_{e})=&(0.1660\textrm{ eV},0.1732\textrm{ eV}),
\end{aligned}
\end{equation}
the corresponding neutrino mass matrix (in eV) is
\begin{equation} \label{22-33_inverted_recons}
M_{\nu}=\left( \begin {array}{ccc}-0.1660 - 0.0001i &  0.0324 - 0.0001i &  0.0377 + 0.0001i\\ \noalign{\medskip}0.0324 - 0.0001i  & 0.1660 + 0.0001i & -0.0074 - 0.0001i
\\ \noalign{\medskip}0.0377 + 0.0001i & -0.0074 - 0.0001i &  0.1660 + 0.0001i
\end {array} \right).
\end{equation}

\begin{figure}[hbtp]
\hspace*{-4cm}
\includegraphics[width=24cm, height=16cm]{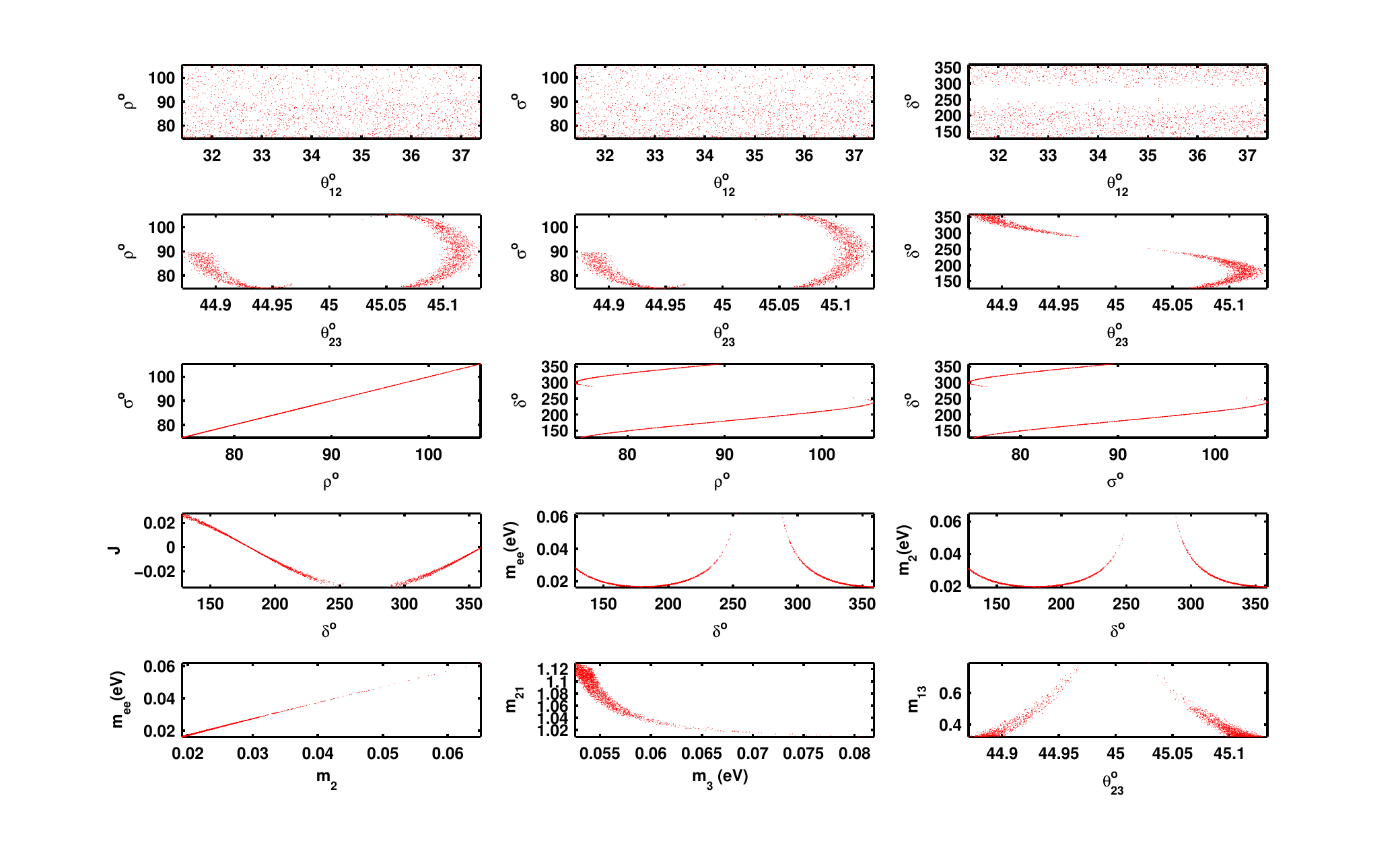}
\caption{The correlation plots for ($\textbf{C}_{22},\textbf{C}_{33}$)$\equiv$ ($M_{ee}+M_{\tau\tau}=0$, $M_{ee}+M_{\mu\mu}=0$) texture, in the normal ordering hierarchy. The first and second row represent the correlations between the mixing angles ($\theta_{12}$,$\theta_{23}$) and the CP-violating phases. The third row introduces the correlations amdist the CP-violating phases, whereas the fourth one represents the correlations between the Dirac phase $\delta$ and each of $J$, $m_{ee}$ and $m_2$ parameters respectively. The last row shows the degree of mass hierarchy plus the ($m_{ee}, m_2$) correlation.}
\label{Tr2233norm}
\end{figure}
\begin{figure}[hbtp]
\hspace*{-4cm}
\includegraphics[width=24cm, height=16cm]{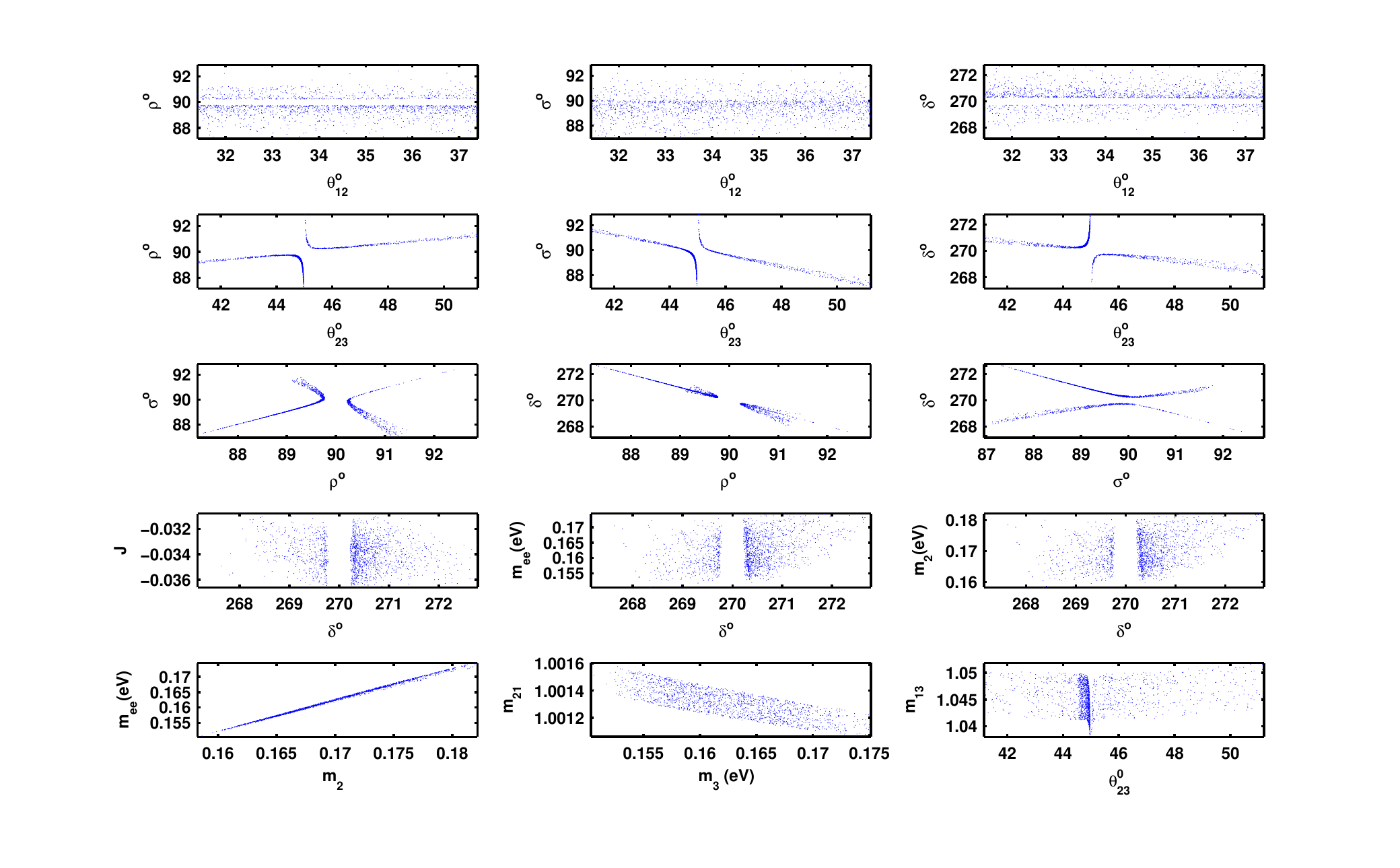}
\caption{The correlation plots for ($\textbf{C}_{22},\textbf{C}_{33}$)$\equiv$ ($M_{ee}+M_{\tau\tau}=0$, $M_{ee}+M_{\mu\mu}=0$) texture, in the inverted ordering hierarchy. The first and second row represent the correlations between the mixing angles ($\theta_{12}$,$\theta_{23}$) and the CP-violating phases. The third row introduces the correlations amdist the CP-violating phases, whereas the fourth one represents the correlations between the Dirac phase $\delta$ and each of $J$, $m_{ee}$ and $m_2$ parameters respectively. The last row shows the degree of mass hierarchy plus the ($m_{ee}, m_2$) correlation.}
\label{Tr2233inverted}
\end{figure}
\newpage


\subsection{Texture($\textbf{C}_{22},\textbf{C}_{12}$)$\equiv$ ($M_{ee}+M_{\tau\tau}=0$, $M_{e\mu}+M_{\tau\tau}=0$) }
The A's and B's are given by
\begin{align}
A_1=&c_x^2c_z^2+(-c_xc_ys_z+s_xs_ye^{-i\delta})^2,~A_2=s_x^2c_z^2+(-s_xc_ys_z-c_xs_ye^{-i\delta})^2,~A_3=s_z^2+c_y^2c_z^2,\nonumber\\
B_1=&c_xc_z(-c_xs_ys_z-s_xc_ye^{-i\delta})+(-c_xc_ys_z+s_xs_ye^{-i\delta})^2,\nonumber\\
B_2=&s_xc_z(-s_xs_ys_z+c_xc_ye^{-i\delta})+(-s_xc_ys_z-c_xs_ye^{-i\delta})^2,\nonumber\\
B_3=&s_yc_zs_z+c_y^2c_z^2.\label{coeffC22C12}
\end{align}

Then $R_{\nu}$ is given by
\begin{equation}
R_{\nu}=\frac{2c_y^4(c_{2x}+2c_yc_xs_xc_\d)}{\bigg|R_2\bigg| \sgn{(R_1)}}+\mathcal{O}(s_z),
\end{equation}
where
\bea
R_1 &=& 4 c_x^2 s_x^2 c_y^2 s_y^2 c_\d^2+2c_yc_xs_xs_y^2(2-c_y^2)c_{2x}c_\d+c_x^2s_x^2c_y^6+c_{2x}^2c_y^4-2c_{2x}^2c_y^2+c_{2x}^2, \\
R_2&=& 8c_x^2s_x^2c_y^2s_y^2c_\d^2+2c_yc_xs_xc_{2x} (4+c_y^4-6c_y^2) c_\d + (-6c_x^2 s_x^2 +1) c_y^4 -4 c_{2x}^2 c_y^2 + 2c_{2x}^2.
\eea

Table \ref{Predictions} shows that ($\textbf{C}_{22},\textbf{C}_{12}$) texture can not accommodate the experimental data in the case of normal ordering, whereas it is viable at the 2-3-$\sigma$ levels for inverted ordering. The allowed experimental ranges of the mixing angles $(\theta_{x},\theta_{y},\theta_{z})$ are covered at all $\sigma$-levels. We find wide disallowed regions for $\delta$ such as, $[236.26^{\circ},322^{\circ}]$ at the 2-$\sigma$-level and $[244.54^{\circ},353^{\circ}]$ at the 3-$\sigma$-level. The phases $\rho(\sigma)$ are bounded to the intervals: $[102.59^{\circ},106.02^{\circ}]([32.23^{\circ},39.70^{\circ}])$ at the 2-$\sigma$ level and $[95.33^{\circ},107.77^{\circ}]([14.34^{\circ},46.26^{\circ}])$ at the 3-$\sigma$ level. One also notes that $m_3$ does not approach a vanishing value, thus the singular mass matrix is not predicted.

\begin{figure}[hbtp]
\hspace*{-3.5cm}
\includegraphics[width=22cm, height=16cm]{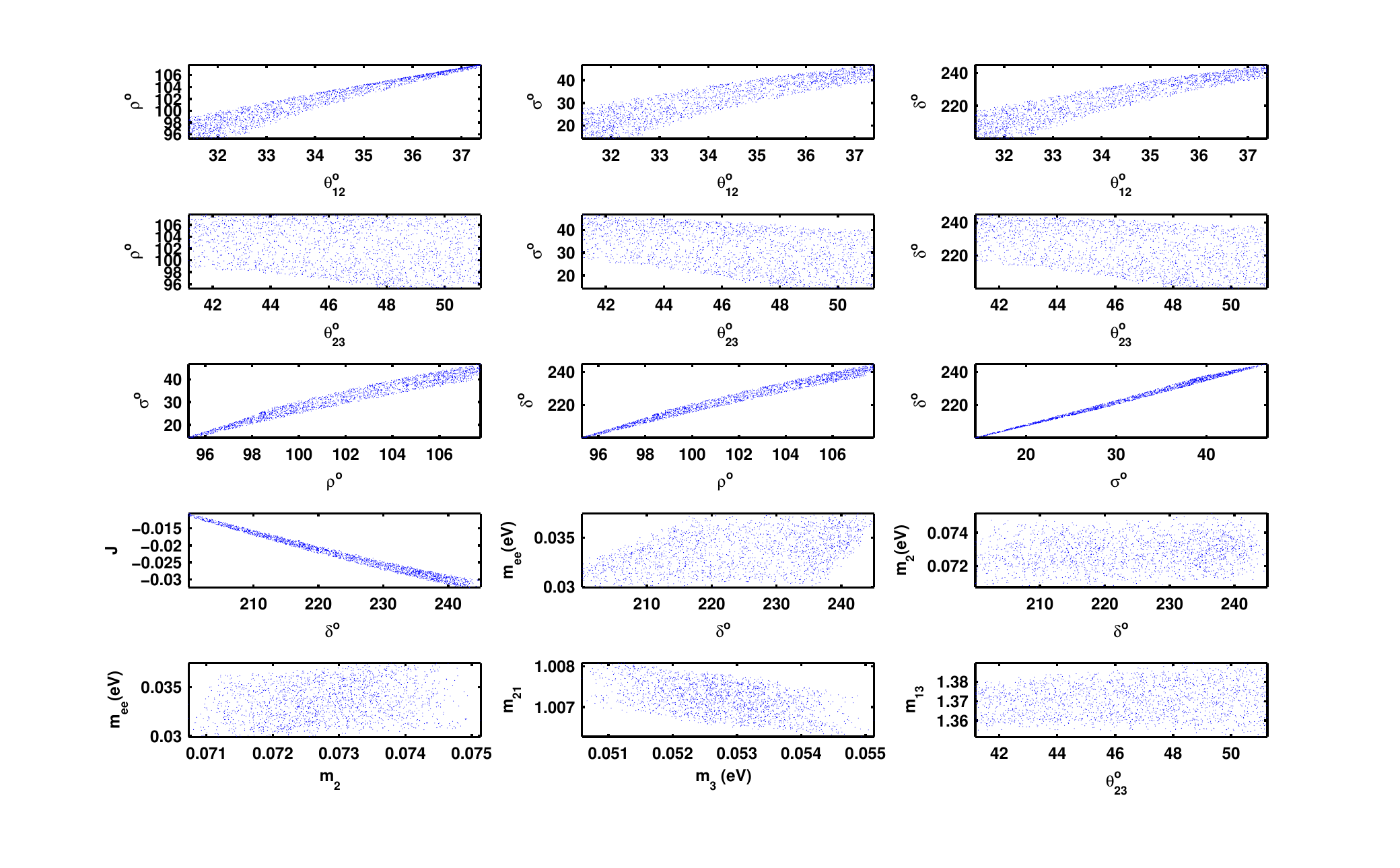}
\caption{The correlation plots for ($\textbf{C}_{22},\textbf{C}_{12}$)$\equiv$ ($M_{ee}+M_{\tau\tau}=0$, $M_{e\mu}+M_{\tau\tau}=0$) texture in the case of inverted hierarchy. The first and second rows represent the correlations between the mixing angles $(\theta_{12},\theta_{23})$ and the CP-violating phases. The third and fourth rows show the correlations amidst the CP-violating phases and the correlations between the Dirac phase $\delta$ and each of $J$, $m_{ee}$ and $m_2$ parameters respectively. The last row shows the degree of mass hierarchy plus the ($m_{ee}, m_2$) correlation.}
\label{Tr2212}
\end{figure}

From Fig. \ref{Tr2212}, we see that $\theta_{x}$ increases when CP-violating phases tend to increase. We also see a strong linear relation for ($\sigma$,$\delta$) correlation as well as  quasi-linear relations for ($\rho$,$\delta$) and ($\rho$,$\sigma$) correlations. We notice that one finds a quasi-degeneracy characterized by $1.35\leq m_{13}\leq 1.39$ and $m_1\approx m_2$.

In order to explain the correlation plots, one computes the mass-squared-difference full and approximate expressions:
\bea
\label{c22c12-full-ms-difference}
m_{23}^2-m_{13}^2 &=& \frac{\mbox{Num}(m_{23}^2-m_{13}^2)}{\mbox{Den}(m_{23}^2-m_{13}^2)}: \\
\mbox{Num}(m_{23}^2-m_{13}^2) &=& s_{2x} c_y c_\d \left[c_z^3 (c_z^2-2) c_y^4 + 2(2c_z^2 -3)c_z (s_z c_z c_y -s_z^2)c_y^2 + s_z^2 (-5c_zs_z^2+2s_ys_z)\right] \nn\\ &&
-c_{2x}\left[  (7c_y^4-2c_y^2-1) c_z^4 + 2s_zs_y(4c_y^2+c_y^4-1) c_z^3 \right.\nn\\ &&
\left. +(2-6c_y^4) c_z^2 -2s_zs_y(-1+3c_y^2)c_z +c_{2y}\right]
\nn\\
\label{c22c12-approx-ms-difference}
m_{23}^2-m_{13}^2 &=&\frac{c_y^4 (c_{2x}+2c_ys_xc_xc_\d)}{4s_x^2c_x^2s_y^2c_y^2c_\d^2+ 2c_y s_y^2 s_xc_xc_{2x}(1+s_y^2)c_\d+c_y^6c_x^2s_x^2+c_{2x}^2c_y^4-c_{2x}^2c_{2y}} +\mathcal{O}(s_z) \nn\\
\eea
We checked that the zeros of $\mbox{Num}(m_{23}^2-m_{13}^2)$ give ``exact correlations'' in excellent agreement with the ``full'' correlations, which are calculated based on numerical exact calculations taking all constraints into consideration. However, the zeros of the leading term of the mass-squared difference, i.e. of ($c_{2x}+2c_ys_xc_xc_\d$), would lead to ``approximate'' correlations which agree mediocrely with the ``full'' and one needs higher orders inclusion in order to have a better agreement. As an illustrative example, we find that the ``full'' range for $m_{13}$, spanned by the allowed points considering all experimental constraints, is $[1.35,1.39]$. Now, if we impose a zero for the ($m_{23}^2-m_{13}^2$)-expression $\left((c_{2x}+2c_ys_xc_xc_\d)-\mbox{expression}\right)$, in terms of ($\t_x,\t_y,\t_z,\d$), then one gets $\d$, say, in terms of $\t_x, \t_y, \t_z$, and so $m_{13}$ is expressed in terms of these mixing angles which, when scanned over their allowable ranges, give the ``exact'' (``approximate'') range for $m_{13}$ found to be $[1.342,1.376]$ ($[1.35,1.75]$). This corresponds to a good (mediocre) approximation, indicating we have an {\bf IH}. Moreover, plugging the zeros of $(m_{23}^2-m_{13}^2)$ in the expression of $m_{13}$ leads to
\bea
\label{c22c12-exact-m13}
m_{13} &=& \sqrt{2+t_y^2} \left(1-\frac{2s_ys_z}{c_y^2(1+c_y^2)}\right)  +\mathcal{O}(s^2_z)
\eea
We can now calculate the ``truncated exact'' range, corresponding to scanning the leading term in Eq. \ref{c22c12-exact-m13}, and we would have found $[1.03,1.32]$ , indicating again a {\bf IH}, albeit the agreement of this correlation with the ``full'' range is again mediocre.

Moreover, one can fix $\t_z \approx 8.5^o$, and for any given $\t_x$ we draw the surface of ($m_{23}^2-m_{13}^2$) varying $\t_y$ and $\d$ over their experimentally allowed regions, then the intersection of this surface with the ($m_{23}^2-m_{13}^2 = 0$) determines an ``exact'' correlation between ($\t_y$ and $\d$). We checked that juxtaposing such curves, upon varying $\t_x$, generates approximately well the ``full'' correlation ($\t_y,\d$). In the left (right) part of Fig. (\ref{C22_C12-delta_theta_y}), we take the minimum (maximum) allowed value of $\t_x=\t_{12}=31.4^o (37.4^o)$, and find that the corresponding `intersection' curves of ($\d, \t_y=\t_{23}$) delimit the corresponding correlation region.

Finally, fixing $\t_x \approx 35^o$, we find from Table \ref{Predictions} that we can take the representative points $\r \simeq 90^0, \s \simeq 30^0$, and so, with $m_1 \sim m_2 \sim 0.07$ eV in $m_{ee} \approx  \left| m_1 \cos^2(35^o)e^{i\pi} + \sin^2(35^o)e^{i\pi/3} + m_3 \sin^2(\t_z)\right|$ we get a partial cancellation of the contributions of ($m_1, m_2$) and we get $m_{ee} \sim 0.04$ eV.

Reconstructing the neutrino mass matrix for inverted ordering, the representative point is taken as following:
\begin{equation}
\begin{aligned}
(\theta_{12},\theta_{23},\theta_{13})=&(34.3208^{\circ},49.2183^{\circ}, 8.5319^{\circ}),\\
(\delta,\rho,\sigma)=&(222.6055^{\circ},101.8830^{\circ},30.2439^{\circ}),\\
(m_{1},m_{2},m_{3})=&(0.0724\textrm{ eV},0.0729\textrm{ eV},0.0527\textrm{ eV}),\\
(m_{ee},m_{e})=&(0.0319\textrm{ eV},0.0721\textrm{ eV}),
\end{aligned}
\end{equation}
and the corresponding neutrino mass matrix (in eV) is
\begin{equation}
M_{\nu}=\left( \begin {array}{ccc} -0.0319 + 0.0003i & -0.0319 + 0.0003i &  0.0564 - 0.0004i\\ \noalign{\medskip} -0.0319 + 0.0003i &  0.0531 - 0.0003i  & 0.0068 + 0.0003i
\\ \noalign{\medskip}  0.0564 - 0.0004i &  0.0068 + 0.0003i  & 0.0319 - 0.0003i
\end {array} \right).
\end{equation}

\newpage


\subsection{Texture($\textbf{C}_{11},\textbf{C}_{12}$)$\equiv$ ($M_{\mu\mu}+M_{\tau\tau}=0$, $M_{e\mu}+M_{\tau\tau}=0$) }
The A's and B's are given by
\begin{align}
A_1=&c_x^2s_z^2+s_x^2e^{-2i\delta},~A_2=s_x^2s_z^2+c_x^2e^{-2i\delta}
,~A_3=c_z^2,\nonumber\\
B_1=&c_xc_z(-c_xs_ys_z-s_xc_ye^{-i\delta})+(-c_xc_ys_z+s_xs_ye^{-i\delta})^2,\nonumber\\
B_2=&s_xc_z(-s_xs_ys_z+c_xc_ye^{-i\delta})+(-s_xc_ys_z-c_xs_ye^{-i\delta})^2,\nonumber\\
B_3=&s_yc_zs_z+c_y^2c_z^2.\label{coeffC11C12}
\end{align}

The $R_{\nu}$ approximate expression will be
\begin{equation}
R_{\nu} = \frac{2\left(s_{2x}c_yc_{\delta}c_{2y}-c_{2x}c^2_{2y}\right)}{\bigg|c^2_{2y}(1-2s_x^2 c_x^2)-s_{2x}c_{2x}c_y c_{2y}c_{\delta}\bigg|}+\mathcal{O}(s_z).
\end{equation}

From Table \ref{Predictions}, we find that ($\textbf{C}_{11},\textbf{C}_{12}$) texture can accommodate the experimental data only at the 3-$\sigma$ level for inverted ordering. We find that the allowed experimental ranges for the mixing angles $(\theta_{x},\theta_{z})$ extend over their allowed experimental ranges. However, the allowed range for $\theta_{y}$ is strongly restricted to the interval $[51.16^{\circ},51.25^{\circ}]$. We also notice that the phases $\delta$, $\rho$ and $\sigma$ are bounded to the intervals $[262.79^{\circ}, 268.92^{\circ}]$, $[5.71^{\circ},9.54^{\circ}]$ and $[168.06^{\circ},173.00^{\circ}]$ respectively. Table (\ref{Predictions}) also reveals that  $m_3$ does not reach a vanishing value. Therefore, the singular mass matrix is not expected for this texture.

\begin{figure}[hbtp]
\hspace*{-3.5cm}
\includegraphics[width=22cm, height=16cm]{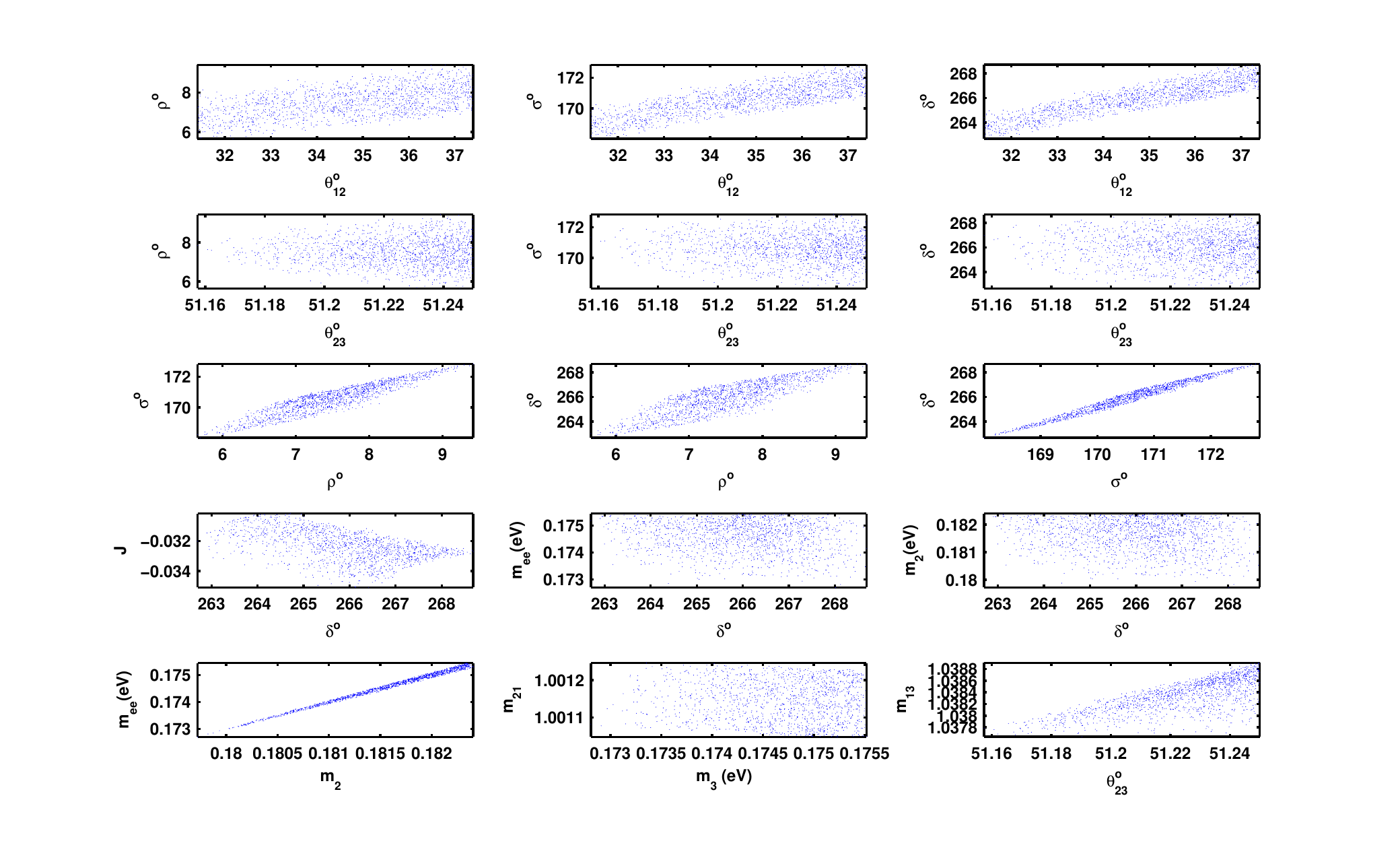}
\vspace*{-1cm}
\caption{The correlation plots for ($\textbf{C}_{11},\textbf{C}_{12}$)$\equiv$ ($M_{\mu\mu}+M_{\tau\tau}=0$, $M_{e\mu}+M_{\tau\tau}=0$) texture in the case of inverted hierarchy. The first and second rows represent the correlations between the mixing angles $(\theta_{12},\theta_{23})$ and the CP-violating phases. The third and fourth rows show the correlations amidst the CP-violating phases and the correlations between the Dirac phase $\delta$ and each of $J$, $m_{ee}$ and $m_2$ parameters respectively. The last row shows the degree of mass hierarchy plus the ($m_{ee}, m_2$) correlation.}
\label{Tr1112inv}
\end{figure}

From Fig. \ref{Tr1112inv}, we see that $\theta_{x}$ increases when the CP-violating phases tend to increase.  We also find a quasi-linear correlation between $\sigma$ and $\delta$. There exists a quasi-degeneracy characterized by $m_1\approx m_2\approx m_3$.

In order to explain the correlation plots, one computes the mass-squared-difference full and approximate expressions:
\bea
\label{c11c12-full-ms-difference}
m_{23}^2-m_{13}^2 &=& \frac{\mbox{Num}(m_{23}^2-m_{13}^2)}{\mbox{Den}(m_{23}^2-m_{13}^2)}: \\
\mbox{Num}(m_{23}^2-m_{13}^2) &=& -c_z^3\{ s_{2x}c_y \left[c_z^2(6c_y^2-5)+c_ys_{2y}s_{2z}+4s_y^2\right]c_\d -c_{2y} c_{2x} (c_zc_{2y}+2s_zs_y) \},
\nn\\
\label{c11c12-approx-ms-difference}
m_{23}^2-m_{13}^2 &=&\frac{s_{2x}c_yc_{2y}c_\d-c_{2y}^2 c_{2x}}{c_y^2 s_x^2 c_x^2}+ \frac{-2s_ys_z}{c_y^3c_x^3s_x^4}
\left[ 2s_x^2c_xc_yc_{2y}c_{2x}c_\d^2 - s_x c_\d  \right. \nn\\&&
\left. \left(c_x^4(-5s_{2y}^2+4)+5s_{2y}^2c_x^2 + c_{2y}\right)  - c_x c_y (4c_y^2-3) c_{2y}s_x^2 c_{2x}\right]
 +\mathcal{O}(s^2_z).
\eea
The zeros of $\mbox{Num}(m_{23}^2-m_{13}^2)$ give ``exact'' correlations in excellent agreement with the ``full'' correlations, and all correlations can be determined from these zeros. However, the zeros of the zeroth order leading term of the mass-squared difference would lead to (zeroth-order) ``approximate'' correlations which do not agree well with the ``full'' ones, and one has to go, say, up to the next-to-leading term in order to get (first-order) ``approximate'' correlations with a better agreement.
Actually, even the zeroth-order leading term of the $(m_{23}^2-m_{13}^2)$-expression can give useful interconnections. For example, from the constraint $m_2>m_1$, we need to have $c_{2y}c_\d>0$, whence, considering the experimental constraints on ($\t_y, \d$), we have the following observed relations
\bea \d >270^o \Rightarrow \t_y < 45^o &,& \d <270^o \Rightarrow \t_y > 45^o.
\eea
Also, we have the (zeroth-order) ``approximate'' correlation:
\bea
c_\d &=& \frac{c_{2y}c_{2x}}{s_{2x}c_y},
\eea
giving an ``approximate'' range ($\d \in[260^o,275^o]$).  Plugging the zeros of $(m_{23}^2-m_{13}^2)$ in the expression of $m_{13}$ leads to an ``exact'' correlation whose ``truncated'' approximations is given by:
\bea
\label{c11c12-exact-m13}
m_{13} = m_{23} &=& \sqrt{ 1+ \frac{c^2_{2y}}{c_y^2}}   +\mathcal{O}(s_z) \geq 1 .
\eea
This ``truncated'' correlation leads $m_{13} \overset{\geq}{\approx} 1$, so the ordering is of {\bf IH} type. Taking $\t_y$ in its allowed range, we see  that the spectrum is quasi degenerate ($m_1 \sim m_2 \sim m_3$) and $\Sigma \approx 3 m_3$.

From Table \ref{Predictions}, we find that in this texture we have ($\s \sim 170^o, \r \sim 7^o$), so in the expression of $m_{ee}= \left| m_1 c_x^2 e^{2i\r}+m_2 s_x^2 e^{2i\s}\right|$, where we put $c_z \sim 1$ and neglect the contribution of $m_3$ as it is proportional to $s_z^2$, we have $m_{ee} \approx m_2$. Similarly, we can show that $m_e \sim m_2$.

Finally, we find that the bounds on $\Sigma \sim 3 m_3$ and $m_{ee} \sim m_3$ in Eq. (\ref{non-osc-cons}) are the severe ones by which, using $m_3^2 = \frac{\d m^2}{m_{23}^2-m_{13}^2}$, most of the $\t_y$ range is excluded, and only a narrow neighbourhood around the value $\t_{y} \approx 51.2^o$ is allowed with $m_{13} \approx 1.04$ (cf. Eq. \ref{c11c12-exact-m13}).

For inverted ordering, the representative point is taken as following:
\begin{equation}
\begin{aligned}
(\theta_{12},\theta_{23},\theta_{13})=&(34.3178^{\circ},51.2346^{\circ},8.5674^{\circ}),\\
(\delta,\rho,\sigma)=&(265.0845^{\circ},6.7720^{\circ},169.8108^{\circ}),\\
(m_{1},m_{2},m_{3})=&(0.1811\textrm{ eV},0.1814\textrm{ eV},0.1745\textrm{ eV}),\\
(m_{ee},m_{e})=&(0.1744\textrm{ eV},0.1811\textrm{ eV}),
\end{aligned}
\end{equation}
the corresponding neutrino mass matrix (in eV) is
\begin{equation}
M_{\nu}=\left( \begin {array}{ccc}0.1742 + 0.0087i &  0.0305 - 0.0002i & -0.0379 - 0.0019i\\ \noalign{\medskip} 0.0305 - 0.0002i &  0.0305 - 0.0002i &  0.1719 + 0.0002i
\\ \noalign{\medskip}-0.0379 - 0.0019i &  0.1719 + 0.0002i & -0.0305 + 0.0002i\end {array} \right).
\end{equation}

\newpage


\subsection{Texture($\textbf{C}_{33},\textbf{C}_{12}$)$\equiv$ ($M_{ee}+M_{\mu\mu}=0$, $M_{e\mu}+M_{\tau\tau}=0$) }
The A's and B's are given by
\begin{align}
A_{1}=&c_x^2c_z^2+(-c_xs_ys_z-s_xc_ye^{-i\delta})^2,~A_2=s_x^2c_z^2+(-s_xs_ys_z+c_xc_ye^{-i\delta})^2,~A_3=s_z^2+s_y^2c_z^2,\nonumber\\
B_1=&c_xc_z(-c_xs_ys_z-s_xc_ye^{-i\delta})+(-c_xc_ys_z+s_xs_ye^{-i\delta})^2,\nonumber\\
B_2=&s_xc_z(-s_xs_ys_z+c_xc_ye^{-i\delta})+(-s_xc_ys_z-c_xs_ye^{-i\delta})^2,\nonumber\\
B_3=&s_yc_zs_z+c_y^2c_z^2.\label{coeffC33C12}
\end{align}
Therefore, $R_{\nu}$ takes a form
\bea
R_{\nu}&=&\frac{2c_{2x}s^2_y(1-3c_y^2)(1+t_{2x}c_yc_{\delta})}{\bigg|R_{2}\bigg| \sgn{(R_1)}} +\mathcal{O}(s_z): \\
R_1&=&-4 c_x^2 c_y^4s_x^2c_\d^2-s_{2x}s_y^2 c_y (1+c_y^2)c_{2x}c_\d + s_y^4 \left(-1+c_x^2(4-c_y^2)s_x^2\right)+\mathcal{O}(s_z), \nonumber\\
R_2&=&-8c_x^2c_y^2s_y^2s_x^2c_\d^2-s_{2x}c_ys_y^2(1+3c_y^2)c_{2x}c_\d+(3-10c_x^2s_x^2)c_y^4-4c_x^2s_x^2c_y^2-1+6c_x^2s_x^2 .\nonumber
\eea

Table \ref{Predictions} shows that ($\textbf{C}_{33},\textbf{C}_{12}$) texture is not viable at all $\sigma$ error levels for normal ordering, whereas it can accommodate the experimental data for inverted ordering at 2-3-$\sigma$ levels. The allowed experimental ranges of the mixing angles $(\theta_{x},\theta_{y},\theta_{z}$) are covered at all allowed $\sigma$-levels. The Dirac phase $\delta$ is bounded to the interval $[231.94^{\circ},248.79^{\circ}]$ at the 2-$\sigma$, and the range tends to be wider at the 3-$\sigma$ level being $[226.89^{\circ},254.21^{\circ}]$. There exist acute restrictions on the phases $\rho(\sigma)$ at the 2-3-$\sigma$ levels. They belong to the intervals $[100.59^{\circ},105.16^{\circ}]([38.47^{\circ},51.29^{\circ}])$ at the 2-$\sigma$ level and $[99.40^{\circ},106.17^{\circ}]([34.65^{\circ},57.53^{\circ}])$ at the 3-$\sigma$ level. Table \ref{Predictions} also reveals that $m_3$ does not reach zero, so the singular mass matrix is not predicted.

\begin{figure}[hbtp]
\hspace*{-3.5cm}
\includegraphics[width=22cm, height=16cm]{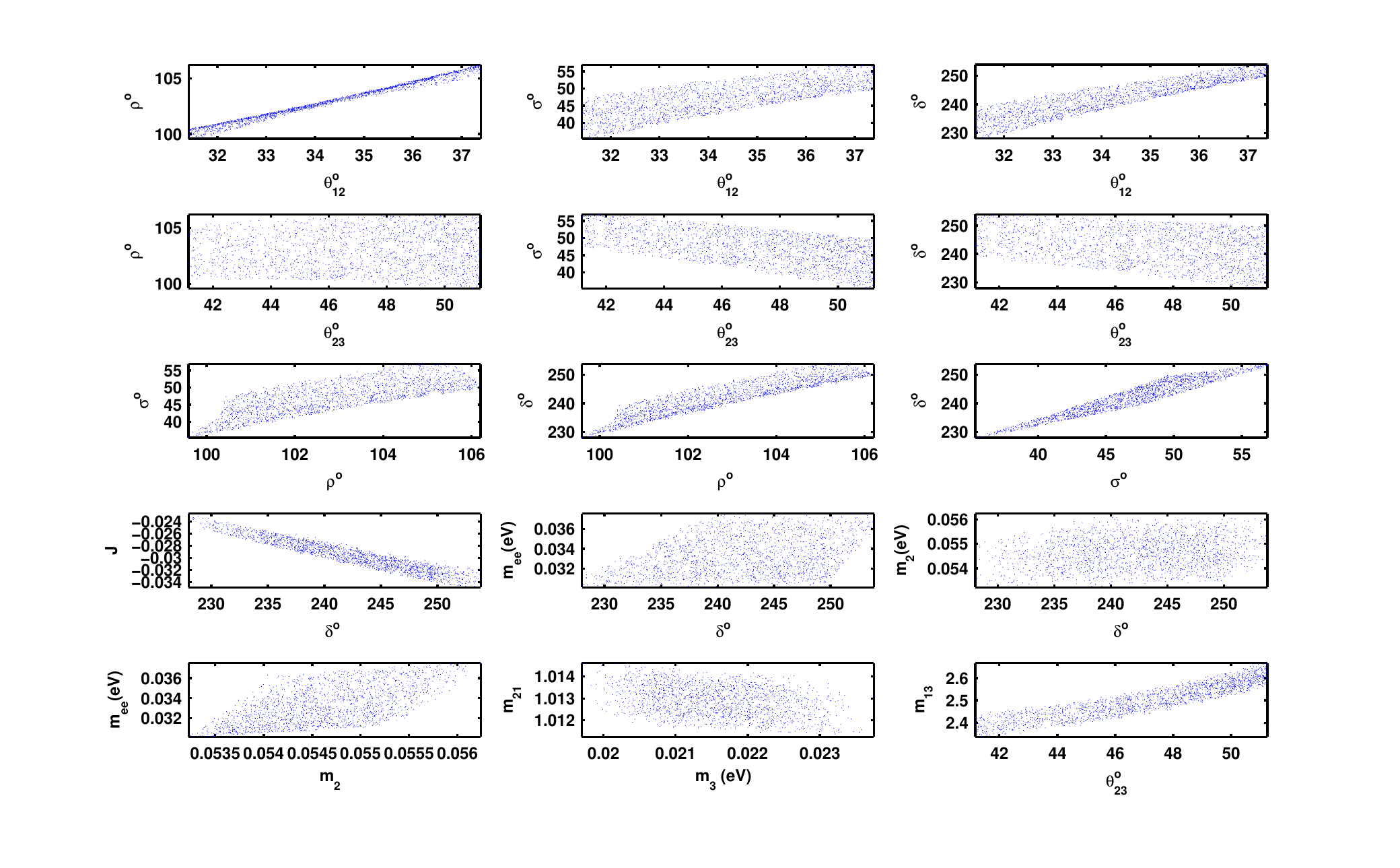}
\caption{The correlation plots for ($\textbf{C}_{33},\textbf{C}_{12}$)$\equiv$ ($M_{ee}+M_{\mu\mu}=0$, $M_{e\mu}+M_{\tau\tau}=0$) texture in the case of inverted hierarchy. The first and second rows represent the correlations between the mixing angles $(\theta_{12},\theta_{23})$ and the CP-violating phases. The third and fourth rows show the correlations amidst the CP-violating phases and the correlations between the Dirac phase $\delta$ and each of $J$, $m_{ee}$ and $m_2$ parameters respectively. The last row shows the degree of mass hierarchy plus the ($m_{ee}, m_2$) correlation.}
\label{Tr3312}
\end{figure}

From Fig. \ref{Tr3312}, we see that $\theta_{x}$ increases when the CP-violating phases tend to increase. We also see a quasi-linear relation for $(\theta_{x},\rho)$ correlation. Fig. \ref{Tr3312} also shows a moderate mass hierarchy characterized by $2.34\leq m_{13} \leq2.67$ together with a quasi-degeneracy characterized by $m_1\approx m_2$.

In order to explain the correlation plots, one computes the mass-squared-difference full and approximate expressions:
\bea
\label{c33c12-full-ms-difference}
m_{23}^2-m_{13}^2 &=& \frac{\mbox{Num}(m_{23}^2-m_{13}^2)}{\mbox{Den}(m_{23}^2-m_{13}^2)}: \\
\mbox{Num}(m_{23}^2-m_{13}^2) &=& s_{2x} c_y c_\d \left[ 3-c_y^2 (-2+c_y^2)c_z^2 + 2 s_z s_y (-2+c_y^2)c_z^4+(-2+12c_y^2+6c_y^4)c_z^3\right. \nn\\ &&
\left. +2s_zs_y(3+c_y^2) c_z^2 -c_{2y} c_z - 2 s_y s_z\right] -c_{2x} \left[ (4+c_y^4-8c_y^2) c_z^4 + 2 c_y^2 s_z s_y \right. \nn \\ && \left.
(-4 + 3 c_y^2) c_z^2 + (2c_y^4 -4 + 6 c_y^2) c_z^2 + 2 c_y^2 s_z s_y c_z - c_{2y} \right]
\nn\\
\label{c33c12-approx-ms-difference}
\mbox{Num}(m_{23}^2-m_{13}^2) &=& s_y^2 (1-3c_y^2) (s_{2x} c_y c_\d +c_{2x})
 +\mathcal{O}(s_z)
\eea
The zeros of $\mbox{Num}(m_{23}^2-m_{13}^2)$ give ``exact correlations'' in excellent agreement with the ``full'' correlations (e.g. the ``exact'' interval for $m_{13}$ is $[2.33,2.64]$ to be compared with the mentioned ``full'' interval $[2.34, 2.67]$ ). Also, the zeros of the leading term of the mass-squared difference numerator (i.e. the zeros of ($s_{2x} c_y c_\d +c_{2x}$)) lead to ``approximate'' correlations which are good, but less, when compared to the ``full'' ones.

Plugging the zeros of $(m_{23}^2-m_{13}^2)$ in the expression of $m_{13}$ leads to an ``exact'' correlation whose ``truncated'' approximation is given by:
\bea
\label{c33c12-exact-m13}
m_{13} &=& \sqrt{1+\frac{1}{c_y^2}} \left(1+ \frac{2s_z}{s_y c_y^2}\right)  +\mathcal{O}(s^2_z)
\eea
Scanning over the allowed values of $\t_y$ and $\t_z$, we find that this ``truncated'' correlation leads to ($1< m_{13} \in [2.12,2.44] $), so the ordering is of {\bf IH} type.
As to $m_{ee}$ we find a value around ( $0.054 \times \left|\cos^2(35^o)e^{2i 103 \pi/180} + \sin^2(35^o)e^{2i 47 \pi/180}\right| \approx 0.0339 $ eV ).

A corresponding benchmark point is taken as:
\begin{equation}
\begin{aligned}
(\theta_{12},\theta_{23},\theta_{13})=&(34.2034^{\circ},49.4333^{\circ},8.5030^{\circ}),\\
(\delta,\rho,\sigma)=&(240.4593^{\circ},102.7186^{\circ},44.6611^{\circ}),\\
(m_{1},m_{2},m_{3})=&(0.0531\textrm{eV},0.0539\textrm{eV},0.0208\textrm{eV}),\\
(m_{ee},m_{e})=&(0.0315\textrm{eV},0.0528\textrm{eV}),
\end{aligned}
\end{equation}
with the reconstructed neutrino mass matrix (in eV) given as:
\begin{equation}
M_{\nu}=\left( \begin {array}{ccc} -0.0314 + 0.0014i & -0.0212 + 0.0012i &  0.0367 - 0.0017i\\ \noalign{\medskip}-0.0212 + 0.0012i &  0.0314 - 0.0014i & -0.0076 + 0.0013i
\\ \noalign{\medskip}0.0367 - 0.0017i & -0.0076 + 0.0013i &  0.0212 - 0.0012i
\end {array} \right).
\end{equation}

\newpage


\subsection{Texture($\textbf{C}_{33},\textbf{C}_{13}$)$\equiv$ ($M_{ee}+M_{\mu\mu}=0$, $M_{e\mu}+M_{\mu\tau}=0$) }
The A's and B's are given by
\begin{align}
A_{1}=&c_x^2c_z^2+(-c_xs_ys_z-s_xc_ye^{-i\delta})^2,~A_2=s_x^2c_z^2+(-s_xs_ys_z+c_xc_ye^{-i\delta})^2,~A_3=s_z^2+s_y^2c_z^2,\nonumber\\
B_1=&c_xc_z(-c_xs_ys_z-s_xc_ye^{-i\delta})+(-c_xs_ys_z-s_xc_ye^{-i\delta})(-c_xc_ys_z+s_xs_ye^{-i\delta}),\nonumber\\
B_2=&s_xc_z(-s_xs_ys_z+c_xc_ye^{-i\delta})+(-s_xs_ys_z+c_xc_ye^{-i\delta})(-s_xc_ys_z-c_xs_ye^{-i\delta}),\nonumber\\
B_3=&s_zs_yc_z+s_yc_z^2c_y.\label{coeffC33C13}
\end{align}
The leading order truncated approximation for $R_{\nu}$ is given by
\bea
R_{\nu}&=&\frac{-2s_y^3}{\bigg| s_{2x}c_{2y}c_{\delta}-c_{2x}s_y(1+c_y^2)\bigg| \sgn(R_1)}+\mathcal{O}(s_z): \\
R_1 &=& -2c_y^4 s_{2x}^2 c_\d^2 + c_y s_{2y} s_{2x} (1+c_y^2)c_{2x}c_\d -2 c_y^2 s_y^2 (-c_y^2 s_x^2 c_x^2 -3 s_x^2 c_x^2 +1). \nonumber
\eea

We see from Table (\ref{Predictions}) that ($\textbf{C}_{33},\textbf{C}_{13}$) texture can accommodate the experimental data in the case of inverted hierarchy at all $\sigma$ levels. However, the texture is not viable for normal hierarchy. We find that the mixing angles $(\theta_{x},\theta_{y},\theta_{z})$ extend over their allowed experimental ranges at all $\sigma$ levels. The Dirac phase $\delta$ is tightly restricted at all $\sigma$-levels, and is bound to be in the range $[262.85^{\circ},267.59^{\circ}]$ at the 3-$\sigma$ level. We notice that the Majorana phases $\rho$($\sigma$) are strongly restricted at all statistical levels to lie in the intervals $[92.56^{\circ},95.16^{\circ}]([78.67^{\circ},84.28^{\circ}])$ at the 3-$\sigma$ level. Table ($\ref{Predictions}$) also reveals that $m_3$ does not reach a vanishing at all $\sigma$-levels. Therefore, the singular texture is not expected with either hierarchy type at all $\sigma$-levels.

\begin{figure}[hbtp]
\hspace*{-3.5cm}
\includegraphics[width=22cm, height=14.2cm]{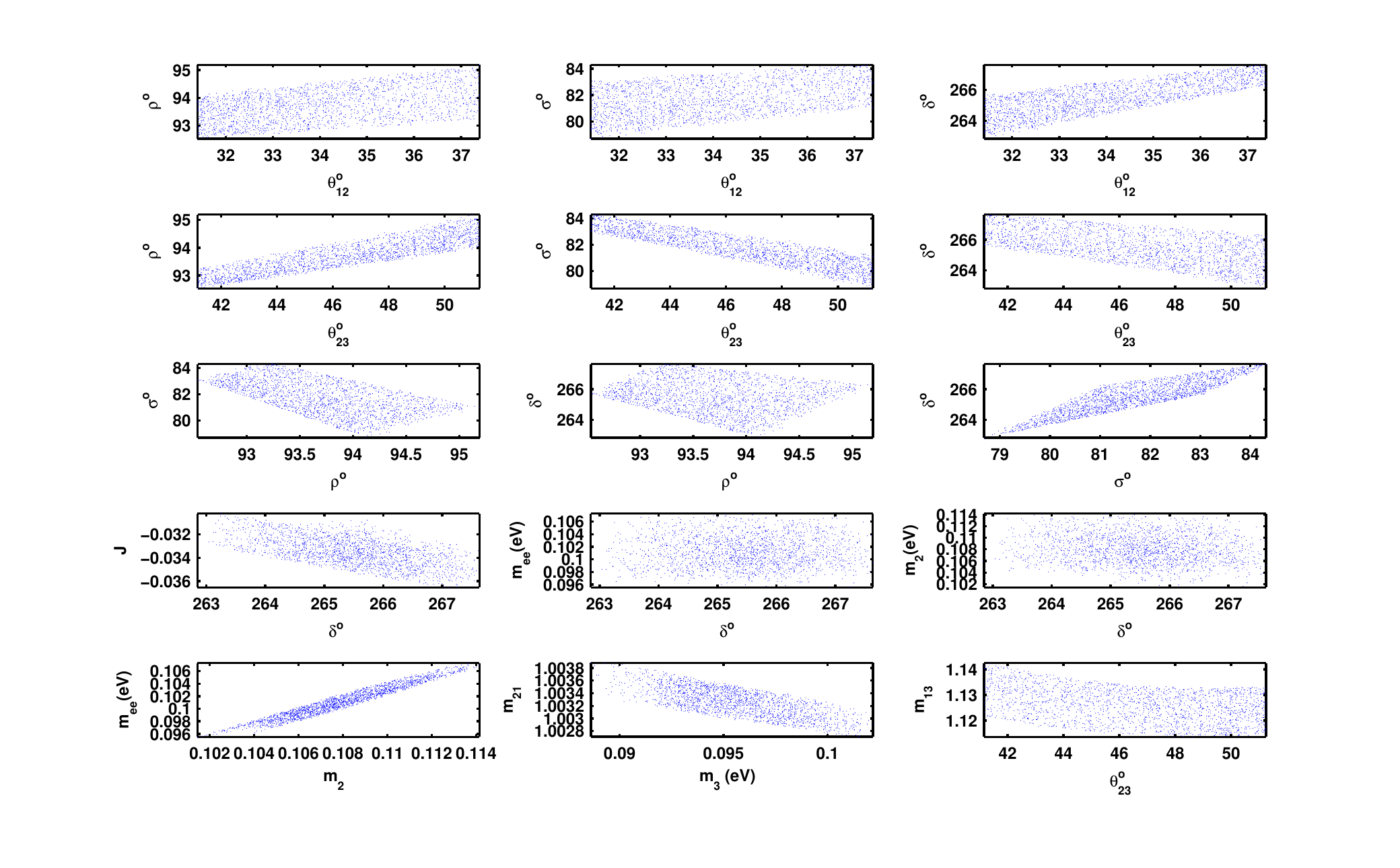}
\caption{The correlation plots for ($\textbf{C}_{33},\textbf{C}_{13}$)$\equiv$ ($M_{ee}+M_{\mu\mu}=0$, $M_{e\mu}+M_{\mu\tau}=0$) texture in the case of inverted hierarchy. The first and second rows represent the correlations between the mixing angles $(\theta_{12},\theta_{23})$ and the CP-violating phases. The third and fourth rows show the correlations amidst the CP-violating phases and the correlations between the Dirac phase $\delta$ and each of $J$, $m_{ee}$ and $m_2$ parameters respectively. The last row shows the degree of mass hierarchy plus the ($m_{ee}, m_2$) correlation.}
\label{Tr3313}
\end{figure}

We see from Fig. \ref{Tr3313} that the mixing angle $\theta_{y}$ increases when the phases $\delta$ and $\sigma$ tend to decrease. However, we notice that $\theta_{y}$ increases when $\rho$ tends to increase. We also see that $\theta_{x}$ increases when the CP-violating phases tend to increase. We find the quasi-degeneracy characterized by $m_{1}\approx m_2$ and $1.11\leq m_{13}\leq1.14$.

In order to explain the correlation plots, one computes the mass-squared-difference full and approximate expressions:
\bea
\label{c33c13-full-ms-difference}
m_{23}^2-m_{13}^2 &=& \frac{\mbox{Num}(m_{23}^2-m_{13}^2)}{\mbox{Den}(m_{23}^2-m_{13}^2)}: \\
\mbox{Num}(m_{23}^2-m_{13}^2) &=& 2c_zs_{2x}s_yc_\d \left[ c_z^4 c_y^2 (c_y^2-3) -2 c_z^3 s_z c_y s_y^2 + c_z^2 (6c_y^2 -2 c_y^4 -1) + 4 s_z c_z c_y s_y^2 -c_{2y}\right]
\nn\\ && +4c_z s_y^2 c_{2x} \left[ c_z^3 c_{2y}-s_z c_z^2 c_y (c_y^2-2)-c_zc_{2y}-s_zc_y\right],
\nn\\
\label{c33c13-approx-ms-difference}
m_{23}^2-m_{13}^2 &=& \frac{c_ys_{2x}s_{2y}s_y^2c_\d}{c_y^4s_{2x}^2c_\d^2 -\frac{1}{2} c_y s_{2x} s_{2y} (1+c_y^2) c_{2x} c_\d + s_y^2 c_y^2 (-c_x^2 s_x^2 c_y^2 -3 s_x^2 c_x^2 +1)}
 +\mathcal{O}(s_z)
\eea
We understand now why $\d$ around $270^o$ is singled out, as this would make ($m_{23}^2 - m_{13}^2$) as small as possible (c.f. Eq. \ref{c33c13-approx-ms-difference}), and, moreover, substituting ($\d \approx 270^0$) in the truncated approximation we get
\[
m_{23}^2-m_{13}^2 \overset{\d\rightarrow 270^o}{\approx} \frac{c_ys_{2x}s_{2y}s_y^2c_\d}{s_y^2 c_y^2 (-c_x^2 s_x^2 c_y^2 -3 s_x^2 c_x^2 +1)}
\]
As the coefficient in front of $c_\d$ in the numerator is positive for allowed $\t_x, \t_y$, whereas the denominator is always positive, we deduce ($\left(m_{23}^2-m_{13}^2 \rightarrow 0^+\right) \Rightarrow \left(\d\rightarrow 270^{o+}\right)  $ ). The higher order terms make ($\d\rightarrow 270^{o-}$).

The zeros of $\mbox{Num}(m_{23}^2-m_{13}^2)$ give ``exact correlations'' in excellent agreement with the ``full'' correlations. However, we found that the zeros of the leading plus next to leading  terms of the mass-squared difference numerator (i.e. the expansion of $m_{23}^2-m_{13}^2$ in the form of a linear form $c_0 + c_1 s_z$) lead to ``approximate'' correlations which are not accurate, when compared to the ``full'' ones, and one needs to go to higher orders to match the ``full'' correlations.

Plugging the zeros of $(m_{23}^2-m_{13}^2)$ in the expression of $m_{13}$ leads to an ``exact'' correlation whose ``truncated'' approximation is given by:
\bea
\label{c33c13-exact-m13}
m_{13} &=& 1+ \frac{2s_z^2}{s_y^2 c_y^2}  +\mathcal{O}(s^3_z)
\eea
Scanning over the allowed values of $\t_y. \t_z$, we find that this ``truncated'' correlation leads to ($m_{13} \in [1.165,1.205] $), whereas the ``exact'' range, coming from the zeros of ($m_{23}^2-m_{13}^2$), is $[1.114,1.143]$, which is very near the ``full'' correct range, so the ordering is of {\bf IH} type.
As to $m_{ee}$, and since we have $\r \approx \s \approx 90^o$ in this pattern, we have  ($m_{ee} \approx m_2$). Actually, at leading order, $\d \approx \frac{3\pi}{2}$ would lead to ($\rho \approx \sigma \approx \frac{\pi}{2}$).

For inverted ordering, the representative point is taken as following:
\begin{equation}
\begin{aligned}
(\theta_{12},\theta_{23},\theta_{13})=&(34.1770^{\circ},49.4202^{\circ},8.6885^{\circ}),\\
(\delta,\rho,\sigma)=&(264.9329^{\circ},94.2458^{\circ},80.6349^{\circ}),\\
(m_{1},m_{2},m_{3})=&(0.1076\textrm{ eV},0.1080\textrm{ eV},0.0955\textrm{ eV}),\\
(m_{ee},m_{e})=&(0.1005\textrm{ eV},0.1074\textrm{ eV}),
\end{aligned}
\end{equation}
the corresponding neutrino mass matrix (in eV) is
\begin{equation}
M_{\nu}=\left( \begin {array}{ccc} -0.1005 + 0.0001i &  0.0076 - 0.0001i & 0.0372 + 0.0001i\\ \noalign{\medskip} 0.0076 - 0.0001i & 0.1005 - 0.0001i & -0.0076 + 0.0001i
\\ \noalign{\medskip} 0.0372 + 0.0001i & -0.0076 + 0.0001i &  0.0957 - 0.0001i \end {array} \right).
\end{equation}

\newpage

\subsection{Texture($\textbf{C}_{11},\textbf{C}_{13}$)$\equiv$ ($M_{\mu\mu}+M_{\tau\tau}=0$, $M_{e\mu}+M_{\mu\tau}=0$) }
The coefficients A's and B's are obtained from Eqs. (\ref{coeffC11C12},\ref{coeffC33C13}).
The A's and B's are given by
\begin{align}
A_1=&c_x^2s_z^2+s_x^2e^{-2i\delta},~A_2=s_x^2s_z^2+c_x^2e^{-2i\delta}
,~A_3=c_z^2,\nonumber\\
B_1=&c_xc_z(-c_xs_ys_z-s_xc_ye^{-i\delta})+(-c_xs_ys_z-s_xc_ye^{-i\delta})(-c_xc_ys_z+s_xs_ye^{-i\delta}),\nonumber\\
B_2=&s_xc_z(-s_xs_ys_z+c_xc_ye^{-i\delta})+(-s_xs_ys_z+c_xc_ye^{-i\delta})(-s_xc_ys_z-c_xs_ye^{-i\delta}),\nonumber\\
B_3=&s_zs_yc_z+s_yc_z^2c_y.\label{coeffC11C13}
\end{align}
The analytical approximate truncated expression for $R_{\nu}$ is
\begin{equation}
R_{\nu}=\frac{4 \left(s_{2x}c_{\delta}-2c_{2x}s_y\right)}{\bigg|s_y(2s_{2x}^2-4)+s_{4x}c_{\delta}\bigg|}+\mathcal{O}(s_z).
\end{equation}

From Table \ref{Predictions}, we find that ($\textbf{C}_{11},\textbf{C}_{13}$) is not viable at all $\sigma$-levels for normal hierarchy. However, it can accommodate the experimental data at all $\sigma$ levels in the case of inverted hierarchy. The mixing angles ($\theta_{x},\theta_{y},\theta_{z}$) extend over their allowed experimental ranges at the all $\sigma$-levels. We find that the allowed range for $\delta$ is very tight at all $\sigma$ levels. It tends to be wider at the 3-$\sigma$ level to be approximately $[292^{\circ},322^{\circ}]$.
 As for the Dirac phase $\delta$, the Majorana phases $\rho(\sigma)$ are strongly restricted at all $\sigma$ levels. They belong to the intervals: $[165.23^{\circ},167.75^{\circ}]([46.81^{\circ},52.37^{\circ}])$ at the 1-$\sigma$ level, $[163.81^{\circ},169.35^{\circ}]([43.84^{\circ},56.57^{\circ}])$ at the 2-$\sigma$ level and $[162.53^{\circ},170.80^{\circ}]([39.98^{\circ},60.41^{\circ}])$ at the 3-$\sigma$ level. One also notes that $m_3$ does not reach a vanishing value. Thus, the singular mass matrix is not expected.

\begin{figure}[hbtp]
\hspace*{-3.5cm}
\includegraphics[width=22cm, height=16cm]{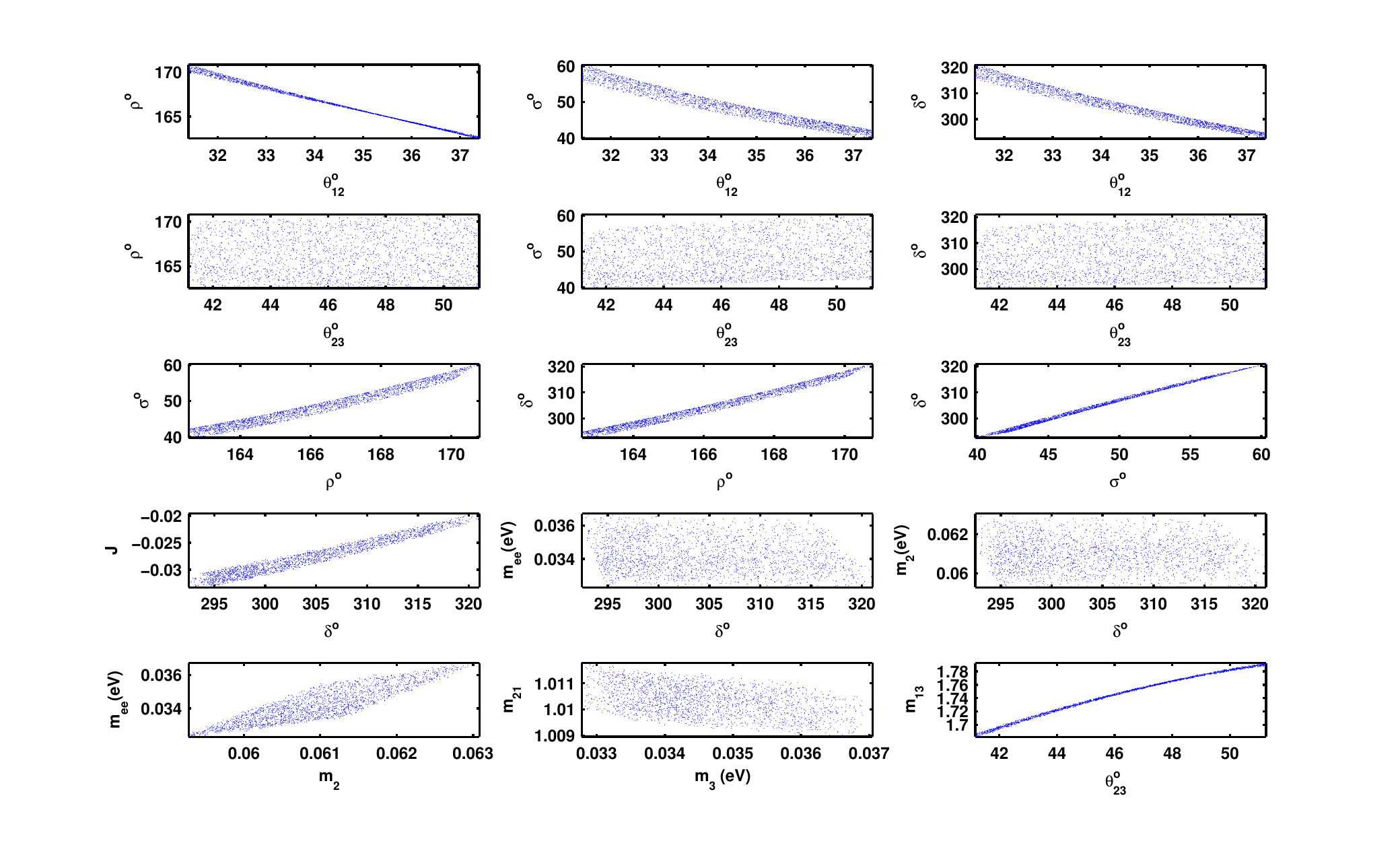}
\caption{The correlation plots for ($\textbf{C}_{11},\textbf{C}_{13}$)$\equiv$ ($M_{ee}+M_{\mu\mu}=0$, $M_{e\mu}+M_{\mu\tau}=0$) texture in the case of inverted hierarchy. The first and second rows represent the correlations between the mixing angles $(\theta_{12},\theta_{23})$ and the CP-violating phases. The third and fourth rows show the correlations amidst the CP-violating phases and the correlations between the Dirac phase $\delta$ and each of $J$, $m_{ee}$ and $m_2$ parameters respectively. The last row shows the degree of mass hierarchy plus the ($m_{ee}, m_2$) correlation.}
\label{Tr1113}
\end{figure}

We see from Fig. \ref{Tr1113} the quasi-linear correlations with negative slope between $\theta_{x}$ and CP-violating phases.  We also see a strong linear relation for the correlation ($\sigma,\delta$) together with quasi-linear relations for ($\rho,\sigma$) and ($\rho,\delta$) correlations. We also find a mild mass hierarchy where $1.68\leq m_{13} \leq 1.79$ as well as a quasi-degeneracy characterized by $m_1\approx m_2$.

In order to explain the correlation plots, one computes the mass-squared-difference full and approximate expressions:
\bea
\label{c11c13-full-ms-difference}
m_{23}^2-m_{13}^2 &=& \frac{\mbox{Num}(m_{23}^2-m_{13}^2)}{\mbox{Den}(m_{23}^2-m_{13}^2)}: \\
\mbox{Num}(m_{23}^2-m_{13}^2) &=& 4c_z^3 \left[ \frac{1}{2} s_{2x}s_y c_\d \left( (1-3c_y^2) c_z^2 -c_y s_{2z} s_y^2 + c_{2y}\right) + c_y s_y^2 c_{2x} (c_y c_z + s_z)\right]
\nn\\
\label{c11c13-approx-ms-difference}
m_{23}^2-m_{13}^2 &=& \frac{2s_y \left(c_\d s_{2x} -2 s_y c_{2x}\right)}{c_x^2 s_x^2}
 +\mathcal{O}(s_z)
\eea

The zeros of $\mbox{Num}(m_{23}^2-m_{13}^2)$ give ``exact correlations'' in excellent agreement with the ``full'' correlations. Likewise, we found that the zeros of the leading term of the mass-squared difference numerator (i.e. of $\left(c_\d s_{2x} -2 s_y c_{2x}\right)$ giving $(c_\d=2s_y\cot_{2x})$) lead to ``approximate'' correlations between the mixing and phase angles which are also good when compared to the ``full'' ones.

Plugging the zeros of $(m_{23}^2-m_{13}^2)$ in the expression of $m_{13}$ leads to an ``exact'' correlation whose `truncated' approximation is given by:
\bea
\label{c11c13-exact-m13}
m_{13} &=& \sqrt{1+4s_y^2} \left(1+ \frac{4 s_y^2 c_{2y} s_z}{1+4s_y^2 }\right)  +\mathcal{O}(s^2_z)
\eea
Scanning over the allowed values of $\t_y, \t_z$, we find that this ``truncated'' correlation leads to ($m_{13} \in [1.68,1.79] $), whereas the ``exact'' range, coming from the zeros of ($m_{23}^2-m_{13}^2$), is $[1.7,1.8]$, which is very near the ``full'' correct range, so the ordering is of {\bf IH} type.
As to $m_{ee}$, and since we have $\r \approx 167^o, \s \approx 50^o$ in this pattern, we have, taking $\t_x \approx 35^o, m_2 \approx 0.06 $ eV, the value ($m_{ee} \approx 0.06 |\cos^2(35^o) e^{2i167\pi/180}+\sin^2(35^o) e^{2i50\pi/180}| \approx 0.032$ eV).

For inverted ordering, the representative point is taken as following:
\begin{equation}
\begin{aligned}
(\theta_{12},\theta_{23},\theta_{13})=&(34.2712^{\circ},49.4721^{\circ},8.6197^{\circ}),\\
(\delta,\rho,\sigma)=&(306.3395^{\circ},166.5813^{\circ},49.8309^{\circ}),\\
(m_{1},m_{2},m_{3})=&(0.0601\textrm{ eV},0.0607\textrm{ eV},0.0338\textrm{ eV}),\\
(m_{ee},m_{e})=&(0.0334\textrm{ eV},0.0598\textrm{ eV}),
\end{aligned}
\end{equation}
the corresponding neutrino mass matrix (in eV) is
\begin{equation}
M_{\nu}=\left( \begin {array}{ccc} 0.0334 + 0.0004i & -0.0322 - 0.0000i &  0.0378 - 0.0001i\\ \noalign{\medskip} -0.0322 - 0.0000i &  0.0127 - 0.0000i  & 0.0322 + 0.0000i
\\ \noalign{\medskip} 0.0378 - 0.0001i &  0.0322 + 0.0000i &-0.0127 + 0.0000i
\end {array} \right).
\end{equation}

\newpage

\subsection{Texture($\textbf{C}_{13},\textbf{C}_{23}$)$\equiv$ ($M_{e\mu}+M_{\mu\tau}=0$, $M_{ee}+M_{\mu\tau}=0$) }
The A's and B's are given by
\begin{align}
A_1=&c_xc_z(-c_xs_ys_z-s_xc_ye^{-i\delta})+(-c_xs_ys_z-s_xc_ye^{-i\delta})(-c_xc_ys_z+s_xs_ye^{-i\delta}),\nonumber\\
A_2=&s_xc_z(-s_xs_ys_z+c_xc_ye^{-i\delta})+(-s_xs_ys_z+c_xc_ye^{-i\delta})(-s_xc_ys_z-c_xs_ye^{-i\delta}),\nonumber\\
A_3=&s_zs_yc_z+s_yc_z^2c_y,\nonumber\\
B_1=&c_x^2c_z^2+(-c_xs_ys_z-s_xc_ye^{-i\delta})(-c_xc_ys_z+s_xs_ye^{-i\delta}),\nonumber\\
B_2=&s_x^2c_z^2+(-s_xs_ys_z+c_xc_ye^{-i\delta})(-s_xc_ys_z-c_xs_ye^{-i\delta}),\nonumber\\
B_3&=s_z^2+s_yc_yc_z^2.
\label{coeffC13C23}
\end{align}
The leading order expression for $R_{\nu}$ is given by
\bea
R_{\nu}=\frac{2s_y^2(c_{2x}+s_{2x}c_yc_{\delta})}{\bigg|s_{2y}s_{2x}^2c_\d^2+s_{2x}c_{2x}s_y (2-s_yc_y)c_\d + (1-6s_x^2 c_x^2) c_y^2 -2 s_{2y} s_x^2 c_x^2 -c_{2x}^2 \bigg| \sgn(R_1)}+\mathcal{O}(s_z) : \\
R_1 = -2 s_{2y}s_x^2c_x^2c_\d^2-s_{2x} c_{2x} s_y (1-s_yc_y)c_\d -s_x^2 c_x^2c_y^4+(5c_x^2s_x^2-1)c_y^2+s_{2y} s_x^2c_x^2+1-3s_x^2c_x^2. \nonumber
\eea

We see from Table \ref{Predictions} that ($\textbf{C}_{13},\textbf{C}_{23}$) texture is viable at all $\sigma$-levels for normal ordering. However, it can not accommodate the experimental data for inverted ordering. The allowed experimental ranges for the mixing angles $(\theta_{x},\theta_{y},\theta_{z})$ can be covered at all $\sigma$-levels. The Dirac phase $\delta$ is bounded to the intervals: $[202.90^{\circ},217.99^{\circ}]$ at the 1-$\sigma$ level, $[152.02^{\circ},232.55^{\circ}]$ at the 2-$\sigma$ level and $[128.01^{\circ},242.79^{\circ}]$ at the 3-$\sigma$ level. We find that $\rho$ is tightly restricted at all $\sigma$ levels, and it's allowed range tends to be wider at the 3-$\sigma$ levels to fall approximately in the interval $[73^{\circ},108^{\circ}]$. For the phase $\sigma$, one notes that there exists a strong restriction at the 1-$\sigma$ level besides  wide forbidden gaps at the 2-3-$\sigma$ levels. The allowed values for the J parameter at the 1-$\sigma$-level are negative, consistent with $\delta$ lying in the third quarter at this $\s$-level. Table \ref{Predictions} also shows that $m_1$ can not reach zero. Thus, a singular mass matrix is not predicted.

\begin{figure}[hbtp]
\hspace*{-3.5cm}
\includegraphics[width=22cm, height=16cm]{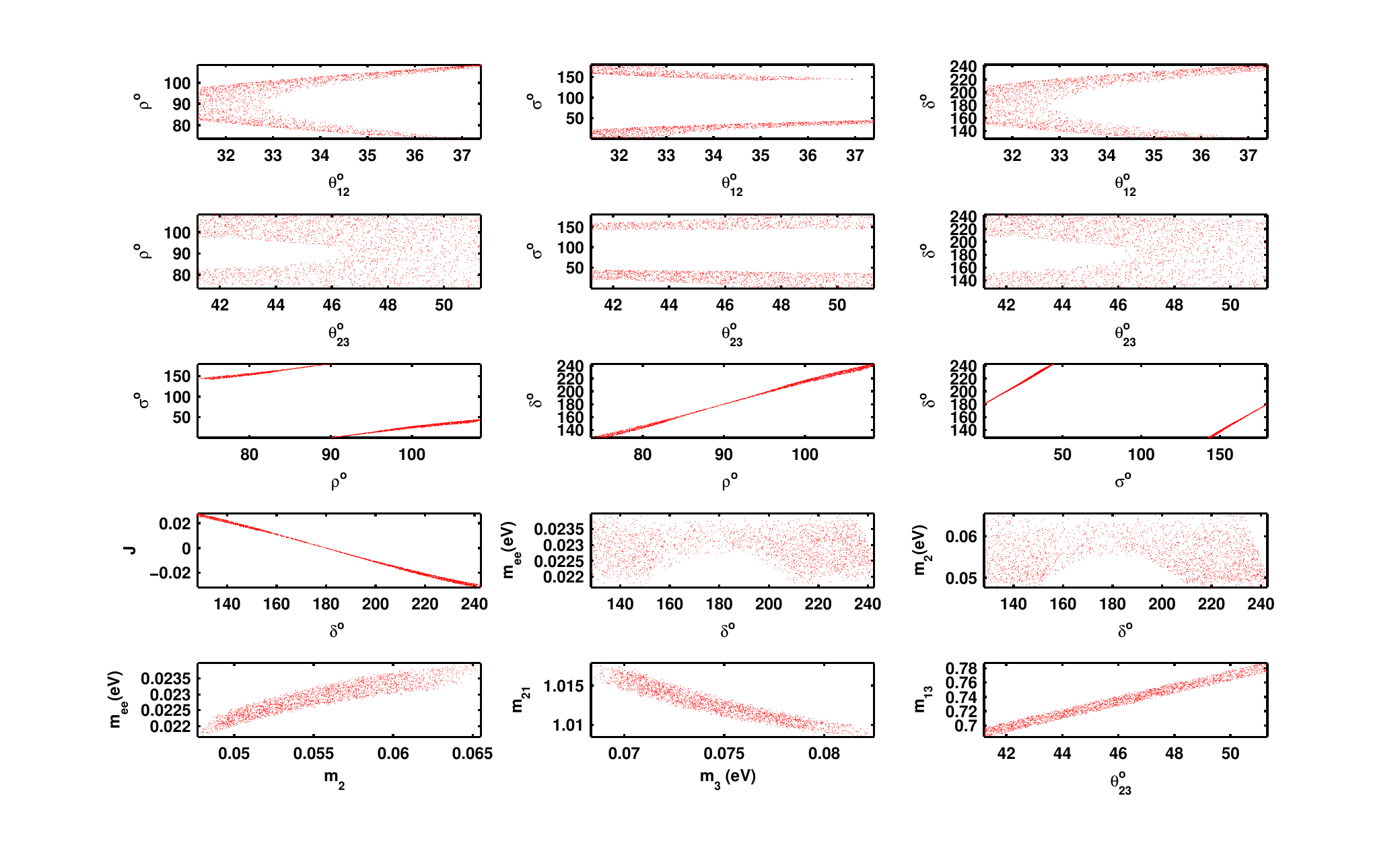}
\caption{The correlation plots for ($\textbf{C}_{13},\textbf{C}_{23}$)$\equiv$ ($M_{e\mu}+M_{\mu\tau}=0$, $M_{ee}+M_{\mu\tau}=0$) texture in the case of normal hierarchy. The first and second rows represent the correlations between the mixing angles $(\theta_{12},\theta_{23})$ and the CP-violating phases. The third and fourth rows show the correlations amidst the CP-violating phases and the correlations between the Dirac phase $\delta$ and each of $J$, $m_{ee}$ and $m_2$ parameters respectively. The last row shows the degree of mass hierarchy plus the ($m_{ee}, m_2$) correlation.}
\label{Tr1323norm}
\end{figure}

From Fig. \ref{Tr1323norm}, we see forbidden gaps in the correlations between the mixing angles ($\theta_{x}$,$\theta_{y}$) and CP-violating phases. We also see the quasi-linear relations for the correlations between the CP-violating phases. Fig. \ref{Tr1323norm} also shows a mild mass hierarchy characterized by $0.68\leq m_{13}\leq0.79$ together with a quasi-degeneracy characterized by $m_1\approx m_2$.

In order to explain the correlation plots, one computes the mass-squared-difference full and approximate expressions:
\bea
\label{c13c23-full-ms-difference}
m_{23}^2-m_{13}^2 &=& \frac{\mbox{Num}(m_{23}^2-m_{13}^2)}{\mbox{Den}(m_{23}^2-m_{13}^2)}: \\
\mbox{Num}(m_{23}^2-m_{13}^2) &=& s_{2x} c_\d \left[ c_z^4 s_y (3c_y^2-1-c_y s_y^3) + s_z s_y^2 c_z^3 (s_{2y} -s_y^2) + c_z^2 \left( (2-5c_y^2) s_y -c_y s_y^2 c_{2y}\right)
\nn \right.\\&&\left. -s_z c_z s_y^2 (s_{2y}+c_{2y}) +s_y c_{2y}\right] + s_y^2 c_{2x} \left[ c_z^3 (s_{2y}+1-3c_y^2)-2c_ys_zc_z^2 (2+s_yc_y)\right. \nn \\
&& \left. +c_z (c_{2y}-s_{2y})+2s_zc_y\right]
\nn\\
 &=& - c_y^2 s_y^2 (c_\d s_{2x}c_y + c_{2x}) +\mathcal{O}(s_z) .\nn
\eea

The zeros of $\mbox{Num}(m_{23}^2-m_{13}^2)$ give ``exact correlations'' in excellent agreement with the ``full'' correlations. Likewise, we found that the zeros of the leading term of the mass-squared difference numerator  giving $(c_\d=-\frac{1}{t_{2x}c_y}$) lead to ``approximate'' correlations between the mixing angles ($\t_x, \t_y$) and the Dirac phase angle $\d$ which are also good when compared to the `full' ones. Moreover, we see that $c_\d <0$, which interprets the observation that $\d$ lies in the second or third quadrant.

Plugging the zeros of $(m_{23}^2-m_{13}^2)$ in the expression of $m_{13}$ leads to an ``exact'' correlation whose ``truncated'' approximation is given by:
\bea
\label{c13c23-exact-m13}
m_{13} &=& \frac{s_y \sqrt{1+c_y^2}}{\sqrt{1+s_{2y}+c_y^2 s_y^2}} \left(1+ \frac{3 (s_y^3 + c_y^3) s_z}{1+6c_y^6 }\right)  +\mathcal{O}(s^2_z)
\eea
One can see, for the allowed values of ($\t_y$), that the zeroth-order leading term ($\frac{s_y \sqrt{1+c_y^2}}{\sqrt{1+s_{2y}+c_y^2 s_y^2}} < 1$), so the ordering is of {\bf NH} type.
Scanning over the allowed values of $\t_y, \t_z$, we find that the ``truncated'' correlation, up to order $\mathcal{O}(s^2_z)$,  leads to ($m_{13} \in [0.69,0.82] $), whereas the ``exact'' range, coming from the zeros of ($m_{23}^2-m_{13}^2$), is $[0.69,0.79]$, which is very near the ``full'' correct range $[0.68,0.79]$.

For normal ordering, the representative point is taken as following:
\begin{equation}
\begin{aligned}
(\theta_{12},\theta_{23},\theta_{13})=&(34.1349^{\circ},49.3654^{\circ},8.5098^{\circ}),\\
(\delta,\rho,\sigma)=&(214.0038^{\circ},100.1902^{\circ},23.9810^{\circ}),\\
(m_{1},m_{2},m_{3})=&(0.0594\textrm{ eV},0.0600\textrm{ eV},0.0777\textrm{ eV}),\\
(m_{ee},m_{e})=&(0.0232\textrm{ eV},0.0600\textrm{ eV}),
\end{aligned}
\end{equation}
the corresponding neutrino mass matrix (in eV) is
\begin{equation}
M_{\nu}=\left( \begin {array}{ccc} -0.0232 - 0.0001i & -0.0232 - 0.0001i &  0.0503 + 0.0002i\\ \noalign{\medskip}-0.0232 - 0.0001i &  0.0624 - 0.0001i &  0.0232 + 0.0001i
\\ \noalign{\medskip}0.0503 + 0.0002i &  0.0232 + 0.0001i &  0.0391 - 0.0002i
\end {array} \right).
\end{equation}

\newpage


\section{Theoretical realization}
We present now some realizations of the texture under study characterized by two vanishing subtraces, irrespective of whether the corresponding texture is viable or not regarding phenomenological data. We present first a symmetry based on the non-abelian group $A_4$ leading to a texture with the related subtraces consisted of the sum of diagonal elements. Second, we present a symmetry based on the non-abelian group $S_4$ where one of the related subtraces corresponded to non-diagonal elements. In the realization, we introduce new scalars, but we have not discussed the question of the scalar potential and finding its general form under the imposed symmetry.  Having these scalars may lead to rich phenomenology at colliders, and asking for just one SM-like Higgs at low scale requires a situation where fine
tuning of the many parameters in the scalar potential, so that to ensure new scalars are out of reach at current experiments, is heavily called upon.

Note that for each presented realizable pattern, there are automatically two other realizable patterns by `transposition'. Thus, by presenting an $S_4$-realization for the hitherto viable but now disallowed ($C_{11}, C_{23}$), then automatically we have, by the transposition ($1\leftrightarrow 3$), a realization for the viable pattern ($C_{33}, C_{21}$), and another realization, by doing now the transposition ($2 \leftrightarrow 3$) on the latter, for the unviable pattern ($C_{22}, C_{31}$). Similarly, the $A_4$-realization of the viable pattern ($C_{22}, C_{33}$) can automatically be translated into a realization of the unviable patterns ($C_{22}, C_{11}$) and ($C_{33},C_{11}$).

\subsection{$A_4$-non abelian group realization}
We present a realization based on the non-abelian group $A_4$ leading to a texture of two vanishing subtraces with the related elements lie on the diagonal.  We summarize the irreducible representations (irreps) of $A_4$ in appendix (\ref{appendix-A4}).
\subsubsection{$A_4$- realization of two equalities: ($M_{\n11}=M_{\n22}=M_{\n33}$)}
We review briefly the setup given in \cite{Dev_2013} leading to a texture of two equalities ($M_{\n11}=M_{\n22}=M_{\n33}$).
Taking the matter content shown in Table \ref{matter-content-A4-equality}, one could form a `neutrino' singlet under $SU(2)_L$-gauge, $A_4$-flavor and Lorentz symmetries as
\bea
\label{Lagrangian-neutrino-A4}
{\cal L} &\ni& Y \left[ \left( D^T_{L\mu} C^{-1} i \tau_2 \Delta_1 D_{L \tau} + D^T_{L\tau} C^{-1} i \tau_2 \Delta_1 D_{L \mu} \right)
+ \left( D^T_{L\tau} C^{-1} i \tau_2 \Delta_2 D_{Le} + D^T_{Le} C^{-1} i \tau_2 \Delta_2 D_{L \tau} \right) \right.\nn \\ &&
+\left. \left( D^T_{Le} C^{-1} i \tau_2 \Delta_3 D_{L \mu} + D^T_{L\mu} C^{-1} i \tau_2 \Delta_3 D_{L e} \right)\right] \nn \\ &&
+ Y' \left[ D^T_{Le} C^{-1} i \tau_2 \Delta_4 D_{L e} +D^T_{L\mu} C^{-1} i \tau_2 \Delta_4 D_{L \mu} + D^T_{L\tau} C^{-1} i \tau_2 \Delta_4 D_{L \tau} \right],
\eea
where $\tau_2$ is the weak isospin matrix, and $\Delta_i = \left(\begin{array} {ccc} \Delta^{+}_i & \sqrt{2}\Delta^{++}_i \\\sqrt{2}\Delta^{o}_i & - \Delta^{+}_i \end{array}\right)$ is the Higgs triplet with $i=1,\ldots,4$ is a family index. When $\Delta_i$ acquires a small vev along the neutral direction $\langle \Delta^0_i\rangle_0$, then we get ($M_{\n11}=M_{\n22}=M_{\n33}$).
As to the charged lepton mass matrix, we have
\bea
\label{Lagrangian-chargedlepton-A4}
{\cal L} &\ni& Y_1  \left( \bar{D}_{Le} e_R + \bar{D}_{L\mu} \mu_R + \bar{D}_{L\tau} \tau_R  \right) \phi_1 \nn \\  && +
 Y_2  \left( \bar{D}_{Le} e_R + \omega \bar{D}_{L\mu} \mu_R + \omega^2 \bar{D}_{L\tau} \tau_R  \right) \phi_2 \nn \\  && +
 Y_3  \left( \bar{D}_{Le} e_R + \omega^2 \bar{D}_{L\mu} \mu_R + \omega \bar{D}_{L\tau} \tau_R  \right) \phi_3.
\eea
When $\phi_i$ acquire a vev then we get a diagonal charged lepton mass:
\bea
\label{charged-Lepton-mass-matrix-A4}
M_{\ell} &=& \mbox{diag} \left( Y_1 \langle\phi_1\rangle_0 + Y_2 \langle\phi_2\rangle_0 + Y_3 \langle\phi_3\rangle_0 , \right. \nn \\ &&\left.
 Y_1 \langle\phi_1\rangle_0 + Y_2\; \omega \;\langle\phi_2\rangle_0 + Y_3 \;\omega^2 \langle\phi_3\rangle_0,
 Y_1 \langle\phi_1\rangle_0 + Y_2 \;\omega^2 \; \langle\phi_2\rangle_0 + Y_3 \;\omega \; \langle\phi_3\rangle_0
 \right).
\eea
The charged lepton matrix has enough free parameters $\{Y_i, \langle \phi_i \rangle_0\}$
 to produce the observed mass hierarchy

 \begin{table}[h]
\caption{matter content and symmetry transformations, leading to texture with two equalities. $i=1,\ldots,3$ is a family index}
\centering
\begin{tabular}{cccccccc}
\hline
\hline
Fields & $D_{L_i}$ & $\ell_{R_i}$ & $\phi_1$ & $\phi_2$ & $\phi_3$ & $\Delta_i$ & $\Delta_4$ \\
\hline
\hline
$SU(2)_L$ & 2 & 1& 2 & 2& 2& 3 & 3 \\
\hline
$A_4$ & ${\bf 3}$ & ${\bf 3}$ & ${\bf 1}$ & ${\bf 1'}$ & ${\bf 1''}$ & ${\bf 3}$ & ${\bf 1}$
 \\
\hline
\end{tabular}
\label{matter-content-A4-equality}
\end{table}

\subsubsection{$A_4$- realization of two anti-equalities: ($-M_{\n11}=M_{\n22}=M_{\n33}$)}
We show here how one can transform the past setup from two equalities into two anti-equalities.

\begin{enumerate}
\item {\bf Strategy of Basis choice:}
Actually, one can consider the two-equalities texture as arising from invariance under symmetry defined by the generators $G$ such that:
\bea
G^T M_\n G = M_\n &\Rightarrow& \mbox{equalities}.
\eea
If one performs a similarity transformation on the generators $G\rightarrow G'\equiv I^{-1} G I$ such that $I$ is unitary ($I^{-1}=I^\dagger$), then we see that the form invariance of $M_\n$ under $G$ is equivalent to the invariance of $M'_\n \equiv I^T M_\n I$ under the generators ($G'$):
\bea
G^T M_\n G = M_\n &\Rightarrow& I^T G^T I^{{T}^{-1}} I^T M_\n I I^{-1} G I = I^T M_\n I \nn\\
&\Rightarrow& G'^T M'_\n G' = M'_\n .
\eea
The question is thus to find $I$ such that equalities in $M_\n$ translate as antiequalities in $M'$. Actually, in order to flip the sign of the element at the entry $(1,1)$ while keeping the signs of the entries $(2,2), (3,3)$ intact, it suffices to take $I=\mbox{diag}(-i,1,1)$, such that :
\bea
M_{\n11}=M_{\n22}=M_{\n33} &\Rightarrow& -M'_{\n11}=M'_{\n22}=M'_{\n33}
\eea

\item {\bf Basis ${\bf B'=(S', T')}$ :} For the irrep {\bf 3}, considering the expressions of $B=(S,T)$ in appendix (\ref{appendix-A4}), we have
\bea
\left(x'_1,x'_2,x'_3\right)^T &=& I \left(x_1,x_2,x_3\right)^T = \left( -i x_1,x_2,x_3\right)^T \label{A4-B'-transformations} ,\nn\\
S' = I^\dagger S I = S &=& \mbox{diag} (1,-1,-1)  , \nn\\
T' = I^\dagger T I &=& \left(
\begin {array}{ccc}
0&i&0\\
0&0&1 \\
-i&0&0
\end {array}
\right)  .
\eea
Note here that the combination ($x'_1y'_1+x'_2y'_2+x'_3y'_3$), whose `unprimed' version appears in the singlet decomposition of ${\bf 3} \otimes {\bf 3}$ in the basis ($S,T$), is not invariant under the basis ($S',T'$). Actually, from Eq. (\ref{A4-B'-transformations}), we find the following:
\bea
\left( \begin {array}{c} x_1y_1+x_2y_2+x_3y_3 \end{array} \right)_{\bf 1} &=& \left( \begin {array}{c} -x'_1y'_1+x'_2y'_2+x'_3y'_3 \end{array} \right)_{\bf 1} \nn\\
 \left( \begin {array}{c} x_1y_1+\omega^2 x_2y_2+\omega x_3y_3 \end{array} \right)_{\bf 1'} &=&  \left( \begin {array}{c} -x'_1y'_1+\omega^2 x'_2y'_2+\omega x'_3y'_3 \end{array} \right)_{\bf 1'} \nn\\
 \left( \begin {array}{c} x_1y_1+\omega x_2y_2+\omega^2 x_3y_3 \end{array} \right)_{\bf 1''}  &=& \left( \begin {array}{c} -x'_1y'_1+\omega x'_2y'_2+\omega^2 x'_3y'_3 \end{array} \right)_{\bf 1''} \label{A4-no_star}
\eea
One can check that when $\left(\begin {array}{ccc}
x'_1, &x'_2, &x'_3
\end {array}
\right)^T$ transforms under $T'$, i.e. under ($x'_1 \rightarrow ix'_2, x'_2 \rightarrow x'_3, x'_3 \rightarrow -i x'_1$), idem for $y'$, then $({\bf 3} \otimes {\bf 3})_{\bf 3_s}\equiv\left( \begin {array}{ccc} x'_2y'_3+x'_3y'_2, &x'_3y'_1+x'_1y'_3, &x'_1y'_2+x'_2y'_1 \end{array} \right)^T$ transforms under $T'^*$. The same applies for $({\bf 3} \otimes {\bf 3})_{\bf 3_a}\equiv \left( \begin {array}{ccc} x'_2y'_3-x'_3y'_2, &x'_3y'_1-x'_1y'_3, &x'_1y'_2-x'_2y'_1 \end{array} \right)^T $.

\item {\bf Basis ${\bf B'^*=(S'^*, T'^*)}$ :} For the irrep {\bf 3}, we have $T$ as a complex matrix in the basis $B'$. This pushes us to consider the  basis $B^*=(S'^*, T'^*)$.
\bea
&S'^*=S', T'^* =\left(
\begin {array}{ccc}
0&-i&0\\
0&0&1 \\
i&0&0
\end {array}
\right) = J^{-1}T' J : J= \mbox{diag} (-1,1,1) \Rightarrow& \nn
\eea
\bea
\left(x'^*_1,x'^*_2,x'^*_3\right)^T &=& J \left(x'_1,x'_2,x'_3\right)^T = \left(-x'_1,x'_2,x'_3\right)^T
\nn \\ &=&   JI \left(x_1,x_2,x_3\right)^T =  \mbox{diag} (i,1,1) \left(x_1,x_2,x_3\right)^T = \left(ix_1,x_2,x_3\right)^T \label{A4-B'*-transformations}
\eea
From Eqs. (\ref{A4-B'-transformations},\ref{A4-B'*-transformations}), we find the following:
\bea
\left( \begin {array}{c} -x'_1y'_1+x'_2y'_2+x'_3y'_3 \end{array} \right)_{\bf 1} &=& \left( \begin {array}{c} x'^*_1y'_1+x'^*_2y'_2+x'^*_3y'_3 \end{array} \right)_{\bf 1},\nn\\
 \left( \begin {array}{c} -x'_1y'_1+\omega^2 x'_2y'_2+\omega x'_3y'_3 \end{array} \right)_{\bf 1'} &=&  \left( \begin {array}{c} x'^*_1y'_1+\omega^2 x'^*_2y'_2+\omega x'^*_3y'_3 \end{array} \right)_{\bf 1'} ,\nn\\
 \left( \begin {array}{c} -x'_1y'_1+\omega x'_2y'_2+\omega^2 x'_3y'_3 \end{array} \right)_{\bf 1''}  &=& \left( \begin {array}{c} x'^*_1y'_1+\omega x'^*_2y'_2+\omega^2 x'^*_3y'_3 \end{array} \right)_{\bf 1''} .\label{A4-B'*}
\eea

\item{\bf Matter content}: It is the same content expressed in Table \ref{matter-content-A4-equality}, but the generators of $A_4$ are taken to be expressed in the $(B'=\{S',T'\})$-basis.
Note here that
\bea \label{symmetric-as-3*}
(D_L^T \otimes D_L)_{\bf 3_s} = (\begin{array}{ccc} D_{L\mu}^TD_{L\tau}+D_{L\tau}^TD_{L\mu},& D_{L\tau}^TD_{Le}+D_{Le}^TD_{L\tau}, & D_{Le}^TD_{L\mu}+D_{L\mu}^TD_{Le}\end{array} )^T
\eea
transforms as ${\bf 3^*}$.

\item {\bf Neutrino mass matrix:} with the Lagrangian:
\bea
\label{Lagrangian-neutrino-A4-antiequality}
{\cal L} &\ni& Y \left[ \left( D^T_{L\mu} C^{-1} i \tau_2 \;\Delta_1\; D_{L \tau} + D^T_{L\tau} C^{-1} i \tau_2 \;\Delta_1\; D_{L \mu} \right)
+ \left( D^T_{L\tau} C^{-1} i \tau_2 \; \Delta_2 \; D_{L e} + D^T_{Le} C^{-1} i \tau_2 \;\Delta_2 \;D_{L \tau} \right) \right.\nn \\ &&
+\left. \left( D^T_{Le} C^{-1} i \tau_2 \;\Delta_3\; D_{L \mu} + D^T_{L\mu} C^{-1} i \tau_2 \;\Delta_3\; D_{L e} \right)\right] \nn \\ &&
+ Y' \left[ -D^T_{Le} C^{-1} i \tau_2 \;\Delta_4\; D_{L e} +D^T_{L\mu} C^{-1} i \tau_2 \;\Delta_4\; D_{L \mu} + D^T_{L\tau} C^{-1} i \tau_2 \;\Delta_4\; D_{L \tau} \right]
\eea
we get, upon acquiring small vevs for $\Delta_i^o, i=1,\ldots,4$, the characteristic constraints ($-M_{\n11}=M_{\n22}=M_{\n33}=$). Note that the $Y$-term represents the trivial singlet expression in Eq. (\ref{A4-B'*}), using Eq. (\ref{symmetric-as-3*}), whereas the $Y'$-term represents the trivial singlet expression of Eq. (\ref{A4-no_star}).

By giving appropriate values to the four vevs ($\Delta_i^o, i=1,\ldots,4$) and to the two couplings ($Y, Y'$), one can reconstruct  the mass matrix of normal type (Eq. \ref{22-33_normal_recons}) leading to the spectrum of mixings of Eq. (\ref{22-33_normal_spectrum}), or of inverted type (Eq. \ref{22-33_inverted_recons}) leading to the spectrum of mixings of Eq. (\ref{22-33_inverted_spectrum}). Thus, we have built an explicit $A_4$-flavor model which predicts the masses, mixing angles and CP phases. Moreover, one should mention that for type II seesaw, the Yukawa couplings ($Y, Y'$) are of order unity and the four vevs ($\Delta_i^o, i=1,\ldots,4$) are quite small compared to the electroweak scale due to the heavy triplet mass term.

\item{\bf Charged Lepton sector:} Note that if $D_{Li}$ transforms under ($B'=(S',T')$), then $\overline{D_{Li}}$ would transform under $B'^*=(S'^*, T'^*)$. Hence, with the expressions representing the singlets of Eqs. (\ref{A4-B'*}) and the rule ${\bf 1' \otimes 1''=1}$ (c.f. appendix \ref{appendix-A4}), the Lagrangian:
\bea
\label{Lagrangian-chargedlepton-A4-anti}
{\cal L} &\ni& Y_1  \left( \overline{D}_{Le} e_R + \overline{D}_{L\mu} \mu_R + \overline{D}_{L\tau} \tau_R  \right) \phi_1 \nn \\  && +
 Y_2  \left( \overline{D}_{Le} e_R + \omega \overline{D}_{L\mu} \mu_R + \omega^2 \overline{D}_{L\tau} \tau_R  \right) \phi_2 \nn \\  && +
 Y_3  \left( \overline{D}_{Le} e_R + \omega^2 \overline{D}_{L\mu} \mu_R + \omega \overline{D}_{L\tau} \tau_R  \right) \phi_3.
\eea
leads, when $\phi_i$ acquire a vev, to a diagonal charged lepton mass:
\bea
\label{charged-Lepton-mass-matrix-A4-anti}
M_{\ell} &=& \mbox{diag} \left( Y_1 \langle\phi_1\rangle_0 + Y_2 \langle\phi_2\rangle_0 + Y_3 \langle\phi_3\rangle_0 , \right. \nn \\ &&\left.
 Y_1 \langle\phi_1\rangle_0 + Y_2 \omega \langle\phi_2\rangle_0 + Y_3 \omega^2 \langle\phi_3\rangle_0,
 Y_1 \langle\phi_1\rangle_0 + Y_2 \omega^2 \langle\phi_2\rangle_0 + Y_3 \omega \langle\phi_3\rangle_0
 \right)
\eea
The charged lepton matrix has enough free parameters $\{Y_i, \langle \phi_i \rangle_0\}$
 to produce the observed mass hierarchy

\end{enumerate}

The method elaborated above allows us to move from any realization imposing a texture involving equalities, to another realization leading to the corresponding texture but with equalities replaced by anti-equalities. Moreover, switching indices, say 1 and 2, allows to move from the texture under study, which is viable even when the strict lower bound of $\Sigma \geq 0.09$ eV is taken, to that characterized by ($M_{\n11}=-M_{\n22}=M_{\n33}$) which can not accommodate data.

\subsection{$S_4$-non abelian group realization of the texture ($M_{\n11}=-M_{\n23}$ and  $M_{\n33}=-M_{\n22}$)}
\label{subsection-s4}
We proceed now with a realization based on the non-abelian group $S_4$ leading to a texture with two vanishing subtraces, where the related elements do not lie all on the diagonal. For completeness,  we summarize the irreps of $(S_n,n=1,\ldots, 4)$ in appendix (\ref{appendix-Sn}). Although the realized texture is unviable vis-\`{a}-vis data, however by switching the indices ($1\leftrightarrow 3$) one has a realization model for the viable texture ($M_{\n33}=-M_{\n21}$ and  $M_{\n11}=-M_{\n22}$) which, as we saw, remains viable when the lower bound of $\Sigma$ reaches $0.2$ eV.

\subsubsection{$S_4$-bases:}
\label{subsubsection-s4-bases}
The symmetric group of order $4$ has two generators, and can be defined minimally as:
\bea
S_4 &=& \langle d, b: d^4=b^3=1,db^2d=b\rangle
= \langle T, S: T^4=S^2=(ST)^3=1 \rangle ,
\eea
leading to ($dbd=bd^2b$ from the first minimal definition, and to $(TS)^3=1$ from the second one), and where one can take ($T=d, ST=b$) linking the two sets of two-generators. $S_4$ has five inequivalent irreps (${\bf 1}, {\bf 1'}, {\bf 2}, {\bf 3}$ and${\bf 3'}$). In appendix (\ref{appendix-s4}), we stated the expressions of the generators in a certain $\tilde{B}$-basis. As was done in the previous subsection, and in order to flip the sign in the texture, we carry out a similarity transformation to go from the $\tilde{B}$-basis to another basis, call it the $B$-basis, where the symmetry assignments for the matter fields will be given, and where the texture of the mass matrix is of the required form. We choose to do this only for the triplet irreps with similarity matrix given by ($U=\mbox{diag}\left(1,-1,1\right)$), whereas the doublet, and evidently the singlets, will remain the same. Thus we have ($d^{(','')},b^{(','')}$ refer to ${\bf 3}({\bf 3'}, {\bf 2})$) (c.f. Eqs. \ref{s4-tilde-basis-2}):
\bea
\label{working-basis-s4}
d=U^\dagger\tilde{d_4}U=\mbox{diag}(-1,-i,i) &,& b=U^\dagger\tilde{b_1}U= =\left( \begin{array}{ccc}0 &\frac{-i}{\sqrt{2}}&\frac{-i}{\sqrt{2}} \\ \frac{-1}{\sqrt{2}}&\frac{-i}{2}&\frac{i}{2}\\ \frac{1}{\sqrt{2}}&\frac{-i}{2}& \frac{i}{2}\end{array}\right), \\
d'=U^\dagger\tilde{d'_4}U=\mbox{diag}(1,i,-i) &,& b'=U^\dagger\tilde{b'_1}U= =\left( \begin{array}{ccc}0 &\frac{-i}{\sqrt{2}}&\frac{-i}{\sqrt{2}} \\ \frac{-1}{\sqrt{2}}&\frac{-i}{2}&\frac{i}{2}\\ \frac{1}{\sqrt{2}}&\frac{-i}{2}& \frac{i}{2}\end{array}\right), \\
d''=\tilde{d''_4}=\mbox{diag}(1,-1) &,& b''=\tilde{b''_1}=\frac{1}{2}\left( \begin{array}{ccc} -1&-\sqrt{3} \\ \sqrt{3}&-1 \end{array}\right)
\eea
One can then check that the following ``symmetry adapted linear combinations" (S.A.L.C.) multiplication rules are valid in the adopted working $B$-basis.
\bea
\label{2x2-s4}
\left( \begin{array}{c} a_1 \\ a_2 \end{array}\right)_{\bf 2} \otimes \left( \begin{array}{c} b_1 \\ b_2\end{array}\right)_{\bf 2} &=& \left(a_1b_1+a_2b_2\right)_{\bf 1} \oplus \left( a_1b_2-a_2b_1 \right)_{\bf 1'} \oplus \left( \begin{array}{c}a_2b_2-a_1b_1 \\ a_1b_2+a_2b_1 \end{array} \right)_{\bf 2}
\eea

\bea
\label{2x3-s4}
\left(\begin{array}{l}
a_{1} \\
a_{2}
\end{array}\right)_{\bf 2} \otimes\left(\begin{array}{l}
b_{1} \\
b_{2} \\
b_{3}
\end{array}\right)_{\bf 3}&=&\left(\begin{array}{c}
a_{1} b_{1} \\
-\frac{\sqrt{3}}{2} a_{2} b_{3}-\frac{1}{2} a_{1} b_{2} \\
-\frac{\sqrt{3}}{2} a_{2} b_{2}-a_{1} b_{3}
\end{array}\right)_{\bf 3} \oplus\left(\begin{array}{c}
-a_{2} b_{1} \\
-\frac{\sqrt{3}}{2} a_{1} b_{3}+\frac{1}{2} a_{2} b_{2} \\
-\frac{\sqrt{3}}{2} a_{1} b_{2}+\frac{1}{2} a_{2} b_{3}
\end{array}\right)_{\bf 3'},\\
\label{2x3'-s4}
\left(\begin{array}{l}
a_{1} \\
a_{2}
\end{array}\right)_{\bf 2} \otimes\left(\begin{array}{l}
b_{1} \\
b_{2} \\
b_{3}
\end{array}\right)_{\bf 3'}&=&\left(\begin{array}{c}
-a_{2} b_{1} \\
-\frac{\sqrt{3}}{2} a_{1} b_{3}+\frac{1}{2} a_{2} b_{2} \\
-\frac{\sqrt{3}}{2} a_{1} b_{2}+\frac{1}{2} a_{2} b_{3}
\end{array}\right)_{\bf 3} \oplus\left(\begin{array}{c}
a_{1} b_{1} \\
-\frac{\sqrt{3}}{2} a_{2} b_{3}-\frac{1}{2} a_{1} b_{2} \\
-\frac{\sqrt{3}}{2} a_{2} b_{2}-a_{1} b_{3}
\end{array}\right)_{\bf 3'} \text {, }
\eea

\bea
\left(\begin{array}{l}
a_{1} \\
a_{2} \\
a_{3}
\end{array}\right)_{\bf 3(3')} \otimes\left(\begin{array}{l}
b_{1} \\
b_{2} \\
b_{3}
\end{array}\right)_{\bf 3 (3')}&=& \left(a_{1} b_{1}-a_{2} b_{3}-a_{3} b_{2}\right)_{\bf 1} \oplus \left(\begin{array}{c}
a_{1} b_{1}+\frac{1}{2}\left(a_{2} b_{3}+a_{3} b_{2}\right) \\
\frac{\sqrt{3}}{2}\left(a_{2} b_{2}+a_{3} b_{3}\right)
\end{array}\right)_{\bf 2} \nn \\
&& \oplus\left(\begin{array}{c}
a_{3} b_{3}-a_{2} b_{2} \\
-a_{1} b_{3}-a_{3} b_{1} \\
a_{1} b_{2}+a_{2} b_{1}
\end{array}\right)_{\bf 3} \oplus\left(\begin{array}{c}
-a_{3} b_{2}+a_{2} b_{3} \\
a_{2} b_{1}-a_{1} b_{2} \\
a_{1} b_{3}-a_{3} b_{1}
\end{array}\right)_{\bf 3'}, \label{3x3-s4} \\
\left(\begin{array}{l}
a_{1} \\
a_{2} \\
a_{3}
\end{array}\right)_{\bf 3} \otimes\left(\begin{array}{l}
b_{1} \\
b_{2} \\
b_{3}
\end{array}\right)_{\bf 3'}&=&\left(a_{1} b_{1}-a_{2} b_{3}-a_{3} b_{2}\right)_{\bf 1'}
\oplus\left(\begin{array}{c} \frac{\sqrt{3}}{2}\left(a_{2} b_{2}+a_{3} b_{3}\right) \\ -a_{1} b_{1}-\frac{1}{2}\left(a_{2} b_{3}+a_{3} b_{2}\right)\end{array}\right)_{\bf 2} \nn\\
&&\oplus\left(\begin{array}{l}
-a_{3} b_{2}+a_{2} b_{3} \\
a_{2} b_{1}-a_{1} b_{2} \\
a_{1} b_{3}-a_{3} b_{1}
\end{array}\right)_{\bf 3} \oplus\left(\begin{array}{c}
a_{3} b_{3}-a_{2} b_{2} \\
-a_{1} b_{3}-a_{3} b_{1} \\
a_{1} b_{2}+a_{2} b_{1}
\end{array}\right)_{\bf 3'},  \label{3x3'-s4}
\eea
Noting that ${\bf v^*}$ transforms according to the irrep ${\cal D^*}$ provided ${\bf v} \sim {\cal D}$ (i.e. ${\bf v} \rightarrow {\cal D} {\bf v}$), which gives ${\bf v}^\dagger \rightarrow {\bf v}^\dagger {\cal D}^\dagger$, and observing that taking trace and taking conjugate commute, which leads to $\cal D$ being equivalent to $\cal D^*$ for $S_4$ where the corresponding character table is real (c.f. Tab. (\ref{characterTableS4})), we state for completeness the rules involving conjugate irreps, stressing the fact that the singlet in, say, (${\bf 3}\otimes{\bf 3}$) changes upon conjugation from ($a_1b_1-a_2b_3-a_3b_2$) to ($a^*_1b_1+a^*_2b_2+a^*_3b_3$) in (${\bf 3^*}\otimes{\bf 3}$) :
\bea
\label{2*x2-s4}
\left( \begin{array}{c} a^*_1 \\ a^*_2 \end{array}\right)_{\bf 2^*} \otimes \left( \begin{array}{c} b_1 \\ b_2\end{array}\right)_{\bf 2} &=& \left(a^*_1b_1+a^*_2b_2\right)_{\bf 1} \oplus \left( a^*_1b_2-a^*_2b_1 \right)_{\bf 1'} \oplus \left( \begin{array}{c}a^*_2b_2-a^*_1b_1 \\ a^*_1b_2+a^*_2b_1 \end{array} \right)_{\bf 2}
\eea
\bea
\label{2*x3-s4}
\left(\begin{array}{l}
a^*_{1} \\
a^*_{2}
\end{array}\right)_{\bf 2^*} \otimes\left(\begin{array}{l}
b_{1} \\
b_{2} \\
b_{3}
\end{array}\right)_{\bf 3}&=&\left(\begin{array}{c}
a^*_{1} b_{1} \\
-\frac{\sqrt{3}}{2} a^*_{2} b_{3}-\frac{1}{2} a^*_{1} b_{2} \\
-\frac{\sqrt{3}}{2} a^*_{2} b_{2}-\frac{1}{2}a^*_{1} b_{3}
\end{array}\right)_{\bf 3} \oplus\left(\begin{array}{c}
-a^*_{2} b_{1} \\
-\frac{\sqrt{3}}{2} a^*_{1} b_{3}+\frac{1}{2} a^*_{2} b_{2} \\
-\frac{\sqrt{3}}{2} a^*_{1} b_{2}+\frac{1}{2} a^*_{2} b_{3}
\end{array}\right)_{\bf 3'},\\
\label{2*x3'-s4}
\left(\begin{array}{l}
a^*_{1} \\
a^*_{2}
\end{array}\right)_{\bf 2^*} \otimes\left(\begin{array}{l}
b_{1} \\
b_{2} \\
b_{3}
\end{array}\right)_{\bf 3'}&=&\left(\begin{array}{c}
-a^*_{2} b_{1} \\
-\frac{\sqrt{3}}{2} a^*_{1} b_{3}+\frac{1}{2} a^*_{2} b_{2} \\
-\frac{\sqrt{3}}{2} a^*_{1} b_{2}+\frac{1}{2} a^*_{2} b_{3}
\end{array}\right)_{\bf 3} \oplus\left(\begin{array}{c}
a^*_{1} b_{1} \\
-\frac{\sqrt{3}}{2} a^*_{2} b_{3}-\frac{1}{2} a^*_{1} b_{2} \\
-\frac{\sqrt{3}}{2} a^*_{2} b_{2}-\frac{1}{2}a^*_{1} b_{3}
\end{array}\right)_{\bf 3'} \text {, }
\eea

\bea
\label{3*x3-s4}
\left(\begin{array}{l}
a^*_{1} \\
a^*_{2} \\
a^*_{3}
\end{array}\right)_{\bf 3^*(3'^*)} \otimes\left(\begin{array}{l}
b_{1} \\
b_{2} \\
b_{3}
\end{array}\right)_{\bf 3 (3')}&=& \left(a^*_{1} b_{1}+a^*_{2} b_{2}+a^*_{3} b_{3}\right)_{\bf 1} \oplus \left(\begin{array}{c}
a^*_{1} b_{1}-\frac{1}{2}\left(a^*_{2} b_{2}+a^*_{3} b_{3}\right) \\
\frac{-\sqrt{3}}{2}\left(a^*_{2} b_{3}+a^*_{3} b_{2}\right)
\end{array}\right)_{\bf 2} \nn \\
&& \oplus\left(\begin{array}{c}
-a^*_{3} b_{2}+a^*_{2} b_{3} \\
a^*_{1} b_{2}-a^*_{3} b_{1} \\
-a^*_{1} b_{3}+a^*_{2} b_{1}
\end{array}\right)_{\bf 3^*} \oplus\left(\begin{array}{c}
a^*_{3} b_{3}-a^*_{2} b_{2} \\
a^*_{1} b_{3}+a^*_{2} b_{1} \\
-a^*_{1} b_{2}-a^*_{3} b_{1}
\end{array}\right)_{\bf 3'^*},
\eea
\bea
\label{3*x3'-s4}
\left(\begin{array}{l}
a^*_{1} \\
a^*_{2} \\
a^*_{3}
\end{array}\right)_{\bf 3^*} \otimes\left(\begin{array}{l}
b_{1} \\
b_{2} \\
b_{3}
\end{array}\right)_{\bf 3'}&=& \left(a^*_{1} b_{1}+a^*_{2} b_{2}+a^*_{3} b_{3}\right)_{\bf 1'} \oplus \left(\begin{array}{c}
-\frac{\sqrt{3}}{2}\left(a^*_{2} b_{3}+a^*_{3} b_{2}\right) \\-a^*_{1} b_{1}+\frac{1}{2}\left(a^*_{2} b_{2}+a^*_{3} b_{3}\right)
\end{array}\right)_{\bf 2} \nn \\
&& \oplus\left(\begin{array}{c}
a^*_{3} b_{3}-a^*_{2} b_{2} \\
a^*_{2} b_{1}+a^*_{1} b_{3} \\
-a^*_{1} b_{2}-a^*_{3} b_{1}
\end{array}\right)_{\bf 3^*} \oplus\left(\begin{array}{c}
-a^*_{3} b_{2}+a^*_{2} b_{3} \\
a^*_{1} b_{2}-a^*_{3} b_{1} \\
-a^*_{1} b_{3}+a^*_{2} b_{1}
\end{array}\right)_{\bf 3'^*},
\eea

\bea
\label{s4-2*x2*}
\left( \begin{array}{c} a^*_1 \\ a^*_2 \end{array}\right)_{\bf 2^*} \otimes \left( \begin{array}{c} b^*_1 \\ b^*_2\end{array}\right)_{\bf 2^*} &=& \left(a^*_1b^*_1+a^*_2b^*_2\right)_{\bf 1^*} \oplus \left( a^*_1b^*_2-a^*_2b^*_1 \right)_{\bf 1'^*} \oplus \left( \begin{array}{c}a^*_2b^*_2-a^*_1b^*_1 \\ a^*_1b^*_2+a^*_2b^*_1 \end{array} \right)_{\bf 2^*}
\eea

\bea
\label{2*x3*-s4}
\left(\begin{array}{l}
a^*_{1} \\
a^*_{2}
\end{array}\right)_{\bf 2^*} \otimes\left(\begin{array}{l}
b^*_{1} \\
b^*_{2} \\
b^*_{3}
\end{array}\right)_{\bf 3^*}&=&\left(\begin{array}{c}
a^*_{1} b^*_{1} \\
-\frac{\sqrt{3}}{2} a^*_{2} b^*_{3}-\frac{1}{2} a^*_{1} b^*_{2} \\
-\frac{\sqrt{3}}{2} a^*_{2} b^*_{2}-\frac{1}{2}a^*_{1} b^*_{3}
\end{array}\right)_{\bf 3^*} \oplus\left(\begin{array}{c}
-a^*_{2} b^*_{1} \\
-\frac{\sqrt{3}}{2} a^*_{1} b^*_{3}+\frac{1}{2} a^*_{2} b^*_{2} \\
-\frac{\sqrt{3}}{2} a^*_{1} b^*_{2}+\frac{1}{2} a^*_{2} b^*_{3}
\end{array}\right)_{\bf 3'^*},\\
\label{2*x3'*-s4}
\left(\begin{array}{l}
a^*_{1} \\
a^*_{2}
\end{array}\right)_{\bf 2^*} \otimes\left(\begin{array}{l}
b^*_{1} \\
b^*_{2} \\
b^*_{3}
\end{array}\right)_{\bf 3'^*}&=&\left(\begin{array}{c}
-a^*_{2} b^*_{1} \\
-\frac{\sqrt{3}}{2} a^*_{1} b^*_{3}+\frac{1}{2} a^*_{2} b^*_{2} \\
-\frac{\sqrt{3}}{2} a^*_{1} b^*_{2}+\frac{1}{2} a^*_{2} b^*_{3}
\end{array}\right)_{\bf 3^*} \oplus\left(\begin{array}{c}
a^*_{1} b^*_{1} \\
-\frac{\sqrt{3}}{2} a^*_{2} b^*_{3}-\frac{1}{2} a^*_{1} b^*_{2} \\
-\frac{\sqrt{3}}{2} a^*_{2} b_{2}-\frac{1}{2}a^*_{1} b^*_{3}
\end{array}\right)_{\bf 3'^*} \text {, }
\eea

\bea
\left(\begin{array}{l}
a^*_{1} \\
a^*_{2} \\
a^*_{3}
\end{array}\right)_{\bf 3^*(3'^*)} \otimes\left(\begin{array}{l}
b^*_{1} \\
b^*_{2} \\
b^*_{3}
\end{array}\right)_{\bf 3^* (3'^*)}&=& \left(a^*_{1} b^*_{1}-a^*_{2} b^*_{3}-a^*_{3} b^*_{2}\right)_{\bf 1^*} \oplus \left(\begin{array}{c}
a^*_{1} b^*_{1}+\frac{1}{2}\left(a^*_{2} b^*_{3}+a^*_{3} b^*_{2}\right) \\
\frac{\sqrt{3}}{2}\left(a^*_{2} b^*_{2}+a^*_{3} b^*_{3}\right)
\end{array}\right)_{\bf 2^*} \nn \\
&& \oplus\left(\begin{array}{c}
a^*_{3} b^*_{3}-a^*_{2} b^*_{2} \\
-a^*_{1} b^*_{3}-a^*_{3} b^*_{1} \\
a^*_{1} b^*_{2}+a^*_{2} b^*_{1}
\end{array}\right)_{\bf 3^*} \oplus\left(\begin{array}{c}
-a^*_{3} b^*_{2}+a^*_{2} b^*_{3} \\
a^*_{2} b^*_{1}-a^*_{1} b^*_{2} \\
a^*_{1} b^*_{3}-a^*_{3} b^*_{1}
\end{array}\right)_{\bf 3'^*},  \label{3*x3*-s4} \\
\left(\begin{array}{l}
a^*_{1} \\
a^*_{2} \\
a^*_{3}
\end{array}\right)_{\bf 3^*} \otimes\left(\begin{array}{l}
b^*_{1} \\
b^*_{2} \\
b^*_{3}
\end{array}\right)_{\bf 3'^*}&=&\left(a^*_{1} b^*_{1}-a^*_{2} b^*_{3}-a^*_{3} b^*_{2}\right)_{\bf 1'^*}
\oplus\left(\begin{array}{c} \frac{\sqrt{3}}{2}\left(a^*_{2} b^*_{2}+a^*_{3} b^*_{3}\right) \\ -a^*_{1} b^*_{1}-\frac{1}{2}\left(a^*_{2} b^*_{3}+a^*_{3} b^*_{2}\right)\end{array}\right)_{\bf 2^*} \nn\\
&&\oplus\left(\begin{array}{l}
-a^*_{3} b^*_{2}+a^*_{2} b^*_{3} \\
a^*_{2} b^*_{1}-a^*_{1} b^*_{2} \\
a^*_{1} b^*_{3}-a^*_{3} b^*_{1}
\end{array}\right)_{\bf 3^*} \oplus\left(\begin{array}{c}
a^*_{3} b^*_{3}-a^*_{2} b^*_{2} \\
-a^*_{1} b^*_{3}-a^*_{3} b^*_{1} \\
a^*_{1} b^*_{2}+a^*_{2} b^*_{1}
\end{array}\right)_{\bf 3'^*},  \label{3*x3'*-s4}
\eea

\subsubsection{Type-II Seesaw Matter Content:}
\label{subsubsection-s4-matter}
We present now a type-II seesaw scenario leading to a neutrino mass matrix of the required form. The matter content is summarized in Table (\ref{matter-content-S4})

 \begin{table}[h]
\caption{matter content and symmetry transformations, leading to texture with two anti-equalities. $i=1,\ldots,3$ is a family index}
\centering
\begin{tabular}{ccccccccc}
\hline
\hline
Fields & $D_{L_i}$ & $\Delta_i$ & $\Delta_4$ & $\ell_{R_i}$ & $\phi_I$ & $\phi_{II}$ & $\phi_{III}$ & $\phi_{III}'$   \\
\hline
\hline
$SU(2)_L$ & 2 & 3 & 3 & 1 & 2 & 2& 2 & 2 \\
\hline
$S_4$ & ${\bf 3}$ & ${\bf 3}$ & ${\bf 1}$ & ${\bf 3}$  & ${\bf 1}$ & ${\bf 2}$ & ${\bf 3}$ & ${\bf 3'}$
 \\
\hline
\end{tabular}
\label{matter-content-S4}
\end{table}
The Lorentz-, gauge- and $S_4$-invariant terms relevant for the neutrino mass matrix are
\bea
\label{Lagrangian-neutrino-S4-antiequality}
{\cal L} &\ni& Y  \left( D^T_{L1} C^{-1} i \tau_2  D_{L1} - D^T_{L2} C^{-1} i \tau_2  D_{L3} -  D^T_{L3} C^{-1} i \tau_2  D_{L2}  \right)  \Delta_4 \nn \\ &&
+ Y' \left[ \left( D^T_{L3} C^{-1} i \tau_2  D_{L3} - D^T_{L2} C^{-1} i \tau_2  D_{L2} \right) \Delta_1 \right.\nn \\ &&
+ \left( D^T_{L1} C^{-1} i \tau_2 D_{L 3} + D^T_{L3} C^{-1} i \tau_2  D_{L1} \right) \Delta_3 \nn \\ &&
\left. - \left( D^T_{L1} C^{-1} i \tau_2 D_{L2} +D^T_{L2} C^{-1} i \tau_2  D_{L1} \right) \Delta_2 \right].
\eea
The $Y(Y')$-term picks up the singlet (triplet) combination from the product of the two triplets ($D^T_{L_i}$ and $D_{L_i}$) (Eq. \ref{3x3-s4}), before multiplying it with the Higgs flavor singlet $\Delta_4$ (triplet $\Delta_i$). We get, upon acquiring small vevs for $\Delta_i^o, i=1,\ldots,4$, the characteristic constraints ($M_{\n33}=-M_{\n22}=Y' \langle \Delta_1^0\rangle$) and ($M_{\n11}=-M_{\n22}= Y \langle \Delta_4^0\rangle$).

\subsubsection{Charged lepton sector:}
\label{subsubsection-s4-charged}
In constructing the charged lepton mass matrix $M_{\ell}$, we did not find a way to construct a non-degenerate diagonal mass matrix. However, we can build a generic mass matrix and impose suitable hierarchy conditions in order to diagonlize $M_\ell$ by rotating infinitesimally the left-handed charged lepton fields. This means that, up to approximations of the order of the charged lepton mass-ratios hierarchies, we are in the `flavor' basis, and the aforementioned phenomenological study is valid, especially that, after all, these corrections due to rotating the fields are not larger than other, hitherto discarded, corrections coming, say, from radiative renormalization group running from the seesaw high scale to the observed data low scale.

Noting that $D_{Li}$ transforming under (${\cal D}$) implies that $\overline{D}_{Li}$ would transform under ${\cal D^*}$, one can use Eq. (\ref{3*x3-s4}) of the product (${\bf 3^*} \otimes {\bf 3}$) and get output irreps of (${\bf 1}$), to be multiplied by a Higgs flavor singlet $\phi_I$, and of (${\bf 2}$), to be multiplied by a Higgs flavor doublet $\phi_{II}$ (c.f. Eq. \ref{2x2-s4}), and of (${\bf 3^*}$), to be multiplied by a Higgs flavor triplet $\phi_{III}$ (c.f. Eq. \ref{3*x3-s4}), and finally of (${\bf 3'^*}$), to be multiplied by another Higgs flavor triplet $\phi_{III'}$ (c.f. Eq. \ref{3*x3-s4}).
The relevant Lagrangian is:
\bea
\label{Lagrangian-chargedlepton-S4-anti}
{\cal L} &\ni& \lambda_1  \left( \overline{D}_{L1} \ell_{R1} + \overline{D}_{L2} \ell_{R2} + \overline{D}_{L3} \ell_{R3}  \right) \phi_I  \\  && +
 \lambda_2  \left[ \overline{D}_{L1} \ell_{R1} \phi_{II_1} -\frac{1}{2} \left( \overline{D}_{L2} \ell_{R2} + \overline{D}_{L3} \ell_{R3}  \right)  \phi_{II_1}   -\frac{\sqrt{3}}{2}  \left( \overline{D}_{L2} \ell_{R3} +  \overline{D}_{L3} \ell_{R2} \right) \phi_{II_2} \right] \nn \\ &&+
\lambda_3  \left[ \left( -\overline{D}_{L3} \ell_{R2} + \overline{D}_{L2} \ell_{R3} \right) \phi_{III_1} + \left( \overline{D}_{L1} \ell_{R2} - \overline{D}_{L3} \ell_{R1} \right) \phi_{III_2} +\left( -\overline{D}_{L1} \ell_{R3} + \overline{D}_{L2} \ell_{R1} \right) \phi_{III_3} \right] \nn \\  && +
\lambda'_3  \left[ \left( \overline{D}_{L2} \ell_{R3} - \overline{D}_{L2} \ell_{R2} \right) \phi_{III'_1} + \left( \overline{D}_{L1} \ell_{R3} + \overline{D}_{L2} \ell_{R1} \right) \phi_{III'_2} -\left( \overline{D}_{L1} \ell_{R2} + \overline{D}_{L3} \ell_{R1} \right) \phi_{III'_3} \right], \nn
\eea
which leads, when $\phi_i, i\in\{I,II,III,III'\}$ acquire a vev, to a  charged lepton mass:
\bea
\label{charged-Lepton-mass-matrix-S4-anti}
M_{\ell} &=& \lambda_1 \left( \begin{array}{ccc}\langle \phi_{I}\rangle_0&0&0\\0&\langle \phi_{I}\rangle_0&0\\0&0&\langle \phi_{I}\rangle_0\end{array}\right)
+\lambda_2 \left( \begin{array}{ccc}\langle \phi_{II_1}\rangle_0&0&0\\0&-\frac{1}{2}\langle \phi_{II_1}\rangle_0&-\frac{\sqrt{3}}{2}\langle \phi_{II_2}\rangle_0\\0&-\frac{\sqrt{3}}{2}\langle \phi_{II_2}\rangle_0&-\frac{1}{2}\langle \phi_{II_1}\rangle_0\end{array}\right)  \\ &&+
\lambda_3 \left( \begin{array}{ccc}0&\langle \phi_{III_2}\rangle_0&-\langle \phi_{III_3}\rangle_0\\\langle \phi_{III_3}\rangle_0&0&\langle \phi_{III_1}\rangle_0\\-\langle \phi_{III_2}\rangle_0&-\langle \phi_{III_1}\rangle_0&0\end{array}\right) +
\lambda'_3 \left( \begin{array}{ccc}0&-\langle \phi_{III_3'}\rangle_0&\langle \phi_{III_2'}\rangle_0\\\langle \phi_{III_2'}\rangle_0&-\langle \phi_{III_1'}\rangle_0&0\\-\langle \phi_{III_3'}\rangle_0&0&\langle \phi_{III_1'}\rangle_0\end{array}\right) \nn
\eea
We state now two ways to get a generic $M_\ell$.
\begin{itemize}
\item We assume a vev hierarchy such that the first components are dominant and comparable ($\langle \phi_I\rangle_0 \approx \langle \phi_{II_1}\rangle_0 \approx \langle \phi_{III_1}\rangle_0 \approx \langle \phi_{III_1'}\rangle_0 \approx v$, whereas other vevs can be neglected). We do not study the Higgs scalar potential, but assume that its various free parameters can be adjusted such that to lead naturally to this assumption.  This leads to a diagonal $M_\ell$:
    \bea
    \label{m-ell-first}
    M_\ell &\approx & v \mbox{ diag} \left(\lambda_1+\lambda_2, \lambda_1-\frac{1}{2}\lambda_2-\lambda'_3, \lambda_1-\frac{1}{2}\lambda_2+\lambda'_3\right)
    \eea
The mass matrix is approximately diagonal with enough parameters to produce the observed charged lepton mass hierarchies by taking:
\bea
&m_e \approx (\lambda_1+\lambda_2) v, m_\mu \approx (\lambda_1-\frac{1}{2}\lambda_2-\lambda'_3)v ,  m_\tau \approx (\lambda_1-\frac{1}{2}\lambda_2+\lambda'_3)v.
\eea
Thus, we are, up to a good approximation of the order of the mass ratio $\leq 10^{-2}$, in the flavor basis. The effect of the `small' neglected non-diagonal terms is to require rotating infinitesimally the left handed charged lepton fields, leading thus to corrections on the observed $V_{\mbox{\tiny PMNS}}$ of the same small order $10^{-2}$.
\item Looking at Eq. (\ref{charged-Lepton-mass-matrix-S4-anti}), we see that we have 9 free vevs and 4 free perturbative coupling constants, appearing in 9 linear combinations, a priori enough to construct the generic $3 \times 3$ complex matrix. Thus,  $M_\ell$ can be casted in the form
   \bea
\label{M_elltype2}
M_{\ell } =  \left( \begin {array}{c}
{\bf a}^T\\{\bf b}^T\\{\bf c}^T
\end {array}
\right) &\Rightarrow&
M_{\ell } M_{\ell}^\dagger = \left(\begin {array}{ccc}
{\bf a.a} &{\bf a.b}&{\bf a.c} \\
{\bf b.a} &{\bf b.b}&{\bf b.c}\\
{\bf c.a} &{\bf c.b}&{\bf c.c}
\end {array} \right)
\eea
where ${\bf a}, {\bf b}$ and ${\bf c}$ are three linearly independent vectors, so taking only the following natural assumption on the norms of the vectors
\bea \parallel {\bf a} \parallel /\parallel {\bf c} \parallel = m_e/m_\tau \sim 3 \times 10^{-4} &,&  \parallel {\bf b} \parallel /\parallel {\bf c} \parallel = m_\mu/m_\tau \sim 6 \times 10^{-2}\eea
one can diagonalize $M_{\ell } M_{\ell}^\dagger$ by an infinitesimal rotation as was done in \cite{Lashin_2012}, which proves that we are to a good approximation in the flavor basis.

\end{itemize}

\section{Summary and Conclusion}
In this study, we carry out a systematic study of the Majorana neutrino mass matrix characterized by two $2\times2$ vanishing subtraces. In light of the recent experimental data for oscillation and non-oscillation parameters, we update the results of the past study \cite{Alhendi_2008}. We introduce the analytical expressions for A's and B's coefficients as given by Eq. (\ref{Coff}), and the leading order term in $s_z$ for the neutrino physical parameter $R_\n$. Moreover, all ``full'' correlations, resulting from the full numerical analysis taking all experimental constraints into consideration, are very well approximated by ``exact'' correlations assuming `zero' solar-to-atmospheric ratio $R_\n$, and in many cases they even do not deviate much from correlations resulting from roots of the leading order of $R_\n$. This helps in studying analytically the 15 textures and justify their viability to accommodate data. Actually, the two vanishing trace conditions put 4 real constraints on $M_{\nu}$, thus we have only 5 free parameters corresponding to the three mixing angles ($\t_x\equiv \theta_{12},\t_y\equiv \theta_{23},\t_z\equiv \theta_{13}$), Dirac phase $\delta$ and the solar neutrino mass difference $\delta m^2$. In contrast to \cite{Alhendi_2008}, we vary the 5 parameters in their allowed experimental range and check whether or not the texture satisfies the bounds of $|\Delta m^2|$ besides those in Eq. (\ref{non-osc-cons}). We find that only 7 textures out of the 15 ones can accommodate the experimental data with only one case viable at both hierarchy types. We notice that neither $m_1$ for normal ordering nor $m_3$ for inverted ordering does reach a vanishing value. Therefore, there are no signatures for the singular textures for all cases at all $\sigma$ levels with either hierarchy type. We find the the phases $\delta$, $\rho$ and $\sigma$ are strongly restricted at all $\sigma$-levels with either hierarchy types. We present 15 correlation plots for each viable texture for both hierarchy types (red and blue plots correspond to normal and inverted orderings respectively) generated
from the accepted points of the neutrino physical parameters at the 3-$\sigma$ level. Moreover, we introduce $M_{\nu}$ for each viable texture for both orderings at one representative point at the 3-$\sigma$ level. The point is chosen to be as close as possible to the best fit values of the mixing and Dirac phase angles.

Finally, we present the symmetry realization for the two-vanishing traces texture, irrespective of whether or not it was accommodating data. We present two examples based on non-abelian groups. The first one uses the alternating group $A_4$ within type-II seesaw scenario to realize a texture where the defining elements lie on the diagonal. The second example uses the symmetry group $S_4$ to find a realization, within type-II seesaw scenario, of a two-vanishing-subtraces texture where the elements defining the texture do not lie all on the diagonal.

We have not discussed the question of the scalar potential
and finding its general form under the imposed symmetry. Nor did we deal with the radiative corrections effect on the phenomenology and whether or not it can spoil the form of the texture while running from the “ultraviolet” scale where the seesaw scale imposes the texture form to the low scale where phenomenology was analyzed.

\section*{{\large \bf Acknowledgements}}
E.I.L and N.C. acknowledge support from ICTP through the Senior Associate programs where a significant part of this work was carried out. N.C. acknowledges support also from the CAS PIFI fellowship and the Humboldt Foundation. E.L.'s work was partially supported by the STDF project 37272.

\vspace{1cm}
\appendix{{\large \bf APPENDICES}}
\section{Failing textures }

\label{appFailing}
We list now all the unviable eight textures, where, for each texture, studying the roots of ($m_{23}^2-m_{13}^2$) gives a justification for the failure to accommodate data.

\subsection{Texture($\textbf{C}_{33},\textbf{C}_{23}$)$\equiv$ ($M_{ee}+M_{\mu\mu}=0$, $M_{ee}+M_{\tau\mu}=0$)}
The A's and B's are given by
\begin{align}
A_{1}=&c_x^2c_z^2+(-c_xs_ys_z-s_xc_ye^{-i\delta})^2,~A_2=s_x^2c_z^2+(-s_xs_ys_z+c_xc_ye^{-i\delta})^2,~A_3=s_z^2+s_y^2c_z^2,\nonumber\\
B_1=&c_x^2c_z^2+(-c_xs_ys_z-s_xc_ye^{-i\delta})(-c_xc_ys_z+s_xs_ye^{-i\delta}),\nonumber\\
B_2=&s_x^2c_z^2+(-s_xs_ys_z+c_xc_ye^{-i\delta})(-s_xc_ys_z-c_xs_ye^{-i\delta}),\nonumber\\
B_3&=s_z^2+s_yc_yc_z^2.
\end{align}
We find
\bea
\label{c33c23-approx-ms-difference}
m_{23}^2-m_{13}^2 &=& \frac{s_y^4 (1-2/t_y)}{c_y^2(1+s_{2y})} +\mathcal{O}(s_z) .\nn
\eea
We find that $\frac{2}{t_y} >1, \forall \t_y \in [41^o,51,3^o]$ implying $m_2 < m_1$, and this result will not be changed by including higher order terms, or by taking the exact result. Actually the ``exact'' result gives always $(m_{23}^2-m_{13}^2)$ as negative and of order unity.  Thus, we deduce that this texture is excluded experimentally.


\subsection{Texture($\textbf{C}_{11},\textbf{C}_{23}$)$\equiv$ ($M_{\mu\mu}+M_{\tau\tau}=0$, $M_{ee}+M_{\tau\mu}=0$)}
The A's and B's are given by
\begin{align}
A_1=&c_x^2s_z^2+s_x^2e^{-2i\delta},~A_2=s_x^2s_z^2+c_x^2e^{-2i\delta}
,~A_3=c_z^2,\nonumber\\
B_1=&c_x^2c_z^2+(-c_xs_ys_z-s_xc_ye^{-i\delta})(-c_xc_ys_z+s_xs_ye^{-i\delta}),\nonumber\\
B_2=&s_x^2c_z^2+(-s_xs_ys_z+c_xc_ye^{-i\delta})(-s_xc_ys_z-c_xs_ye^{-i\delta}),\nonumber\\
B_3&=s_z^2+s_yc_yc_z^2.
\end{align}
We find
\bea
\label{c11c23-approx-ms-difference}
m_{23}^2-m_{13}^2 &=& \frac{c_{2y}^2}{c_{2x}} + \frac{s_{2x}s_{2y} c_{2y} c_\d (-1+s_{2y})s_z}{c_{2x}^2} +\mathcal{O}(s^2_z).
\eea
We find that at order $\mathcal{O}(s_z)$, we have $(m_{23}^2-m_{13}^2\geq0)$. However, we checked that by including the order $\mathcal{O}(s_z^2)$, the sign would be inverted ($m_{23}^2-m_{13}^2\leq 0$), such that higher orders, voire the exact result, will no longer change this sign. So, the texture is excluded experimentally.


\subsection{Texture($\textbf{C}_{22},\textbf{C}_{23}$)$\equiv$ ($M_{ee}+M_{\tau\tau}=0$, $M_{ee}+M_{\tau\mu}=0$)}
The A's and B's are given by
\begin{align}
A_1=&c_x^2c_z^2+(-c_xc_ys_z+s_xs_ye^{-i\delta})^2,~A_2=s_x^2c_z^2+(-s_xc_ys_z-c_xs_ye^{-i\delta})^2,~A_3=s_z^2+c_y^2c_z^2,\nonumber\\
B_1=&c_x^2c_z^2+(-c_xs_ys_z-s_xc_ye^{-i\delta})(-c_xc_ys_z+s_xs_ye^{-i\delta}),\nonumber\\
B_2=&s_x^2c_z^2+(-s_xs_ys_z+c_xc_ye^{-i\delta})(-s_xc_ys_z-c_xs_ye^{-i\delta}),\nonumber\\
B_3&=s_z^2+s_yc_yc_z^2.
\end{align}
We find
\bea
\label{c22c23-approx-ms-difference}
m_{23}^2-m_{13}^2 &=& \frac{c_y^4 (1-2t_y)}{s_y^2 c_{2x}(1+s_{2x})} +\mathcal{O}(s_z).
\eea
We find that $2 t_y >1, \forall \t_y \in [41^o,51,3^o]$ implying $m_2 < m_1$, and this result will not be changed by including higher order terms, or by taking the ``exact'' result which shows that $\d m^2$ is negative and of order unity. No zeros were found for the ($m_{23}^2-m_{13}^2$)-expression. Thus, we deduce that this texture is excluded experimentally.


\subsection{Texture($\textbf{C}_{33},\textbf{C}_{11}$)$\equiv$ ($M_{ee}+M_{\mu\mu}=0$, $M_{\mu\mu}+M_{\tau\tau}=0$)}
The A's and B's are given by
\begin{align}
A_{1}=&c_x^2c_z^2+(-c_xs_ys_z-s_xc_ye^{-i\delta})^2,~A_2=s_x^2c_z^2+(-s_xs_ys_z+c_xc_ye^{-i\delta})^2,~A_3=s_z^2+s_y^2c_z^2,\nonumber\\
B_1=&c_x^2s_z^2+s_x^2e^{-2i\delta},~B_2=s_x^2s_z^2+c_x^2e^{-2i\delta}
,~B_3=c_z^2.
\end{align}
We find
\bea
\label{c33c11-approx-ms-difference}
m_{23}^2-m_{13}^2 &=& \frac{c_{2y}^4 }{c_{2x}} +\mathcal{O}(s_z).
\eea

We find that ($m_{23}^2- m_{13}^2$)-leading term is positive but of order unity, for all the allowed values of ($\t_x, \t_y$). This fact remains intact in the case of exact result for the ($m_{23}^2- m_{13}^2$)-expression, such that no zeros for this expression for all allowed ($\t_x, \t_y, \t_z, \d$). Whence, the texture is excluded.

\subsection{Texture($\textbf{C}_{22},\textbf{C}_{11}$)$\equiv$ ($M_{ee}+M_{\tau\tau}=0$, $M_{\mu\mu}+M_{\tau\tau}=0$)}
The A's and B's are given by
\begin{align}
A_1=&c_x^2c_z^2+(-c_xc_ys_z+s_xs_ye^{-i\delta})^2,~A_2=s_x^2c_z^2+(-s_xc_ys_z-c_xs_ye^{-i\delta})^2,~A_3=s_z^2+c_y^2c_z^2,\nonumber\\
B_1=&c_x^2s_z^2+s_x^2e^{-2i\delta},~B_2=s_x^2s_z^2+c_x^2e^{-2i\delta}
,~B_3=c_z^2.
\end{align}
We find
\bea
\label{c22c11-approx-ms-difference}
m_{23}^2-m_{13}^2 &=& \frac{s_{2y}^2 }{c_{2x}} +\mathcal{O}(s_z).
\eea

We find that $m_{23}^2- m_{13}^2$-leading term is positive but of order unity, for all the allowed values of ($\t_x, \t_y, \t_z, \d$). This fact remains intact in the case of exact result for the ($m_{23}^2- m_{13}^2$)-expression, such that no zeros for this expression. Actually, for the exact result we have:
\bea
\label{c12c11-full-ms-difference}
m_{23}^2-m_{13}^2 &=& \frac{\mbox{Num}(m_{23}^2-m_{13}^2)}{\mbox{Den}(m_{23}^2-m_{13}^2)}: \nn\\
\mbox{Num}(m_{23}^2-m_{13}^2) &=& -4c_z^2 c_y (c_y^2 c_z^2 -c_{2z}) (-c_y c_{2x}+ s_{2x}s_z s_y c_\d)
\eea
The zeros of ($\mbox{Num}(m_{23}^2-m_{13}^2)$) give $c_\d = \frac{1}{t_y t_{2x} s_z} >1$ for all acceptable values of ($\t_x, \t_y, \t_z$). Thus, the texture is excluded phenomenologically.


\subsection{Texture($\textbf{C}_{13},\textbf{C}_{12}$)$\equiv$ ($M_{\mu e}+M_{\tau\mu}=0$, $M_{\mu e}+M_{\tau\tau}=0$)}
The A's and B's are given by
\begin{align}
A_1=&c_xc_z(-c_xs_ys_z-s_xc_ye^{-i\delta})+(-c_xs_ys_z-s_xc_ye^{-i\delta})(-c_xc_ys_z+s_xs_ye^{-i\delta}),\nonumber\\
A_2=&s_xc_z(-s_xs_ys_z+c_xc_ye^{-i\delta})+(-s_xs_ys_z+c_xc_ye^{-i\delta})(-s_xc_ys_z-c_xs_ye^{-i\delta}),\nonumber\\
A_3=&s_zs_yc_z+s_yc_z^2c_y,\nonumber\\
B_1=&c_xc_z(-c_xs_ys_z-s_xc_ye^{-i\delta})+(-c_xc_ys_z+s_xs_ye^{-i\delta})^2,\nonumber\\
B_2=&s_xc_z(-s_xs_ys_z+c_xc_ye^{-i\delta})+(-s_xc_ys_z-c_xs_ye^{-i\delta})^2,\nonumber\\
B_3=&s_yc_zs_z+c_y^2c_z^2.
\end{align}
We find
\bea
\label{c13c12-approx-ms-difference}
m_{23}^2-m_{13}^2 &=& \frac{c_y^4 s_{2x}s_y (1-t_y)c_\d-c_y^2s_y^2c_{2x}}{s_x^2 c_x^2 s_y^2 c_y^2 (1+s_{2y})} +\mathcal{O}(s_z) .
\eea

We find that the zeros of $m_{23}^2- m_{13}^2$-leading term should satisfy ($c_\d = \frac{t_y^2}{t_{2x}s_y(1-t_y)} \notin [-1,+1]$) for acceptable ($\t_x, \t_y$), and so no zeros at $\mathcal{O}(s_z)$.  Actually, we could by scanning over allowed values of ($\t_x, \t_y, \t_z, \d$) that ($m_{23}^2-m_{13}^2 < 0$). Thus texture is rejected.

\subsection{Texture($\textbf{C}_{13},\textbf{C}_{22}$)$\equiv$ ($M_{\mu e}+M_{\tau\mu}=0$, $M_{e e}+M_{\tau\tau}=0$)}
The A's and B's are given by
\begin{align}
A_1=&c_xc_z(-c_xs_ys_z-s_xc_ye^{-i\delta})+(-c_xs_ys_z-s_xc_ye^{-i\delta})(-c_xc_ys_z+s_xs_ye^{-i\delta}),\nonumber\\
A_2=&s_xc_z(-s_xs_ys_z+c_xc_ye^{-i\delta})+(-s_xs_ys_z+c_xc_ye^{-i\delta})(-s_xc_ys_z-c_xs_ye^{-i\delta}),\nonumber\\
A_3=&s_zs_yc_z+s_yc_z^2c_y,\nonumber\\
B_1=&c_x^2c_z^2+(-c_xc_ys_z+s_xs_ye^{-i\delta})^2,~B_2=s_x^2c_z^2+(-s_xc_ys_z-c_xs_ye^{-i\delta})^2,~B_3=s_z^2+c_y^2c_z^2.
\end{align}
We find
\bea
\label{c22c13-full-ms-difference}
m_{23}^2-m_{13}^2 &=& \frac{\mbox{Num}(m_{23}^2-m_{13}^2)}{\mbox{Den}(m_{23}^2-m_{13}^2)}: \nn \\
\mbox{Num}(m_{23}^2-m_{13}^2) &=& -2c_z (c_y^2 c_z^2 -s_z^2) \left[ s_{2x} s_y (c_z^2 s_y^2 + c_{2y}) c_\d + s_z s_y s_{2y} c_{2x}\right].
\eea
The zeros of ($\mbox{Num}(m_{23}^2-m_{13}^2)$) give, when plugged in $m_{13}, m_{23}$ the approximation ($m_{13} = m_{23} = 1$). As mentioned before, this leads to rejection of the texture. This comes because we have at these zeros a degenerate spectrum ($m_1 \approx m_2 \approx m_3$), and so $m_3^2 - m_1^2 \approx m_2^2 -m_1^2 $, thus forming ($R_\n \approx \frac{m_2^2 - m_1^2}{m_3^2 - m_1^2} \approx  1 $), which is rejected.

\subsection{Texture($\textbf{C}_{23},\textbf{C}_{12}$)$\equiv$ ($M_{e e}+M_{\mu\tau}=0$, $M_{\mu e}+M_{\tau\tau}=0$)}
The A's and B's are given by
\begin{align}
A_1=&c_x^2c_z^2+(-c_xs_ys_z-s_xc_ye^{-i\delta})(-c_xc_ys_z+s_xs_ye^{-i\delta}),\nonumber\\
A_2=&s_x^2c_z^2+(-s_xs_ys_z+c_xc_ye^{-i\delta})(-s_xc_ys_z-c_xs_ye^{-i\delta}),\nonumber\\
A_3&=s_z^2+s_yc_yc_z^2,\nonumber\\
B_1=&c_xc_z(-c_xs_ys_z-s_xc_ye^{-i\delta})+(-c_xc_ys_z+s_xs_ye^{-i\delta})^2,\nonumber\\
B_2=&s_xc_z(-s_xs_ys_z+c_xc_ye^{-i\delta})+(-s_xc_ys_z-c_xs_ye^{-i\delta})^2,\nonumber\\
B_3=&s_yc_zs_z+c_y^2c_z^2.
\end{align}
We find
\bea
\label{c23c12-full-ms-difference}
m_{23}^2-m_{13}^2 &=& \frac{\mbox{Num}(m_{23}^2-m_{13}^2)}{\mbox{Den}(m_{23}^2-m_{13}^2)}: \nn \\
\mbox{Num}(m_{23}^2-m_{13}^2) &=& (s_z^2 -c_z^2 c_y^2) \{ s_{2x} \left[ c_z^3 (c_y^3 - s_y^3) -s_z c_z^2 (1-3s_y c_y) -c_{2y} (c_y+s_y) c_z   \right. \nn
\\ &&  \left.  -s_z s_{2y}\right] c_\d -c_{2x} \left[ c_z^2 (4c_y^2 -2) + s_z s_{2y} (s_y + c_y) c_z -c_{2y} \right] \}.
\eea
The zeros of ($\mbox{Num}(m_{23}^2-m_{13}^2)$) give, when plugged in $m_{13}, m_{23}$ the approximation ($m_1 \approx m_2 \approx m_3$), which -like the previous pattern- is rejected phenomenologically, as it cannot accommodate a `small' value for $R_\n$ .

\section{Majorana phases}
\label{appCP}
We state here for each of the viable patterns the leading orders, in powers of $s_z$, of the Majorana phases, up to multiples of $\pi/2$. Any constraint on ($\t_x, \t_y$ and $\d$) steming from meeting the acceptable value of ($R_\n \sim 10^{-2}$) would be reflected as a constraint on $\rho$ and $\sigma$.

\begin{itemize}

\item{\bf Texture($\textbf{C}_{22},\textbf{C}_{33}$)}
\bea
\label{c33c22-approx-rhosigma}
\rho=\frac{1}{2}\tan^{-1}\bigg(\frac{s_{2\delta}s_x^2}{1-2s_x^2s_{\delta}^2}\bigg)+\mathcal{O}(s_z) &,&  \sigma= \frac{1}{2}\tan^{-1}\bigg(\frac{s_{2\delta}c_x^2}{1-2c_x^2s_{\delta}^2}\bigg)+\mathcal{O}(s_z)
.\eea

\item{\bf Texture($\textbf{C}_{22},\textbf{C}_{12}$)}
\begin{equation}
\rho=\frac{1}{2}\tan^{-1}\bigg(\frac{\rho_{N3}}{\rho_{D3}}\bigg)+\mathcal{O}(s_z)~~~,~~~\sigma=\frac{1}{2}
\tan^{-1}\bigg(\frac{\sigma_{N3}}{\sigma_{D3}}\bigg)+\mathcal{O}(s_z),
\end{equation}
where
\begin{align}
\rho_{N3}=&4s_xs_{\delta}\big[\big((c_{\delta}+\frac{1}{4})c_y^2-c_{\delta}^2-\frac{1}{2}\big)c_yc_x^3-\frac{1}{2}s_xs_y^2(c_y^2-2)c_{\delta}c_x^2+(c_ys_y^2c_{\delta}^2+\frac{1}{4}c_y)c_x-\frac{1}{2}s_xs_y^2c_{\delta}\big],\nonumber\\
\sigma_{N3}=&2s_{\delta}\big[(c_y^4-3c_y^2+2)c_{\delta}c_x^3-\big((2c_{\delta}^2+\frac{1}{2})c_y^2-2c_{\delta}^2-1\big)s_xc_yc_x^2-s_y^4c_{\delta}c_x-\frac{1}{2}s_xc_ys_y^2\big],\nonumber\\
\rho_{D3}=&\big[c_{2\delta}c_y^4+2(1-3c_{\delta}^2)c_y^2+2c_{2\delta}\big]c_x^4-4s_xc_yc_{\delta}\big[(c_{\delta}^2-\frac{1}{4})c_y^2-c_{\delta}^2\big]c_x^3-\big[c_{2\delta}c_y^4+(3-8c_{\delta}^2)c_y^2+3c_{2\delta}\big]c_x^2\nonumber\\
+&2s_xc_yc_{\delta}\big[c_{2\delta}c_y^2-2c_{\delta}^2+\frac{1}{2}\big]c_x+c_{2\delta}s_y^2,\nonumber\\
\sigma_{D3}=&\big[c_{2\delta}c_y^4+2(1-3c_{\delta}^2)c_y^2+2c_{2\delta}\big]c_x^3-4s_xc_yc_{\delta}\big[(c_{\delta}^2-\frac{1}{4})c_y^2-c_{\delta}^2\big]c_x^2-\big[c_{2\delta}c_y^4+(1-4c_{\delta}^2)c_y^2+c_{2\delta}\big]c_x\nonumber\\-&s_xs_y^2c_yc_{\delta}
.\end{align}

\item{\bf \bf Texture($\textbf{C}_{11},\textbf{C}_{12}$)}
\begin{align}
\rho=\frac{1}{2}\tan^{-1}\bigg[\frac{(c_xc_{2y}-2s_xc_yc_{\delta})s_{\delta}}{c_xc_{2y}c_{\delta}-s_xc_yc_{2\delta}}\bigg]+\mathcal{O}(s_z)&,&
\sigma=\frac{1}{2}\tan^{-1}\bigg[\frac{(s_xc_{2y}+2c_xc_yc_{\delta})s_{\delta}}{s_xc_{2y}c_{\delta}+c_xc_yc_{2\delta}}\bigg]+\mathcal{O}(s_z).
\end{align}

\item{Texture($\textbf{C}_{33},\textbf{C}_{12}$)}
\begin{equation}
\rho=\frac{1}{2}\tan^{-1}\bigg(\frac{\rho_{N2}}{\rho_{D2}}\bigg)+\mathcal{O}(s_z)~~~,~~~
\sigma=\frac{1}{2}\tan^{-1}\bigg(\frac{\sigma_{N2}}{\sigma_{D2}}\bigg)+\mathcal{O}(s_z),
\end{equation}
where
\begin{align}
\rho_{N2}=&-s_xc_yc_{\delta}\big[2s_xc_x^2c_y^5c_{\delta}+(c_x^3+4c_xs_x^2c_{\delta}^2)c_y^4-2s_x(c_{2x}+c_x^2)c_y^3c_{\delta}+c_x(s_x^2-c_y^2)+2s_xc_{2x}c_yc_{\delta}\big],\nonumber\\
\sigma_{N2}=&2c_xc_ys_{\delta}\big[(c_y^5-3c_y^3+2c_y)c_x^3c_{\delta}+2s_x(c_y^4c_{\delta}^2-\frac{1}{4}c_y^4+\frac{1}{4})c_x^2-c_ys_y^4c_{\delta}c_x-\frac{1}{8}s_xs_{2y}^2\big],\nonumber\\
\rho_{D2}=&-\frac{1}{4}s_{2x}^2c_{2\delta}c_y^6+s_{2x}c_{\delta}\big[(2c_{\delta}^2-\frac{3}{2})c_x^2-c_{2\delta}\big]c_y^5+\big[(-6c_{\delta}^2+8)c_x^4+(8c_{\delta}^2-7)c_x^2-c_{2\delta}\big]c_y^4\nonumber\\
&+\frac{1}{2}s_{2x}c_{\delta}(1-4c_x^2)c_y^3\big[(4c_{\delta}^2-9)c_x^4+(-6c_{\delta}^2+7)c_x^2+c_{2\delta}\big]c_y^2+\frac{1}{2}s_{2x}(c_{2x}+c_x^2)c_{\delta}c_y+c_x^2c_{2x},\nonumber\\
\sigma_{D2}=&\big[(2c_y^6-6c_y^4+4c_y^2)c_{\delta}^2-c_y^6+8c_y^4-9c_y^2+2\big]c_x^4+4s_xc_yc_{\delta}\big[c_y^4c_{\delta}^2-\frac{3}{4}c_y^4-c_y^2+\frac{3}{4}\big]c_x^3\nonumber\\
+&\big[(-2c_y^6+4c_y^4-2c_y^2)c_{\delta}^2+c_y^6-9c_y^4+11c_y^2-3\big]c_x^2+c_ys_xc_{\delta}(c_y^4+3c_y^2-2)c_x+2c_y^4-3c_y^2+1.
\end{align}

\item{\bf Texture($\textbf{C}_{33},\textbf{C}_{13}$)}
\begin{equation}
\rho=\frac{1}{2}\tan^{-1}\bigg(\frac{\rho_{N1}}{\rho_{D1}}\bigg)+\mathcal{O}(s_z)~~~,~~~
\sigma=\frac{1}{2}\tan^{-1}\bigg(\frac{\sigma_{N1}}{\sigma_{D1}}\bigg)+\mathcal{O}(s_z),
\end{equation}
where
\begin{align}
\rho_{N1}=&s_ys_x^2s_{2\delta}\big[s_y\big(-1+c_x^2(c_y^2+2)\big)-s_{2x}c_y^2c_{\delta}\big],\nonumber\\
\rho_{D1}=&\big(c_{2\delta}c_y^4-2c_y^2s_{\delta}^2-4c_{\delta}^2+3\big)c_x^4-s_{2x}c_x^2s_yc_{\delta}\big(1-c_y^2c_{2\delta}\big)\nn \\ &+c_x^2s_y^2\big(c_y^2+3\big)c_{2\delta}-c_{2\delta}s_y\big(s_y+s_{2x}c_y^2c_{\delta}\big),\nonumber\\
\sigma_{N1}=&s_{2\delta}c_x^2\big[c_x^2c_y^2(1-t_x^2c_y^2)-c_{2x}+s_{2x}s_yc_y^2c_{\delta}\big],\nonumber\\
\sigma_{D1}=&\big(c_{2\delta}c_y^4-2c_y^2s_{\delta}^2-4c_{\delta}^2+3\big)c_x^4-s_{2x}c_x^2s_yc_{\delta}\big(1-c_y^2c_{2\delta}\big)\nn \\ & -\big(c_{2\delta}c_y^4-2c_y^2-2c_{\delta}^2+3\big)c_x^2+s_y\big(s_y+s_{2x}c_{\delta}\big).
\end{align}

\item{\bf Texture($\textbf{C}_{11},\textbf{C}_{13}$)}
\begin{align}
\rho=\frac{1}{2}\tan^{-1}\bigg[\frac{s_{2x}s_{\delta}(s_y-t_xc_{\delta})}{s_{2x}s_yc_{\delta}-s_x^2c_{2\delta}}\bigg]+\mathcal{O}(s_z)&,&
\sigma=\frac{1}{2}\tan^{-1}\bigg[\frac{2c_{x}s_{\delta}(c_{\delta}+t_xs_{y})}{2s_{x}s_yc_{\delta}+c_xc_{2\delta}}\bigg]+\mathcal{O}(s_z).
\end{align}

\item{\bf Texture($\textbf{C}_{13},\textbf{C}_{23}$)}
\begin{equation}
\rho=\frac{1}{2}\tan^{-1}\bigg(\frac{\rho_{N4}}{\rho_{D4}}\bigg)+\mathcal{O}(s_z)~~~,~~~
\sigma=\frac{1}{2}\tan^{-1}\bigg(\frac{\sigma_{N4}}{\sigma_{D4}}\bigg)+\mathcal{O}(s_z),
\end{equation}
where
\begin{align}
\rho_{N4}=&4s_xs_ys_{\delta}\big[-s_yc_xc_ys_x^2c_{\delta}^2+\frac{1}{2}s_xs_y(c_x^2c_y^2-c_{2x})c_{\delta}+\frac{1}{4}c_x(s_yc_x^2c_y+s_x^2)\big],\nonumber\\
\sigma_{N4}=&2s_ys_{\delta}\big[\big(c_x^3(c_y^2-2)c_{\delta}-2s_xc_x^2c_y(c_{\delta}^2+\frac{1}{4})+c_xs_y^2c_{\delta}+\frac{1}{2}s_xc_y\big)s_y+\frac{1}{2}s_xc_x^2\big],\nonumber\\
\rho_{D4}=&-4s_xc_xc_ys_y^2s_x^2c_{\delta}^3+2s_y^2s_x^2(c_x^2c_y^2-c_{2x})c_{\delta}^2+s_xc_x\big[s_x^2s_y-c_ys_y^2(c_x^2-2)\big]c_{\delta}\nonumber\\+&s_x^2\big[c_x^2c_y^4+(1-3c_x^2)c_y^2-s_yc_x^2c_y+c_{2x}\big],\nonumber\\
\sigma_{D4}=&\big[s_yc_y+(c_y^4-3c_y^2+2)c_{2\delta}\big]c_x^3-s_xc_{\delta}\big[s_y-c_ys_y^2(4c_{\delta}^2-1)\big]c_x^2\nn \\ & -\big[s_yc_y+(c_y^4-2c_y^2+2)c_{2\delta}\big]c_x-s_xc_ys_y^2c_{\delta}.
\end{align}

\end{itemize}

\section{$A_4$-Irreps}
\label{appendix-A4}
$A_4$ is the group of even permutations of four objects. It is defined in terms of two generators $(S, T)$ such that
\bea
\label{app-a4-def}
S^2=T^3=(ST)^3=1,
\eea
leading also to $(TS)^3=1$. It has four irreps ($\bf 1$, $\bf 1'$, $\bf 1''$, $\bf 3$ )
\begin{itemize}
\item $\bf 1$:
$S=1$, $T=1$
\item $\bf 1'$:
$S=1$, $T=\omega: \omega^3=1$
\item $\bf 1''$:
$S=1$, $T=\omega^2$
\item $\bf 3$:
$S=\left(
\begin {array}{ccc}
1&0&0\\
0&-1&0 \\
0&0&-1
\end {array}
\right)$, $T=\left(
\begin {array}{ccc}
0&1&0\\
0&0&1 \\
1&0&0
\end {array}
\right)\equiv(123)$
\end{itemize}
We have
\bea
\label{app-a4-directproduct}
{\bf 1'}\otimes{\bf 1'}={\bf 1''},{\bf 1''}\otimes{\bf 1'}={\bf 1}&,&{\bf 1''}\otimes{\bf 1''}={\bf 1'},{\bf 1}\otimes{\bf 1}={\bf 1}, \\
{\bf 3}\otimes{\bf 3}&=&{\bf 1} \oplus {\bf 1'} \oplus {\bf 1''}\oplus {\bf 3_s}\oplus {\bf 3_a}:\\
\left(\begin {array}{ccc}
x_1, &x_2, &x_3
\end {array}
\right)^T
\otimes
\left(\begin {array}{ccc}
y_1,&y_2,&y_3
\end {array}
\right)^T &=&
\left( \begin {array}{c} x_1y_1+x_2y_2+x_3y_3 \end{array} \right)_{\bf 1}  \\ &&
\oplus \left( \begin {array}{c} x_1y_1+\omega^2 x_2y_2+\omega x_3y_3 \end{array} \right)_{\bf 1'} \nn \\ &&
 \oplus \left( \begin {array}{c} x_1y_1+\omega x_2y_2+\omega^2 x_3y_3 \end{array} \right)_{\bf 1''}\nn \\ &&
 \oplus \left( \begin {array}{ccc} x_2y_3+x_3y_2,&x_3y_1+x_1y_3,&x_1y_2+x_2y_1 \end{array} \right)^{T}_{\bf 3_s}\nn \\ &&
 \oplus \left( \begin {array}{ccc} x_2y_3-x_3y_2,&x_3y_1-x_1y_3,&x_1y_2-x_2y_1 \end{array} \right)^{T}_{\bf 3_a} \nn
\eea

\section{$S_n$-Irreps}
\label{appendix-Sn}
The symmetric group of order $n$, $S_n$, is the group of permutations of $N_n=\{1,\ldots.n\}$. It is of order $n!$, and for $n\geq 3$, it is non-abelian. Any permutation can be decomposed as a product of cycles with disjoint supports, which in turn can be decomposed as a product of transpositions. $S_1$ is the trivial group consisting of just one element. Any group $G$ is divided into conjugacy classes according to the equivalence relation $(a \sim b \Leftrightarrow \exists c \in G: b=c^{-1}ac)$. The number of equivalence classes is equal to the number of inequivalent unitary irreps, which is depicted by the character table showing, for each irrep $\cal D$, listed in upper line of the Table, and each equivalence class $C$, listed in the leftmost column of the table, the trace $(\chi_{\cal D}(g))$ of the irrep $\cal D$ evaluated at one representative member $g$ of the class $C$.

In order to construct, for a group $G$ of order $n_G$,  the character table, for $n_c$ Classes (the class $s C_h=[g]$ includes $s$ elements $g$ all of order $h$\footnote{the order of an element $g$ is the order of the subgroup generated by this element and is equal to ($\min\{n\in N\backslash\{0\}: g^n=1\}$). For a permutation written as a product of disjoint cycles, the order is the least common multiplier of the cardinalities of these cycles' supports.}) and $n_r$ inequivalent unitary irreps (the number of inequivalent unitary $n$-dimenional irrep ${\cal D}_n$ is $m_n$), one usually uses the following rules, :
\bea
  n_c = n_r &,& \sum_{n\in N} m_n n^2 = n_G, \label{irreps-count}\\
  \sum_{a\in G} \chi_\a(a) \chi_\b^*(a) &=& n_G \d_{\a\b}, \label{orthogonality} \\
  \sum_{\a \in \mbox{irreps}} \chi_\a(a) \chi_\a^*(b) &=& \frac{n_G}{\mbox{card}[a]} \d_{[a][b]}, \label{orthogonality_tr} \\
  \chi_{\a \otimes \b}(g) &=& \chi_\a(g)\chi_\b(g), \forall g\in G, \label{direct product} \\
  {\cal D} = \bigoplus_{\a \in \mbox{irreps}} m_\a \a &\Rightarrow& m_\a=\frac{1}{n_G} \sum_{g\in G}\chi_\a^* (g) \chi_{\cal D}(g), \label{decomposition} \\
  P_\a \phi &=& \sum_{g\in G} \chi_\a^*(g) g.\phi \label{salc}\;\;\;.
\eea
The `orthogonality' relations (eq. \ref{orthogonality}) means that the columns of the character table are orthogonal and that the inner product of each column with itself is the cardinality of the group. Since the product of the character table matrix with its conjugate is a scalar matrix, then the rows of the character table are as well orthogonal with squared-norm equal to $n_G$ (eq. \ref{orthogonality_tr}). The 'Direct Product' rule (Eq. \ref{direct product}) gives the character for a direct product of irreps, whereas Eq. (\ref{decomposition}) gives the number $m_\a$ the irrep $\a$ appears in the decomposition of the reducible representation ${\cal D}$. In order to find the linear combination corresponding to a given symmetry characterized by an irrep, or what the chemists call ``the symmetry adapted linear combination (S.A.L.C.)", one uses Eq. (\ref{salc}) which gives the projection of the `basis function' $\phi$ onto the subspace transforming under the irrep $\a$.

\subsection{$S_2$}
\label{appendix-s2}
It has two elements: the identity $E$, and the transposition $A=(12)$, with $A^2=1$.   We have two classes: ($1C_1=\{E\}, 1C_2=\{A\}$). There are two singlet irreps (${\bf 1, 1'}$), with character table
\begin{table}[h]
\caption{Character table of $S_2$}
\centering
\begin{tabular}{ccc}
\hline
\hline
classes$/$irreps & $\chi_{\bf 1}$ & $\chi_{\bf 1'}$  \\
\hline
\hline
$1C_1$ & 1 &1 \\
\hline
$1C_2$ & 1 &- 1 \\
\hline
\end{tabular}
\label{characterTableS2}
\end{table}

Taking $(x_1, x_2)^T$ as the defining (fundamental) representation transforming under ($E=\mbox{diag}(1,1), A\equiv(12)=\left(\begin{array}{cc}0&1\\1&0\end{array}\right)$), then applying Eq. ($\ref{salc}$), we find
\bea
x_1 + x_2 &\sim & {\bf 1} \nn \\
x_1 - x_2 &\sim & {\bf 1'}
\eea
\subsection{$S_3$}
\label{appendix-s2}
In terms of cycles' notation, we have the 6-elements symmetry group of order $3$: $S_3 = \{E,A=(23), B=(13), C=(12), D =(132), F=(123)\}$ which can be divided into three classes $\left(1C_1=\{E\}, 3C_2=\{A,B,C\}, 2C_3=\{D,F\}\right)$, so we have three unitary inequivalent irreps, and by applying Eqs. (\ref{irreps-count}):
\bea \sum_{n\in N}m_n=3, \sum_{n\in N}m_nn^2=6 &\Rightarrow&  m_1=2, m_2=1.\eea
Applying Eqs. (\ref{orthogonality}, \ref{orthogonality_tr}), we have the character table of $S_3$ (Table \ref{characterTableS3}).
\begin{table}[h]
\caption{Character table of $S_3$}
\centering
\begin{tabular}{cccc}
\hline
\hline
classes$/$irreps & $\chi_{\bf 1}$ & $\chi_{\bf 1'}$ & $\chi_{\bf 2}$   \\
\hline
\hline
$1C_1$ & 1 &1 & 2 \\
\hline
$2C_3$ & 1 & 1  & -1 \\
\hline
$3C_2$ & 1 & -1  & 0 \\
\hline
\end{tabular}
\label{characterTableS3}
\end{table}

One can apply Eq. (\ref{salc}) in order to find the S.A.L.C., but, sometimes, it turns out to be more illuminating to investigate directly the irreps.
Concretely, if one takes the defining $3$-dim representation of the permutations acting on ($x_1,x_2,x_3$), then one can check that its character values for the three classes ($1C_1, 2C_3, 3C_2$) are respectively ($3,0,1 $) which, by applying (Eq. \ref{decomposition}), gives (${\bf 3}= {\bf 1} \oplus {\bf 2}$). Now, one can see directly that the combination $W \equiv (x_1+x_2+x_3)/\sqrt{3}$ is invariant under all permutations expressing, thus, the irrep ${\bf 1}$. The corresponding orthogonal subspace, spanned by, say, ($V\equiv (x_3-x_2)/\sqrt{2}$, and $W \equiv (x_2+x_3-2x_1)/\sqrt{3}$) is also invariant under the action of the permutations representation,  which give the S.A.L.C. for the irrep ${\bf 2}$. The symmetry group $S_3$ is generated by two elements, like $(A, C)$ or $(B,F)$. In the space $\langle V,W\rangle$ spanned by the basis ($V,W$), we have \bea A=\left(\begin{array}{cc} -1&0\\0&1 \end{array} \right) &,& C=\frac{1}{2}\left(\begin{array}{cc}1 & \sqrt{3} \\ \sqrt{3}& -1 \end{array} \right).
\eea
We see here the advantage of taking ($A,C$) as generators since both belong to the same conjugacy class, having thus common character, and that $A$ is diagonal in the basis ($V,W$), which makes its action evident. For example, if we take two defining irreps on $\langle V,W\rangle$ with four linear combinations:
\bea
\left(x_1,x_2\right)^T \sim {\bf 2} &,& \left(y_1,y_2\right)^T \sim {\bf 2}: \nn \\
L_1=(x_1y_1 + x_2y_2), L_2=(x_1y_1 - x_2y_2) &,& L_3=(x_1y_2 + x_2y_1), L_4=(x_1y_2 - x_2y_1) \label{2x2-S3-four-combinations}
\eea then we have
\bea
\left(L_1 \overset{A,C}{\rightarrow} L_1\right) &\Rightarrow& L_1 \sim {\bf 1}   \label{S3-2x2 -includes-1} ,\nn\\
\left( L_4 \overset{A,C}{\rightarrow} -L_4\right)  &\Rightarrow& L_4 \sim {\bf 1'}    \label{S3-2x2 -includes-1'} ,\nn\\
\left( L_3, L_2\right)^T \overset{ A(C)}{\rightarrow}  A(C) \left( L_3, L_2\right)^T &\Rightarrow& \left( L_3, L_2\right)^T \sim {\bf 2}  \label{S3-2x2 -includes-2} .
\eea

Moreover, if we assume $y' \sim {\bf 1'}$ then it is immediate to check that $\left( y'x_2, -y'x_1 \right)^T \sim {\bf 2}$ so we get  $({\bf 1'} \otimes {\bf 2} = {\bf 2})$. Similarly, one checks that $({\bf 1} \otimes {\bf 1} \sim {\bf 1} , {\bf 1'} \otimes {\bf 1'} \sim {\bf 1} , {\bf 1'} \otimes {\bf 1} \sim {\bf 1'} , {\bf 1} \otimes {\bf 2} \sim {\bf 2} )$.

One could look at the basis $(X, Y, Z)$ as resulting from applying onto the canonical basis a similarity transformation defined by the unitary matrix $U$:
\bea
U&=& \left(\begin{array}{ccc}\frac{1}{\sqrt{3}}&0&\frac{\sqrt{2}}{\sqrt{3}}\\ \frac{1}{\sqrt{3}}&\frac{1}{\sqrt{2}}&\frac{-1}{\sqrt{6}} \\ \frac{1}{\sqrt{3}}&\frac{-1}{\sqrt{2}}&\frac{-1}{\sqrt{6}} \end{array}\right)\label{U}
,\nn \\
A^{\mbox{can}}=(23)=\left(\begin{array}{ccc}1&0&0\\0&0&1 \\ 0&1&0 \end{array}\right) &\Rightarrow&A=  U^\dagger .A^{\mbox{can}}. U= \left( \begin{array}{ccc}1&0&0\\ 0&-1&0 \\ 0&0&1 \end{array} \right)
, \nn \\
C^{\mbox{can}}= (12)=\left(\begin{array}{ccc}0&1&0\\1&0&0 \\ 0&0&1 \end{array}\right) &\Rightarrow&C =  U^\dagger .C^{\mbox{can}}. U= \left(\begin{array}{ccc}1&0&0\\ 0&\frac{1}{2}& \frac{\sqrt{3}}{2} \\ 0&\frac{\sqrt{3}}{2}&\frac{-1}{2} \end{array}\right),
\eea
which shows explicitly that ${\bf 3} = {\bf 1} \oplus {\bf 2}$. Actually, one can look at $(U^\dagger g U)_{ij}$ as the inner product of the $i$-th and $j$-th columns of the matrix $U$ using $g$ as metric. Another common similarity transformation, when the generators are taken as ($B, F$) is given by $U_\omega$:
 \bea
U_\omega&=& \frac{1}{\sqrt{3}}\left(\begin{array}{ccc}1&1&1\\ 1&\omega&\omega^2 \\ 1&\omega^2&\omega \end{array}\right): \omega = e^{2i\pi/3}\label{U_omega} ,\nn\\
B^{\mbox{can}}=(13)=\left(\begin{array}{ccc}0&0&1\\0&1&0 \\ 1&0&0 \end{array}\right) &\Rightarrow&B= U_\omega^\dagger .B^{\mbox{can}}. U_\omega= \left( \begin{array}{ccc}1&0&0\\ 0&0&\omega \\ 0&\omega^2&0 \end{array} \right), \nn \\
F^{\mbox{can}}= (123)=\left(\begin{array}{ccc}0&1&0\\0&0&1 \\ 1&0&0 \end{array}\right) &\Rightarrow& F= U_\omega^\dagger .F^{\mbox{can}}. U_\omega= \left(\begin{array}{ccc}1&0&0\\ 0&\omega& 0 \\ 0&0&\omega^2 \end{array}\right).
\eea
One notes again that, in this basis, the decomposition ${\bf 3} = {\bf 1} \oplus {\bf 2}$ is explicit. Moreover, since $F$ is diagonal then its action on any defining representation is easy to compute. Concretely, we have
\bea
\left(t_1,t_2\right)^T \sim {\bf 2}, t=x,y &\Rightarrow& \nn \\
\left(x_1y_2+x_2y_1 \overset{B,F}{\rightarrow} x_1y_2+x_2y_1\right) &\Rightarrow& x_1y_2+x_2y_1 \sim {\bf 1}   ,\nn \\
\left( x_1y_2-x_2y_1 \overset{B(F)}{\rightarrow} -(+)(x_1y_2-x_2y_1)\right)  &\Rightarrow& x_1y_2-x_2y_1 \sim {\bf 1'}    ,\nn  \\
\left( x_2y_2, x_1y_1\right)^T \overset{ B(F)}{\rightarrow}  B(F) \left( x_2y_2, x_1y_1\right)^T &\Rightarrow& \left( x_2y_2, x_1y_1\right)^T \sim {\bf 2}  \label{complex-S3-2x2 -includes-2} ,
\eea
showing ${\bf 2} \otimes {\bf 2}= {\bf 1} \oplus {\bf 1'} \oplus {\bf  2}$.

It is easier in this `complex' basis to find rules involving conjugate irreps. For example,
\bea
\left(x^*_1,x^*_2\right)^T \sim {\bf 2^*}, \left(y_1,y_2\right)^T \sim {\bf 2} &\Rightarrow& \nn \\
\left(x^*_1y_1+x^*_2y_2 \overset{B,F}{\rightarrow} x^*_1y_1+x^*_2y_2\right) &\Rightarrow& x^*_1y_1+x^*_2y_2 \sim {\bf 1}   , \nn \\
\left(x^*_1y_1-x^*_2y_2 \overset{B(F)}{\rightarrow} -(+)(x^*_1y_1-x^*_2y_2)\right) &\Rightarrow& x^*_1y_1-x^*_2y_2 \sim {\bf 1'}    ,\nn \\
\left( x^*_1y_2, x^*_2y_1 \right)^T \overset{ B(F)}{\Rightarrow} B(F) \left( x^*_1y_2, x^*_2y_1\right)^T &\Rightarrow& \left( x^*_1y_2, x^*_2y_1\right)^T \sim {\bf 2},
\label{complex-S3-2x2 -includes-2}
\eea
showing ${\bf 2^*} \otimes {\bf 2} = {\bf 1} \oplus {\bf 1'} \oplus {\bf 2}$.

\subsection{$S_4$}
\label{appendix-s4}
In terms of cycles' notation, we have the 24-elements symmetry group of order $4$: $S_4 = \{a_1=e,a_2=(12)(34), a_3=(13)(24), a_4=(14)(23), b_1=(243), b_2=(142), b_3=(123), b_4=(134), c_1=(234), c_2=(132), c_3=(143), c_4=(124), d_1 =(34), d_2=(12), d_3=(1423), d_4=(1324), e_1=(23), e_2=(1342), e_3=(1243), e_4=(14), f_1=(24), f_2=(1432), f_3=(13), f_4=(1234)\}$ which can be divided into five classes\footnote{One can find the order of a product of disjoint cycles as being equal to the least common multiplier of the cardinalities of cycles' supports}
\bea & 1C_1=\{e\}, 3C_2=\{a_2, a_3, a_4\}, 6C_2=\{d_1, d_2, e_1, e_4, f_1, f_3\},  \nn \\  &  8C_3=\{b_1, b_2, b_3, b_4, c_1, c_2, c_3, c_4\},
 6C_4=\{d_3, d_4, e_2, e_3, f_2, f_4\},\eea so we have five unitary inequivalent irreps, and by applying Eqs. (\ref{irreps-count}):
\bea \sum_{n\in N}m_n=5, \sum_{n\in N}m_nn^2=24 &\Rightarrow&  m_1=2, m_2=1, m_3=2\eea
Applying Eqs. (\ref{orthogonality}, \ref{orthogonality_tr}), we have the character table of $S_4$ (Table \ref{characterTableS4}).
\begin{table}[h]
\caption{Character table of $S_4$}
\centering
\begin{tabular}{cccccc}
\hline
\hline
classes$/$irreps & $\chi_{\bf 1}$ & $\chi_{\bf 1'}$ & $\chi_{\bf 2}$ & $\chi_{\bf 3}$ & $\chi_{\bf 3'}$  \\
\hline
\hline
$1C_1$ & 1 &1 & 2 & 3 & 3 \\
\hline
$3C_2$ & 1 & 1  & 2 & -1 & -1 \\
\hline
$6C_2$ & 1 & -1  & 0 & 1 & -1 \\
\hline
$6C_4$ & 1 & -1  & 0 & -1 & 1 \\
\hline
$8C_3$ & 1 & 1  & -1 & 0 & 0 \\
\hline
\end{tabular}
\label{characterTableS4}
\end{table}
It has two generators, and actually it can be defined as :
\bea
S_4 &=& \langle D, B: D^4=B^3=1,DB^2D=B\rangle ,\nn \\
&=& \langle T, S: T^4=S^2=(ST)^3=1 \rangle,
\eea
with the first (second) definition leading to $DBD=BD^2B$ ($(TS)^3=1$). One can take ($D=d_4, B= b_1$) for the first set of generators, or ($T=D, S=BD^{-1}$) for the second set.
In the canonical basis ($x_1, x_2, x_3, x_4$), the linear combination ($x_1+x_2+x_3+x_4$) is invariant under the action of the permutations representation. Thus, the orthogonal subsapce spanned by
\bea
\label{A-space}
\left( \begin{array}{c} A_x \\ A_y\\ A_z\end{array} \right) &=& \left( \begin{array}{c} x_1 +x_2-x_3-x_4 \\ x_1 -x_2 + x_3 -x_4 \\ x_1 -x_2 -x_3 + x_4\end{array}\right)
\eea
is also invariant. The restriction of the permutations representation onto the $3$-dim $A$-space is the ${\bf 3}$ irrep given, in this $A$-basis, by:
\bea
\label{A-basis-3}
b_1^{\mbox{\tiny can.}} = \left(\begin{array}{cccc} 1&0&0&0\\ 0&0&0&1\\0&1&0&1 \\0&0&1&0 \end{array}\right) &\Rightarrow& b_1^{\mbox{\tiny A}} = \left(\begin{array}{ccc} 0&0&1\\ 1&0&0\\0&1&0 \end{array}\right) \\
d_4^{\mbox{\tiny can.}} = \left(\begin{array}{cccc} 0&0&1&0\\ 0&0&0&1\\0&1&0&0\\ 1&0&0&0 \end{array}\right) &\Rightarrow& d_4^{\mbox{\tiny A}} = \left(\begin{array}{ccc} -1&0&0\\ 0&0&-1\\0&1&0 \end{array}\right),
\eea
whereas the ${\bf 3'}$ irrep is given, again in this $A$-basis, by:
\bea
 {b'}_1^{\mbox{\tiny A}} = \left(\begin{array}{ccc} 0&0&1\\ 1&0&0\\0&1&0 \end{array}\right) &,&
{d'}_4^{\mbox{\tiny A}} = \left(\begin{array}{ccc} 1&0&0\\ 0&0&1\\0&-1&0 \end{array}\right) \label{A-basis-3'}, \\
\eea
and the ${\bf 2}$ irrep is:
\bea
 {b''}_1^{\mbox{\tiny A}} = \left(\begin{array}{cc} \omega&0\\ 0&\omega^2 \end{array}\right) &,&
{d''}_4^{\mbox{\tiny A}} = \left(\begin{array}{cc} 0&1\\ 1&0 \end{array}\right) \label{A-basis-2},
\eea
where $\omega = e^{i2\pi/3}$, and one can compute the corresponding $T^{\mbox{\tiny(','')A}} =d_4^{\mbox{\tiny(','')A}}, S^{\mbox{\tiny(','')A}}: S^{\mbox{\tiny(','')A}}T^{\mbox{ \tiny(','')A}}=b_1^{\mbox{\tiny(','')A}}$ in these irreps.

Another common basis is the $\tilde{B}$-basis given by the unitary similarity matrices $U_{\mbox{\tiny doublet}}, U_{\mbox{\tiny triplet}}$:
\bea
\label{s4-from-A-to-tilde}
U_{\mbox{\tiny doublet}}= \frac{1}{\sqrt{2}} \left( \begin{array}{cc}1&i\\1&-i\end{array}\right) &,&
U_{\mbox{\tiny triplet}}= \frac{1}{\sqrt{2}} \left( \begin{array}{ccc}\sqrt{2}&0&0\\0&1&1 \\ 0& i& -i\end{array}\right),
\eea
so we have:
\bea
\tilde{b}_1   =U_{\mbox{\tiny triplet}}^\dagger b_1^{\mbox{\tiny  A}} U_{\mbox{\tiny triplet}} =\left( \begin{array}{ccc}0 &\frac{i}{\sqrt{2}}&\frac{-i}{\sqrt{2}} \\ \frac{1}{\sqrt{2}}&\frac{-i}{2}&\frac{-i}{2}\\ \frac{1}{\sqrt{2}}&\frac{i}{2}& \frac{i}{2}\end{array}\right) &,&
\tilde{d}_4   =U_{\mbox{\tiny triplet}}^\dagger d_4^{\mbox{\tiny  A}} U_{\mbox{\tiny triplet}} =\mbox{diag}\left(-1,-i,i\right) ,
\nn \\
\tilde{b'}_1   =U_{\mbox{\tiny triplet}}^\dagger {b'}_1^{\mbox{\tiny  A}} U_{\mbox{\tiny triplet}} =\left( \begin{array}{ccc} 0&\frac{i}{\sqrt{2}}&\frac{-i}{\sqrt{2}} \\ \frac{1}{\sqrt{2}}&\frac{-i}{2}&\frac{-i}{2} \\ \frac{1}{\sqrt{2}}&\frac{i}{2}&\frac{i}{2} \end{array}\right) &,&
\tilde{d'}_4   =U_{\mbox{\tiny triplet}}^\dagger {d'}_4^{\mbox{\tiny  A}} U_{\mbox{\tiny triplet}} =\mbox{diag}\left(1,i,-i\right) ,\nn \\
\tilde{b''}_1   =U_{\mbox{\tiny doublet}}^\dagger {b''}_1^{\mbox{\tiny  A}} U_{\mbox{\tiny doublet}} =\frac{1}{2}\left( \begin{array}{ccc} -1&-\sqrt{3} \\ \sqrt{3}&-1 \end{array}\right) &,&
\tilde{d''}_4   =U_{\mbox{\tiny doublet}}^\dagger {d''}_4^{\mbox{\tiny  A}} U_{\mbox{\tiny doublet}} =\mbox{diag}\left(1,-1\right) ,\label{s4-tilde-basis-2}
\eea
and one can compute $\tilde{T},\tilde{T'},\tilde{T''},\tilde{S}, \tilde{S'}, \tilde{S''}$.

One can find the multiplication rules, as was done in the previous subsection in any basis. However, we refer the reader to \cite{Kobayashi:2022moq} for the $\tilde{B}$-basis rules, whereas we state explicitly in subsubsection (\ref{subsubsection-s4-bases}) the corresponding rules in the $B$-basis adopted to define the texture and the matter fields symmetry assignments.

\bibliographystyle{unsrt}

\end{document}